\documentclass[11pt,a4paper]{article}
\pdfoutput=1
\usepackage{amsmath}
\usepackage[left=2cm,right=2cm,top=2cm,bottom=2cm]{geometry}
\usepackage{linearA}
\usepackage{subfigure}
\usepackage{graphics}
\usepackage{float}
\usepackage{xcolor}

\usepackage{jcappub}
\allowdisplaybreaks

\def\lsim{~\rlap{$<$}{\lower 1.0ex\hbox{$\sim$}}}
\def\bsim{~\rlap{$>$}{\lower 1.0ex\hbox{$\sim$}}}

\def\hmsun{\ {\rm M_\odot/{\it h}}}

\def\ln{{\rm ln}}

\def\det{{\rm det}}

\def\mathbi#1{\textbf{\em #1}}

\def\vk{\mathrm{\bf k}}

\def\vq{\mathrm{\bf q}}

\def\vx{\mathrm{\bf x}}
\def\vy{\mathrm{\bf y}}

\newcommand{\kL}{k_{\mathrm{l}}}
\newcommand{\ks}{k_{\mathrm{s}}}

\newcommand{\vkL}{{\bf k_{\mathrm{l}}}}
\newcommand{\vks}{{\bf k_{\mathrm{s}}}}
\newcommand{\veta}{\boldsymbol{\eta}}

\newcommand{\esp}{\textnormal{\small \textsc{esp}}}

\def\npk{n_{\rm pk}}
\def\nesp{n_\esp}

\def\bnesp{\bar{n}_\esp}

\def\desp{\delta_\esp}

\def\be{\begin{equation}}
\def\ee{\end{equation}}
\def\bea{\begin{eqnarray}}
\def\eea{\end{eqnarray}}
\def\ba{\begin{align}}

\def\bi{\begin{itemize}}
\def\ei{\end{itemize}}

\newcommand{\Ddel}{\delta_{\rm D} }
\newcommand{\mb}{\mathbf}
\newcommand{\fnl}{ f_{\rm NL} }
\newcommand{\MpcOh}{ \,  \mathrm{Mpc}  \, h^{-1} }
\newcommand{\hOMpc}{ \,  \mathrm{Mpc}^{-1}  \, h  }
\newcommand{\nn}{ \nonumber }
\newcommand{\Msun}{ \,   M_{\odot}  {h}^{-1}   }
\newcommand{\beq}{\begin{equation}}
\newcommand{\eeq}{\end{equation}}
\newcommand{\beqa}{\begin{eqnarray}}
\newcommand{\eeqa}{\end{eqnarray}}

\title{Squeezing the halo bispectrum: a test of bias models}

\author[a]{Azadeh Moradinezhad Dizgah,}
\author[b,a]{Kwan Chuen Chan,}
\author[a]{Jorge Nore\~na,}
\author[a]{Matteo Biagetti}
\author[a]{and Vincent Desjacques}

\affiliation[a]{D\'epartement de Physique Th\'eorique
and 
Center for Astroparticle Physics, 
Universit\'e de Gen\`eve, CH-1211 Gen\`eve, Switzerland}
\affiliation[b]{Institut de Ci\`encies de l'Espai, IEEC-CSIC, Campus UAB,
Carrer de Can Magrans, s/n, 08193 Bellaterra, Barcelona, Spain}

\emailAdd{azadeh.moradinezhad@unige.ch, chan@ice.cat, jorge.norena@unige.ch}

\abstract{
We study the halo-matter cross bispectrum in the presence of primordial non-Gaussianity of the local type.
We restrict ourselves to the squeezed limit, for which the calculation are straightforward, and perform
the measurements in the initial conditions of N-body simulations, to mitigate the contamination induced
by nonlinear gravitational evolution. Interestingly, the halo-matter cross bispectrum is not trivial even 
in this simple limit as it is strongly sensitive to the scale-dependence of the quadratic and third-order 
halo bias. Therefore, it can be used to test biasing prescriptions. 
We consider three different prescription for halo clustering: excursion set peaks (ESP), local bias and 
a model in which the halo bias parameters are explicitly derived from a peak-background split.
In all cases, the model parameters are fully constrained with statistics other than the cross bispectrum.
We measure the cross bispectrum involving one halo fluctuation field and two mass overdensity fields for
various halo masses and collapse redshifts. We find that the ESP is in reasonably good agreement with the
numerical data, while the other alternatives we consider fail in various cases. 
This suggests that the scale-dependence of halo bias also is a crucial ingredient to the squeezed limit 
of the halo bispectrum.
}

\begin{document}

\maketitle
\flushbottom

\setcounter{footnote}{0}

\section{Introduction}

Cosmological observations have the potential to test fundamental physics beyond what is accessible with 
laboratories on Earth. In particular, cosmological perturbations are believed to have been seeded during 
an inflationary phase which may have occurred at an energy scales potentially as high as $10^{14}$ GeV.
The measurement of correlation functions beyond the two-point function can teach us about the interaction 
of the inflaton and the field content of the universe during that period by constraining the primordial 
non-Gaussianity (PNG). 
The Planck satellite \cite{Ade:2015ava} has already put constraints on PNG, but there is still much space 
for interesting phenomenology, especially if we can access the regime where the amplitude of the PNG is 
an order of magnitude smaller than the current limits. 
 
Forthcoming surveys of the large scale structure (LSS) of the Universe are one of our best hopes for 
improving the current cosmic microwave background (CMB) limits on PNG. 
Much effort has already been devoted to constrain PNG from a scale dependence in the galaxy bias 
\cite{Dalal:2007cu, Matarrese:2008nc, Slosar:2008hx}. 
Current LSS limits are at the level of the CMB pre-Planck constraints 
\cite{Giannantonio:2013uqa, Leistedt:2014zqa}, 
and they shall improve by 1 - 2 orders of magnitude with the advent of large redshift surveys
\cite{Agarwal:2013qta,dePutter:2014lna,Raccanelli:2014kga,Camera:2014bwa,Alonso:2015sfa,Raccanelli:2015oma,
Dore:2014cca}.

If inflation generated a physical coupling between short and long-wavelength perturbations of the gravitational 
potential, this would induce a characteristic scale dependence on the halo bias that cannot be mimicked by 
astrophysical effects, since the latter do not generate a mode coupling in the gravitational potential. 
Single-field models of inflation predict that this effect is absent
\cite{Maldacena:2002vr, Creminelli:2004yq,Creminelli:2011rh,Creminelli:2012ed}. 
However, they could generate a potentially large PNG of the equilateral type for instance \cite{Babich:2004gb} 
which would not show up in the $k$-dependence of the large scale galaxy power spectrum, yet leave an imprint
in the galaxy bispectrum.
Clearly, higher order clustering statistics such as galaxy bispectrum or 3-point function, which has recently
been measured in
\cite{Gaztanaga:2008sq, McBride:2010zn, Marin:2013bbb, Gil-Marin:2014sta, Slepian:2015hca},
contain additional information on PNG. 
Therefore, they are natural observables to constrain different PNG shapes, while they also provide a 
consistency check for the constraints obtained with the power spectrum 
\cite{Scoccimarro:2003wn, Sefusatti:2007ih, Jeong:2009vd, Pollack:2011xp,Schmittfull:2012hq, Figueroa:2012ws, 
Sefusatti:2012ye, Tasinato:2013vna, Baghram:2013lxa, Saito:2014qha, Byun:2014cea}.

However, modelling the scale and shape dependence of the galaxy bispectrum is very challenging
(see e.g. \cite{Nishimichi:2006vn, Nishimichi:2009fs, Sefusatti:2010ee, Baldauf:2010vn, Sefusatti:2011gt, Tellarini:2015faa, Lazanu:2015rta} 
for recent attempts in the context of PNG).
Any non-linearity in the description of the galaxy number over-density induces non-Gaussianity in its distribution. 
One important source of such non-linearity arises from biasing, i.e. the fact that galaxies do not follow the dark 
matter (DM) distribution perfectly. Furthermore, the nonlinear bias of LSS tracers generates scale-dependence and 
stochasticity which complicate the interpretation of the measurements. 
An accurate understanding of galaxy bias will therefore be necessary in order to hunt for PNG signatures in LSS
data.

In this work, we will study the distribution of halos rather than galaxies since the former are more amenable to 
analytic modelling. Furthermore, halo clustering can be easily simulated with low computational cost.
In general, the formation of a halo depends on several variables, and not only the density.  
Therefore, any realistic model should take this into account. 
Here, we will use the excursion set peaks (ESP) approach to halo clustering \cite{Paranjape:2012jt, Desjacques:2012eb},
which we briefly summarise in Section \ref{sec:esp} and Appendix \ref{app:esp}.
In this approach, the Lagrangian position of each halo is approximated as the position of a peak of the initial density field. 
To illustrate the complications brought by the scale-dependence of halo bias, we will focus on the squeezed limit 
of the halo-matter cross-bispectrum $B_\text{hmm}$ in the presence of a local-type PNG. 
Moreover, we will present measurements at the initial conditions only, to avoid the contamination induced by 
nonlinear clustering. 
The squeezed limit is particularly convenient because expressions greatly simplify, and the calculation becomes 
very tractable. 
Surprisingly, we find that is not trivial to get a good agreement with the data even in this simple limit, 
which can thus be used to discriminate between different models of halo bias. 

Using the ESP and the formalism of integrated perturbation theory \cite{Matsubara:2011ck}, we derive expressions 
for the halo bispectrum in Lagrangian space $B_\text{hhh}$ and the bispectrum involving a halo and two matter 
over-density modes $B_\text{hmm}$, including the contribution of PNG. 
This calculation is presented in Section \ref{sec:bispec}, with details left to Appendix \ref{app:bispectra}. 
We compare our theoretical predictions to a measurements of $B_\text{hmm}$ in the initial conditions of a series 
of N-body simulation with a large local-type PNG. These are described in Section \ref{sec:sims}. 
In the squeezed limit, the signature of PNG is clean and very sensitive to the second- and third-order biasing. 
We find the measurements to be in reasonably good agreement with the ESP, while a simplistic local bias model 
fails to even qualitatively describe the simulations. 
For comparison, we also use a model in which the bias parameters are explicitly derived from a peak-background 
split, as described in Appendix \ref{sec:ESbias_PBS}. This model works better than local biasing, but does not 
capture the correct behaviour of the bispectrum when the wavemode corresponding to the matter fluctuation field
is squeezed. 
Our conclusions and suggestions for extending this work are summarised in Section \ref{sec:conclusions}.

\section{Excursion set peaks: a proxy for  dark matter halos}
\label{sec:esp}

We consider initial density peaks as a proxy for the formation sites of dark matter (DM) halos, assuming that 
the virialization proceeds according to the usual spherical collapse prescription. 
This indeed is a good approximation for halos with mass $M\gtrsim M_\ast$ \cite{Ludlow:2010xd}, where 
$M_\ast$ is the typical mass of halos collapsing at redshift $z_\ast$.
As a result, the clustering of the density peaks is fully specified by the value of $\delta_c$ and the halo mass 
$M$, which sets the scale $R$ at which the density field should be filtered. To ensure that no halo forms inside 
a bigger halo (and, thus, avoid the cloud-in-cloud problem), another constraint is added on the slope of the 
linear, smoothed density field $\delta_R$ w.r.t. the filter scale.
Henceforth, we shall refer to the Lagrangian patches that collapse into DM halos as proto-halos.

\subsection{Halo bias as constraints in the initial conditions}\label{sec:PC}

Let $\delta_L$ be the initial density field linearly extrapolated to the collapse redshift $z_\ast$ of the halos. 
For conciseness, we will omit the explicit dependence of $\delta_L$ and related statistics on $z_\ast$ throughout
Sec.~\S\ref{sec:esp}.
Furthermore, let $\delta_R\equiv  W_R \star  \delta_L $ be the initial density field convolved with a window $W_R$.
Excursion set peaks are easily defined in terms of the normalised variables
\bea
\nu(\vx) &\equiv& \frac{1}{\sigma_0} \delta_R(\vx) \\
\eta_i(\vx) &\equiv&  \frac{1}{\sigma_1} \partial_i\delta_R(\vx) \\
\zeta_{ij}(\vx) &\equiv& \frac{1}{\sigma_2}\partial_{ij} \delta_R(\vx) \\
\mu(\vx) &\equiv & -\frac{1}{\sigma_\mu} \frac{d\delta_R}{dR}(\vx) \;,
\eea
The $\sigma_i$ are the spectral moments
\be
\sigma_n^2(R) \equiv \frac{1}{2\pi^2} \int_0^\infty \!\! dk \, k^{2(n+1)} P_0(k) \tilde{W}_R^2(k)\;,
\ee
where $\tilde{W}_R$ is the Fourier transform of the spherically symmetric filter $W_R$, $P_0(k)$ is the power 
spectrum of the initial density field linearly extrapolated to $z_\ast$, and
\begin{equation}
\sigma_\mu^2(R)\equiv \frac{1}{2\pi^2}\int_0^\infty\!\!dk\,k^2 P_0(k) \Big(\frac{d\tilde{W}}{dR}(k)\Big)^2
\end{equation}
is the variance of the fluctuation field $\mu(\vx)$.
Invariance under rotations implies that the peak clustering depends only on the scalar functions $\nu(\vx)$, 
$\mu(\vx)$, $J_1(\vx)=-{\rm tr}\big(\zeta_{ij}(\vx)\big)$,
the chi-square quantity $\eta^2(\vx)=\sum_i \eta_i^2(\vx)$, and the jointly distributed variables
\begin{equation}
J_2(\vx) = \frac{3}{2}{\rm tr}\big(\bar{\zeta}_{ij}^2(\vx)\big) \qquad
J_3(\vx) = \frac{9}{2}{\rm tr}\big(\bar{\zeta}_{ij}^3(\vx)\big) \;.
\end{equation}
Here, $\bar{\zeta}_{ij}$ is the traceless part of $\zeta_{ij}$. The density peak constraint translates into
\begin{gather}
\label{eq:peak}
\nu(\vx)=\nu_c \;, \qquad \eta^2(\vx) = 0 \;, \qquad  J_1(\vx)>0 \\
J_2(\vx)<J_1^2(\vx) \;,\qquad -1 < x_3 < {\rm min}\Big[1,(y/2)(y^2-3)\Big]  
\nonumber \;,
\end{gather}
where $x_3\equiv J_3/J_2^{3/2}$ and $y\equiv J_1/J_2^{1/2}$ \cite{Lazeyras:2015giz}.
The peak height is $\nu_c=\delta_c/\sigma_0$, with $\delta_c=1.68$ being the critical threshold for spherical
collapse.
The first-crossing condition is imposed through the requirement that density peaks on a given smoothing scale 
$R$ are counted only if the conditions $\delta(R) > B(R)$ and $\delta(R+\Delta R) < B(R)$ are satisfied
\citep{Bond:1989fh, appel:1990, Bond:1993we, Paranjape:2012jt}. Here, $B(R)$ is the effective barrier for 
collapse. Following \cite{Biagetti:2013hfa}, we will assume that each halo ``sees''  a constant (flat) barrier
with a value $B(R)$ varying from halo to halo. Therefore, we implement the first-crossing condition as
\begin{equation}
\label{eq:crossing}
\mu > 0 \;.
\end{equation}
The constraints Eqs. (\ref{eq:peak}) and (\ref{eq:crossing}) define the excursion set peaks. To incorporate the
triaxiality of halo collapse, we do not assume a flat barrier $B(R)=\delta_c$ but, rather, a stochastic 
barrier distributed around a mean value increasing with decreasing halo mass (note that this differs from the diffusing barrier approach of \cite{Maggiore:2009rw}). In practice, we consider a 
square-root stochastic barrier
\be
\label{eq:barrier}
B(R)= \delta_c + \beta \sigma_0 \;,
\ee
wherein the stochastic variable $\beta$ closely follows a lognormal distribution with mean and variance
\be
\big\langle\beta\big\rangle = 0.5 \;,\qquad {\rm Var}(\beta)=0.25 \;.
\ee
This furnishes a good description of the critical collapse threshold of actual halos \citep{Robertson:2008jr} 
(once it is implemented in the peak approach) and the halo mass function and biases \cite{Paranjape:2012jt}.
Therefore, $B(R)$ is the only ingredient of our approach that is not derived from first principles. However,
once the values of $\big\langle\beta\big\rangle$ and Var$(\beta)$ are fixed, there is no free parameter left 
in the model.

The excursion set peak constraint defines a ``localised'' number density $\nesp$ which selects the Lagrangian  
points that correspond to the position of proto-halo centres,
\be
\label{eq:nesp}
\nesp(\vy) = -\frac{\mu}{\nu\sigma_0'} \theta_H(\mu)\,\npk(\vy) \;,
\ee
where the prime designates a derivative w.r.t. $R$, and $\npk(\vy)$ is the usual BBKS peak number density 
\cite{Bardeen:1985tr},
\be
\label{eq:npk}
\npk(\vy) = \frac{3^{3/2}}{R_\star^3} \left|\det\zeta_{ij}\right| 
\delta_D\!\left(\pmb{\eta}\right) \theta_H\!\left(\lambda_3\right) 
\delta_D\!\left(\nu-\nu_c\right)
\ee
Here, $\mathbi{y}=(\nu,J_1,\mu,3\eta^2,5J_2,J_3)$ and $R_\star=\sqrt{3}\sigma_1/\sigma_2$ is the characteristic 
radius of a density maximum.

\subsection{Bias factors of excursion set peaks}\label{sec:biases}

The bias coefficients of excursion set peaks are given by (1-point) ensemble average of derivative operators,
or, equivalently, of orthonormal polynomials. The type of these polynomials depend on the nature of the
variables. Namely, the scalars $\nu$, $\mu$ and $J_1$ are associated with Hermite polynomials while the chi-square 
variable $\eta^2$ corresponds to Laguerre polynomials $L_q^{(1/2)}(3\eta^2/2)$. The jointly distributed 
scalars $(J_2,J_3)$ generate the polynomials $F_{\ell m}(5 J_2,J_3)$ 
\cite{Desjacques:2012eb,Lazeyras:2015giz}. 
For the variables $(\nu,J_1,\mu)$, the bias factors are
\begin{align}
\sigma^i_0 \sigma^j_2 \sigma_\mu^k b_{ijk} 
&= \frac{1}{\bnesp}
\left\langle\frac{\partial^{i+j+k}\nesp}{\partial\nu^i\partial J_1^j\partial \mu^k}\right\rangle \\
&= \frac{1}{\bnesp} \int\! d\vy\,\nesp(\vy)\, H_{ijk}(\nu,J_1,\mu) P(\vy) \nonumber \;.
\end{align}
Here, $H_{ijk}$ are tri-variate Hermite polynomials.
Since these derivatives can be re-summed into a shift operators, the $b_{ijk}$ can also be written as
\be
\sigma^i_0\sigma^j_2 \sigma_\mu^k b_{ijk} = \frac{1}{\bnesp}
\frac{\partial^{i+j+k}\bnesp}{\partial\nu_l^i\partial J_{1l}^j\partial \mu_l^k} \;,
\ee
where it is understood that the long-mode perturbations $\nu_l$, $u_l$ and $\mu_l$ shift the mean of the 1-point 
PDF ${\cal N}(\nu,J_1,\mu)$, where ${\cal N}$ is a Normal distribution. Similar expressions arise for $\eta^2$,
$J_2$ and $J_3$. Namely, the bias factors $\chi_k$ associated with $\eta^2$ are defined as
\begin{align}
\sigma_1^{2q} \chi_q
&= \frac{(-1)^q}{\bnesp} \int\! d\vy\,\nesp(\vy)\, L_q^{(1/2)}\!(3\eta^2/2) P(\vy) \\
&= \frac{1}{\bnesp}\frac{\partial^q\bnesp}{\partial(\eta_l^2)^q} \nonumber
\end{align}
where, owing to rotational symmetry, derivatives are taken w.r.t. the modulus squared $\eta_l^2=\veta_l\cdot\veta_l$ 
of the long-wavelength perturbation $\veta_l=(\eta_{l1},\eta_{l2},\eta_{l3})$. 
As shown in \cite{Desjacques:2012eb}, the long mode perturbation $\veta_l$ shift the chi-square distribution for 
$\veta(\vx)$ into a non-central chi-square PDF that can be expanded in Laguerre polynomials $L_q^{(\alpha)}$ with
$k=2(\alpha+1)=3$ degrees of freedom.
Finally, the bias $\omega_{\ell m}$ 
which correspond to $J_2$, $J_3$ are given by
\begin{align}
\sigma_2^{2\ell+3m}\omega_{lm} &= 
\frac{1}{\bnesp} \int\! d\vy\,\nesp(\vy)\, F_{\ell m}(5J_2,J_3) P(\vy) \\
&= \frac{1}{\bnesp}\frac{\partial^{\ell+m}\bnesp}{\partial J_{2l}^\ell\partial J_{3l}^m}\;, \nonumber
\end{align}
where $J_{2l}$ and $J_{3l}$ are long-wavelength perturbations to the second- and third-order invariant traces 
of $\bar{\zeta}_{ij}$. The polynomials $F_{\ell m}$ are given by
\begin{equation}
\label{eq:Flm}
F_{\ell m}(5J_2,J_3) 
= (-1)^\ell \sqrt{\frac{\Gamma(5/2)}{2^{3m}\Gamma(3m+5/2)}}\,L_l^{(3m+3/2)}\!(s/2)\,P_m(x_3)\;,
\end{equation}
where $s=5J_2$ and $P_m(x)$ are Legendre polynomials.
Again, the factor of $(-1)^\ell$ ensures that the term with highest power always has positive sign.
In general however, the bias coefficients do not factorise into a product of $b_{ijk}$,
$\chi_q$ and $\omega_{\ell m}$, and take the generic form
\be
c_{ijkq\ell m} = 
\frac{\Big\langle\nesp\, H_{ijk}(\nu,J_1,\mu) L_q^{(1/2)}\!(3\eta^2/2) F_{lm}(5J_2,J_3)\Big\rangle}
{\sigma_0^i\sigma_1^{2q}\sigma_2^{j+2l+3m}\sigma_\mu^k\,\bnesp} \;.
\ee
We refer the reader to \cite{Lazeyras:2015giz} for details about the construction of perturbative bias expansions.

\subsection{Perturbative bias expansion for excursion set peaks}\label{sec:perturb_bias}

The bias factors $c_{ijkq\ell m}$ are the coefficients of the ESP Lagrangian perturbative bias expansion $\desp^L(\vx)$. 
For the sake of completeness, $\desp^L(\vx)$ is explicitly given in Appendix \S\ref{app:esp} up to third order in the 
initial density field and its derivatives. Most importantly, the bias coefficients $c_{ijkq\ell m}$ also multiply 
orthonormal polynomials \cite{Desjacques:2012eb,Desjacques:2013qx,Lazeyras:2015giz}.
This series can be used to compute the whole hierarchy of $N$-point correlation functions of ESP in Lagrangian space.
Furthermore, it is also valid in the presence of (weak) primordial non-Gaussianity.

In Fourier space, this perturbative bias expansion takes the compact form
\begin{align}
\desp^L(\vx,z_\ast) &= \sum_{n=1}^\infty \frac{1}{n!}
\int\!\frac{d^3k_1}{(2\pi)^3}\dots\frac{d^3k_n}{(2\pi)^3}\,
c_n^L(\vk_1,\dots,\vk_n;z_\ast) \\
& \qquad \times \Big[\delta(\vk_1,z_\ast)\dots\delta(\vk_n,z_\ast)+\dots \Big]\, e^{i(\vk_1+\dots+\vk_n)\cdot\vx}
\nonumber \;,
\end{align}
where the dots stand for the terms of order $n-2$, $n-4$ etc. at each order $n$. They arise from the fact 
that the $c_{ijq\ell m}$ are the coefficients of an expansion in orthogonal polynomials. 
The $n$-order Lagrangian bias functions $c_n^L(\vk_1,\dots,\vk_n;z_\ast)$ sum over all the possible combinations 
of rotational invariants that can be generated from $n$ powers of the initial density field $\delta_0$ and/or its 
derivatives. We use the notation $c_n^L$ to emphasise that the Fourier space ESP bias factors correspond exactly 
to the renormalised Lagrangian bias functions of the ``integrated perturbation theory'' (iPT) 
\cite{Matsubara:2011ck,Matsubara:2013ofa},
\be
\label{eq:cdef}
\left < 
\frac{\delta^n\delta_\esp^L(\vk,z_\ast)}{\delta \delta_L(\vk_1,z_\ast)\dots\delta\delta_L(\vk_n,z_\ast)}
\right >  
= 
(2\pi)^{3-3n}\delta_D^3(\vk -\vk_{1...n})\ c_n^L(\vk_1,\dots,\vk_n; z_\ast) \;.
\ee
Here, $\delta_\esp^L(\vk,z_\ast)$ is the Fourier transform of the effective overabundance $\desp^L(\vx,z_\ast)$ of the 
biased tracers (the excursion set peaks) in Lagrangian space. 
We have momentarily re-introduced the explicit redshift dependence to remind the reader that, in this section, all 
quantities are evaluated at the virialization redshift $z_\ast$.
We also note the important caveat that the ESP constraint involves variables other than $\delta_L$, so that all the 
Lagrangian bias functions $c_n^L$ are scale-dependent.

The first order ESP Lagrangian bias function is
\begin{align}
c_1^L(k) &= \biggl(b_{100}+b_{010} k^2- b_{001}\frac{d\ln\tilde{W}_R}{dR}\biggr) \tilde{W}_R(k)
\end{align}
while, at second-order, we have
\begin{align}
c_2^L(\vk_1,\vk_2) &= \biggl\{b_{200} + b_{110} \left(k_1^2+k_2^2\right) 
+ b_{020} k_1^2 k_2^2 +b_{002}\frac{d\ln\tilde{W}_R}{dR}(k_1)\frac{d\ln\tilde{W}_R}{dR}(k_2) \\
&\qquad - b_{101} \bigg[\frac{d\ln\tilde{W}_R}{dR}(k_1)+\frac{d\ln\tilde{W}_R}{dR}(k_2)\bigg]
-b_{011}\bigg[k_1^2\frac{d\ln\tilde{W}_R}{dR}(k_2)+k_2^2\frac{d\ln\tilde{W}_R}{dR}(k_1)\bigg]
\nonumber \\
&\qquad -2 \chi_1 \left(\vk_1\cdot\vk_2\right)
+\omega_{10} \biggl[3\left(\vk_1\cdot\vk_2\right)^2 -k_1^2 k_2^2\biggr]\biggr\}\, 
\tilde{W}_R(k_1) \tilde{W}_R(k_2) \;. \nonumber
\end{align}
The third-order ESP Lagrangian bias function is spelt out in Appendix \S\ref{app:esp}.
At this point, we should stress that, while we have assumed a unique smoothing kernel $\tilde{W}$ for conciseness, 
the ESP implementation of \cite{Paranjape:2012jt} which we adopt here involves two different windows: a tophat 
filter for $\nu$ and $\mu$, and a Gaussian filter for $\eta_i$ and $\zeta_{ij}$. Therefore, the overall 
multiplicative factor of $\tilde{W}$ should be replaced by $\tilde{W}_T$ and $\tilde{W}_G$ wherever appropriate.
Note, however, that it should be possible to write down a model with a unique smoothing kernel which, if required, 
can be measured directly from simulations \cite{Dalal:2008zd,Baldauf:2014fza,Chan:2015zjt}.

\section{Halo bispectra in the presence of primordial non-Gaussianity}
\label{sec:bispec}

The main quantity of interest is the bispectrum $B_\text{hhh}$, i.e. the ensemble average of three halo number 
overdensity fields, since it is directly related to the bispectrum of galaxy number counts that can be 
extracted from galaxy survey data. However, there are some difficulties with this statistics: its measurement 
in simulations can be very noisy; it is affected by stochasticity; and we will see that, in the presence of 
PNG, there is no clear way to compute it from a Lagrangian bias expansion.
Therefore, we will instead focus on the bispectrum $B_\text{hmm}$ involving one halo number overdensity and two 
DM overdensity fields, which suffers much less from those problems. We will perform the calculation 
in Lagrangian space where the Lagrangian halo biases are established. 
We will consider specifically the effect of a local-type PNG on these bispectra and illustrate its sensitivity 
to the biasing model. For convenience, we decompose $B_\text{hmm}=B_\text{hmm}^G+\Delta B_\text{hmm}^{NG}$ 
into a contribution generated by Gaussian initial conditions and by PNG.

\subsection{Local primordial non-Gaussianity}

The local-type PNG model is conveniently expressed in terms of the Bardeen's potential $\Phi$ deep in the 
matter era (i.e. immediately after matter-radiation equality)
\cite{Salopek:1990re,Gangui:1993tt,Komatsu:2001rj}.
The initial density field is related to the potential $\Phi$ through
\be
\delta_L(\vk,z) = \mathcal{M}(k)D(z) \Phi(\vk) \;,
\ee
where $\mathcal{M}(k)$ is determined by the transfer function and the Poisson equation as
\be
\mathcal{M}(k) = \frac{2}{3}\frac{k^2T(k)}{H_0^2\Omega_m}\;.
\ee
As is common practice, the transfer function $T(k)$ tends towards unity at large scales, while the linear growth 
rate $D(z)$ is normalised to unity at $z=0$ in the Einstein-de Sitter universe. Moreover, $H_0$ and $\Omega_m$ are
the present-day value of the Hubble rate and matter density, respectively. 
For a local-type non-Gaussianity, the bispectrum and trispectrum of $\Phi$ are given by
\bea
B_\Phi(k_1,k_2,k_3) &=& 2 f_{\rm NL} [P_\Phi(k_1)P_\Phi(k_2)+2 \ {\rm cyc}] \\
T_\Phi(\vk_1,\vk_2,\vk_3, \vk_4) &=& 6 \ g_{\rm NL}[P_\Phi(k_1)P_\Phi(k_2)P_\Phi(k_3) + 3  \ {\rm cyc}] 
\nonumber \\
&+& \frac{25}{18} \tau_{\rm NL} [ P_\Phi(k_1)P_\Phi(k_2)\{P_\Phi(k_{13})+P_\Phi(k_{14})\}+11 \ {\rm cyc}] 
\eea
where $k_{ij} = | \vk_i+\vk_j  |$. For simple models of inflation $\tau_{\rm NL} \propto f_{\rm NL}^2$.

\subsection{Calculational strategy}

In order to organise the calculation of correlation functions, we take advantage of the connection that exists 
between the iPT and the peak approach. In the spirit of iPT, Eq.~(\ref{eq:cdef}) can be generalised to Eulerian space 
upon defining a multi-point propagator for biased tracers \cite{Matsubara:2011ck},
\be
\left < \frac{\delta^n\delta_\esp(\vk,z)}{\delta \delta_L(\vk_1,z_\ast)\dots\delta\delta_L(\vk_n,z_\ast)}\right >  
= (2\pi)^{3-3n}\delta_D^3(\vk -\vk_{1...n})\,\Gamma_\esp^{(n)}(\vk_1,\dots,\vk_n; z),
\ee
In Lagrangian space, i.e. in the limit $z\to\infty$, they match exactly our renormalised ESP bias functions, 
\be
\label{eq:GamEspApprox}
\Gamma_\esp^{(n)}(\vk_1,\dots,\vk_n; z\to\infty)~ \to~ c_n^L(\vk_1,\dots,\vk_n; z_\ast) \;.
\ee
Moreover, they are similar to the multi-point propagators $\Gamma_m^{(n)}$ of the matter distribution employed in 
\cite{Bernardeau:2008fa}. Therefore, similar diagrammatic and counting rules apply.

The propagators $\Gamma_m^{(n)}$ and $\Gamma_\esp^{(n)}$ can be used to calculate $N$-point correlation functions of
matter fields or biased tracers at any redshift.
For instance, \cite{Bernardeau:2010md} derived expression for the bispectrum of the matter field in the presence 
of PNG. Similarly, \cite{Yokoyama:2013mta} computed the halo bispectrum $B_\text{hhh}$ within the iPT framework, 
using the multi-point propagator introduced in \cite{Matsubara:2012nc}. 
Here, we followed the same strategy to derive both $B_\text{hhh}$ and $B_\text{hmm}$ at 1-loop. We also checked that, 
in Lagrangian space, the results  agree with a calculation based on ESP perturbative expansion.

In all our calculations, we keep track of terms proportional to the bispectrum $B_0$ and trispectrum $T_0$ of the 
initial density field since we are interested in the effect of PNG. For the sake of completeness, the full expressions 
of $B_\text{hmm}$ and $B_\text{hhh}$ at 1-loop are presented in Appendix \ref{app:bispectra}. 
We will now summarise the relevant theoretical expressions.

\subsection{Squeezed bispectra in the initial conditions in the presence of PNG}\label{sec:sqb}

As mentioned previously, the focus of this work is on the bispectra $B_\text{hhh}$ and $B_\text{hmm}$. We compare the theoretical prediction for the squeezed limit bispectra against that measured on N-body simulation at the initial redshift $z_i\sim {\cal O}(100)$.
Given that the measurement is done at high redshifts, we are sensitive to the local PNG without being too concerned about the modelling of non-Gaussianities induced by the gravitational evolution. 
Following Eq.~(\ref{eq:GamEspApprox}), we will hereafter approximate $\Gamma_\esp^{(n)}$ by $c_n^L$, 
yet include the small nonlinearities in the matter through the usual PT kernels, $\Gamma_m^{(n)}=F_m^{(n)}$. We can ignore the non-linear evolution of matter in the multi-point propagator $\Gamma_\esp^{(n)}$ since it is suppressed with respect to the non-linear biasing by powers of the growth factor, which is small at high redshifts, as we will see below. Note, however, that our model includes the highly non-linear evolution of the small-scale modes that collapse into halos through the spherical collapse approximation.

Let $B_\text{hmm}$ and $B_\text{mhm}$ be the cross-bispectra when the squeezed Fourier mode $k_1=k_l$ corresponds to the 
halo field $\delta_h(\vk,z)$ and matter field $\delta_m(\vk,z)$, respectively. 
Retaining only the terms with the strongest divergent behaviour in the squeezed limit, the contribution induced by
PNG reads
\begin{align}
&\lim_{\kL\rightarrow 0} \Delta B_\text{hmm}^{NG}(k_{\rm l},k_{\rm s},k_{\rm s};z_i) \approx  
\left(\frac{D(z_i)}{D(z_\ast)}\right)^2 c^L_1(\kL)\,B_0(\kL,\ks,\ks)  \nonumber \\ 
& \qquad + \left(\frac{D(z_i)}{D(z_\ast)}\right)^3 P_0(\ks)\, 
F^{(2)}_m(\vks,\vkL)\! \int \!\frac{d^3q}{(2\pi)^3}\, c_2^L(\vq,\!-\!\vq)\, B_0(\kL,q, q)\nonumber\\
& \qquad + \frac{1}{2} \left(\frac{D(z_i)}{D(z_\ast)}\right)^2 \, \!  \int\!
\frac{d^3q}{(2\pi)^3}\,c_2^L(\vq,-\vq)\, T_0(\vks,-\vks,\vq,-\!\vq)\,,
\label{eq:Hsq}
\end{align}
where $\vks \equiv \vk_2 \simeq -\vk_3$ is the short mode and $\vkL \equiv \vk_1$ is the long mode.
Moreover, the initial matter power spectrum $P_0$, bispectrum $B_0$ and trispectrum $T_0$ are evaluated at the redshift 
$z_\ast$ of halo collapse, like the Lagrangian bias coefficients $c_n^L$. Therefore to relate them to the quantities at the initial conditions of simulation at which the measurement is done, there would be factors of $D(z_i)/D(z^*)$ per matter density and a factor of $[D(z_*)/D(z_i)]^n$ for the Lagrangian biases at order $n$. The powers of growth factors account for the redshift evolution of $B_\text{hmm}$. 
Since the initial conditions of the simulations are typically laid down at redshift $z_i\sim {\cal O}(100)$, and the 
collapse redshift of halo  is in the range $0\leq z_\ast \leq 2$, each factor of $D(z_i)/D(z_\ast)$ represents a two 
orders of magnitude suppression. 
This corresponds to the intuition that, at $z_i$, the non-linear evolution of matter is small. 
We have kept next to leading order terms in this small parameter since, for very small squeezed Fourier mode, the stronger 
divergence of the trispectrum can compensate for this suppression. 
We call Eq.~(\ref{eq:Hsq}) the ``halo squeezed'' bispectrum. 

Analogously, for the halo matter matter bispectrum, when we take the limit of one of the Fourier modes corresponding to a 
matter overdensity field to zero, we obtain
\begin{align}
&\lim_{\kL\rightarrow 0} \Delta B_\text{mhm}^{NG}(k_{\rm l},k_{\rm s},k_{\rm s};z_i) \approx  
\left(\frac{D(z_i)}{D(z_\ast)}\right)^2 c_1^L(\ks)\,B_0(\kL,\ks,\ks)  \nonumber \\ 
& \qquad + \frac{1}{2}\left(\frac{D(z_i)}{D(z_\ast)}\right)^3 \ 
P_0(\ks)c_2^L(-\vks,\vkL)\! \int \!\frac{d^3q}{(2\pi)^3}\, F^{(2)}_m(\vq,\!-\!\vq)\,
B_0(\kL,q, q)\nonumber\\
& \qquad + \left(\frac{D(z_i)}{D(z_\ast)}\right)^3\int\!\frac{d^3q}{(2\pi)^3}\, 
c_2^L(\vq,\vks \!-\!\vq)\, F^{(2)}_m(\vq,\vks\!-\!\vq)\, P_0(|\vks\!-\!\vq|)\,
B_0(\kL,q,q)  \nonumber \\
& \qquad+  \frac{1}{2}\left(\frac{D(z_i)}{D(z_\ast)}\right)^2 \,P_0(\ks)\! \int\! \frac{d^3q}{(2\pi)^3}\, 
c_3^L(-\vks,\vq,-\!\vq)\, B_0(\kL,q,-q)  \nonumber \\
& \qquad + \frac{1}{2} \ \left(\frac{D(z_i)}{D(z_\ast)}\right)^2 
\int\! \frac{d^3q}{(2\pi)^3}\,c_2^L(\vq,\vks-\vq)\, T_0(-\vks,\vkL,\vq,\vks-\!\vq)\,.
\label{eq:Msq}
\end{align}
In this particular limit, the third order bias contribution is not suppressed relative to the tree level contribution. 
We call Eq.~(\ref{eq:Msq}) the ``matter squeezed'' bispectrum.

Although we have stopped at one loop, the loop expansion does not in principle correspond to an expansion in a small 
parameter. However, higher loops will necessarily involve either higher powers of $\Phi$ (such as two insertions of 
the primordial bispectrum, or a non-zero primordial 5-point function), or higher order contributions arising from the 
non-linear evolution of matter. The former are suppressed since the potential perturbations are $\Phi \sim 10^{-5}$, 
but they may become important at extremely large scales as they diverge with higher powers of $1/\kL$. 
The latter will be suppressed by factors of $D(z_i)$ corresponding to the smallness of nonlinearities in the evolution 
of matter at $z\sim z_i$.

\subsection{The specific case of a local type PNG}\label{sec:locB}

\begin{figure}
\centering
\begin{minipage}[b]{0.49\textwidth}
\hspace{.16in}\includegraphics[width=\textwidth]{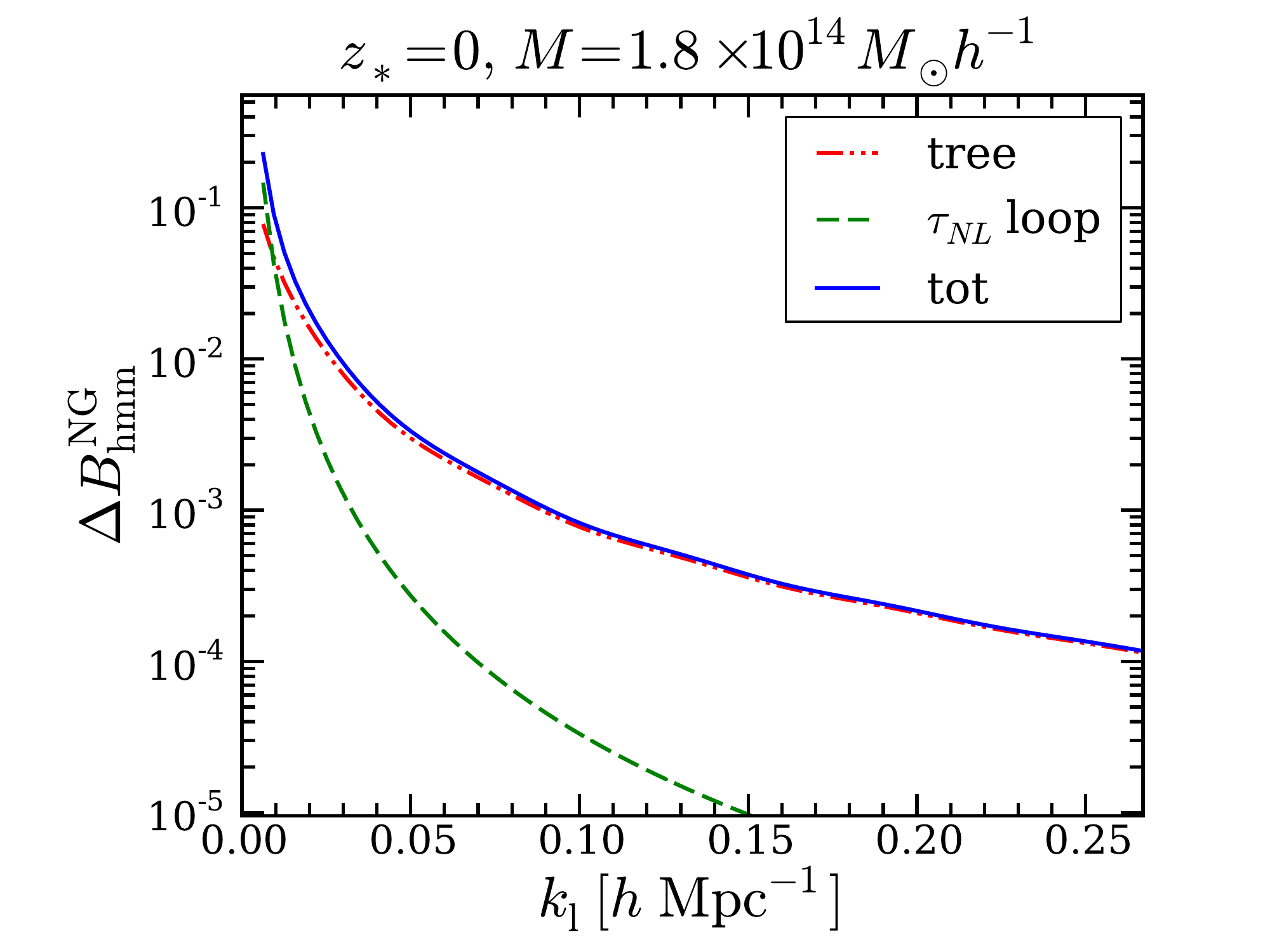}
\end{minipage}
\begin{minipage}[b]{0.49\textwidth}
\hspace{-.16in}\includegraphics[width=\textwidth]{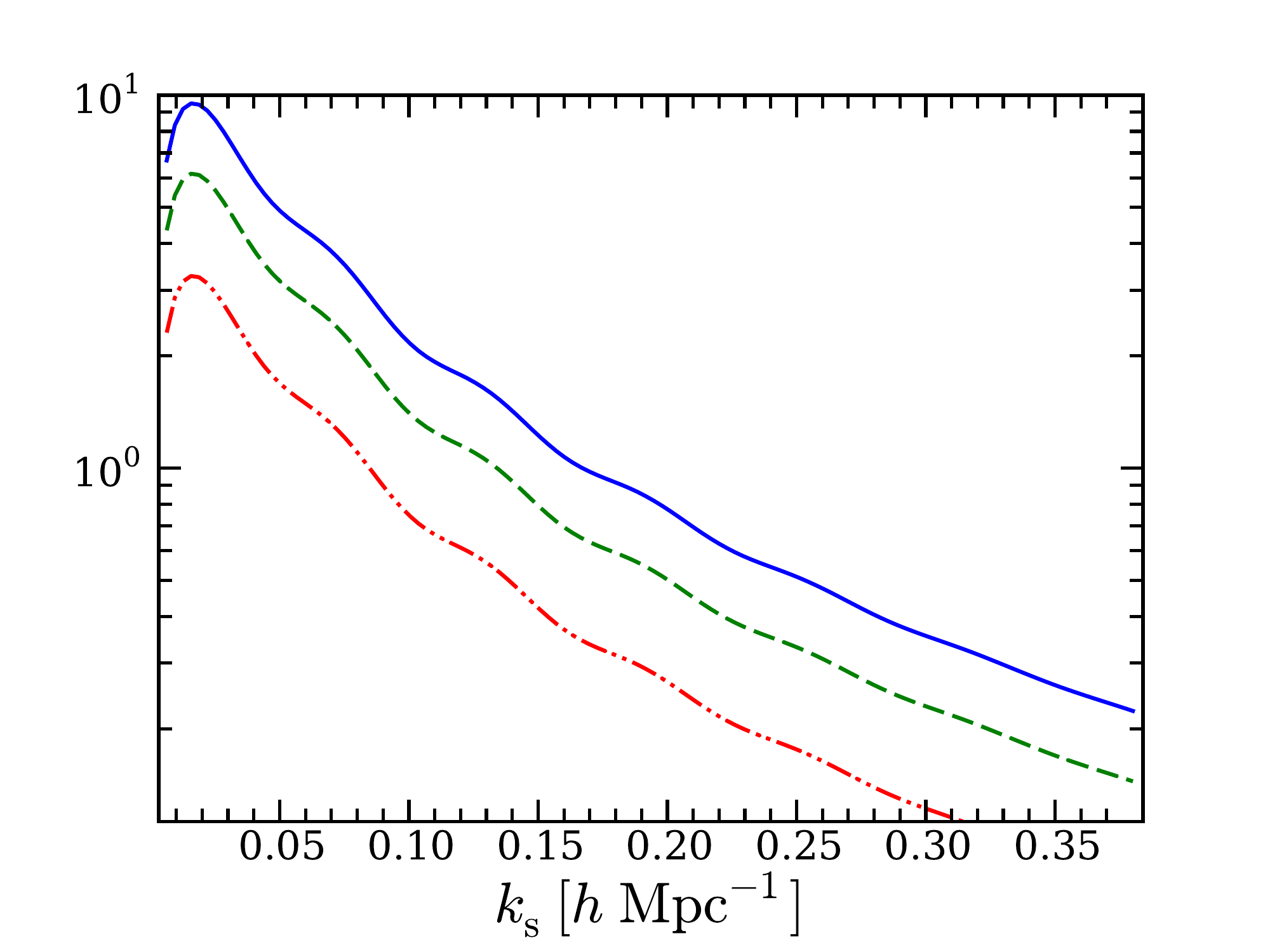}
\end{minipage}
\caption{The halo-matter-matter bispectrum, where the squeezed Fourier mode corresponds to the halo overdensity. The primordial non-Gaussianity is parameterised by $f_{\rm NL} = 250$, $g_{\rm NL} =0$ and $\tau_{\rm NL}=\left(6 f_{\rm NL}/5\right)^2$. The triangular configuration is chosen to have two sides equal to $k_{\rm s}$ and the third equal to $k_{\rm l}$. The bispectrum is shown for halos of mass $M=1.8 \times 10^{14} \Msun$ at redshift $z_\ast = 0$. On the left we plot the bispectra as a function of $k_{\rm l}$, fixing $k_{\rm s} = 0.38 \ {\rm Mpc}^{-1}h$ while on the left we fix $k_{\rm l} = 0.006 \ {\rm Mpc}^{-1}h$ .  Different lines correspond to the dominant contributions given in Eq.~(\ref{eq:HSQ}). }
\label{fig:hmm1_th}
\end{figure}

\begin{figure}
\centering
\begin{minipage}[b]{0.49\textwidth}
\hspace{.16in}\includegraphics[width=\textwidth]{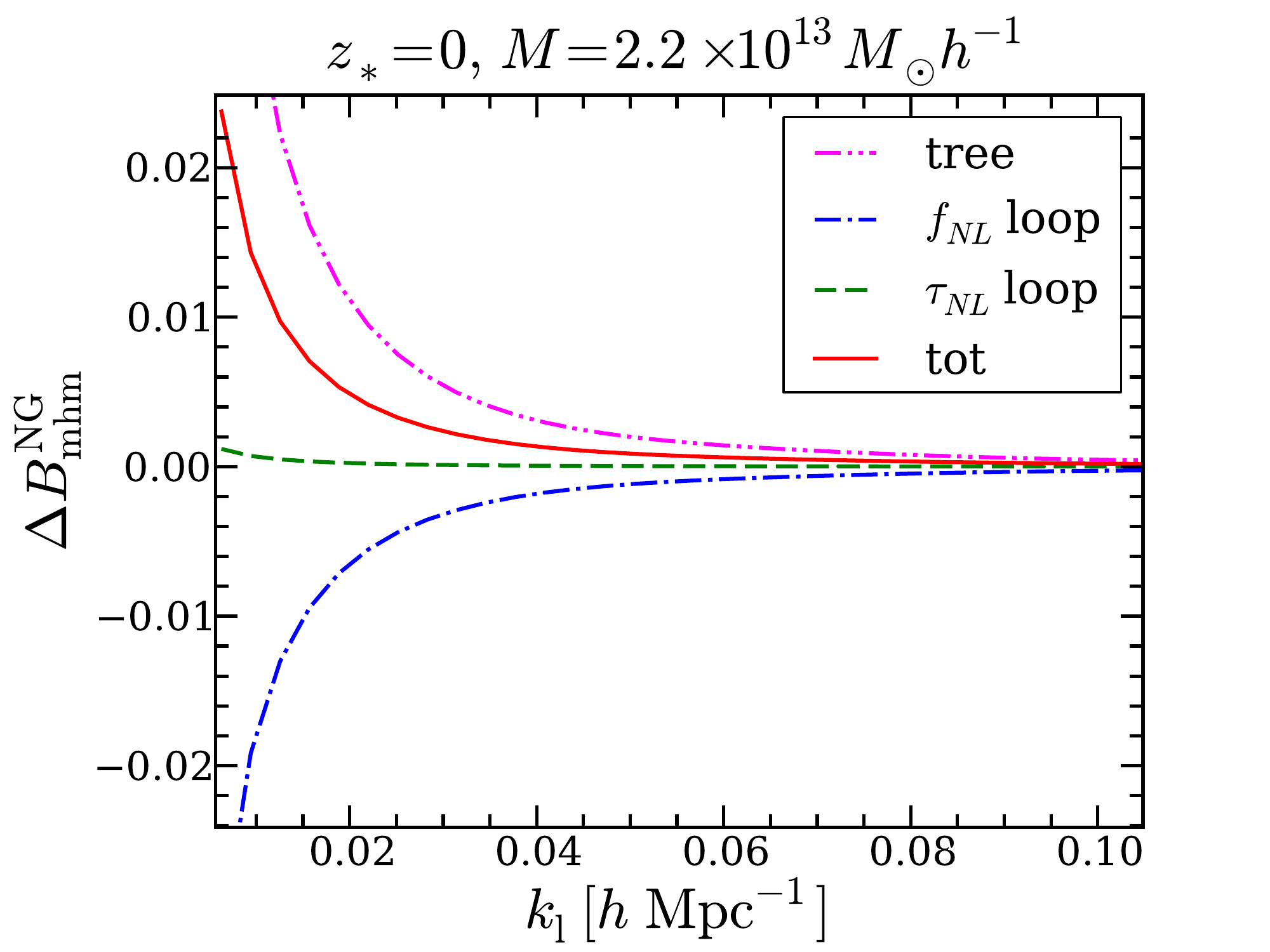}
\end{minipage}
\begin{minipage}[b]{0.49\textwidth}
\hspace{-.16in}\includegraphics[width=\textwidth]{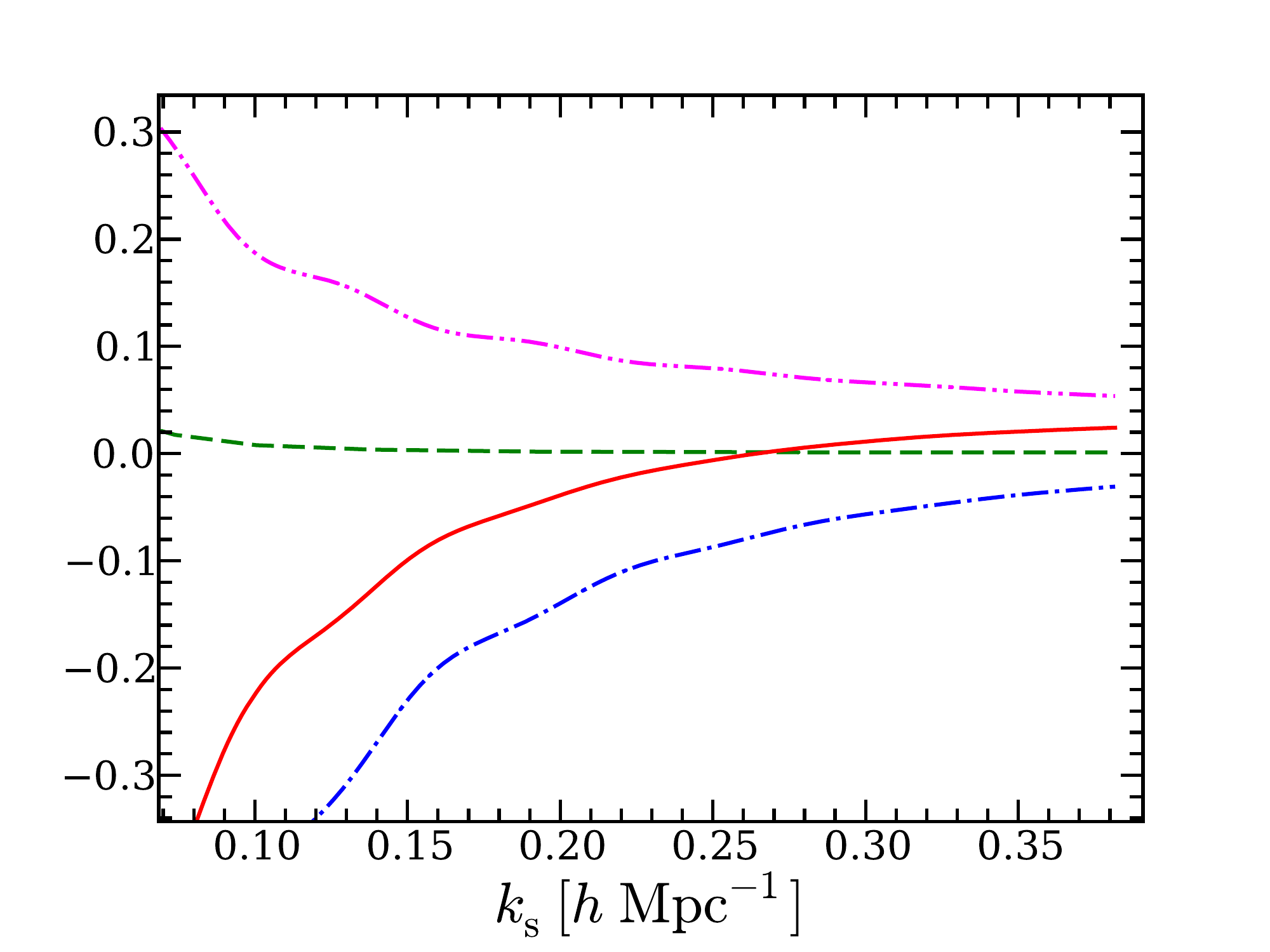}
\end{minipage}
\begin{minipage}[b]{0.49\textwidth}
\hspace{.16in}\includegraphics[width=\textwidth]{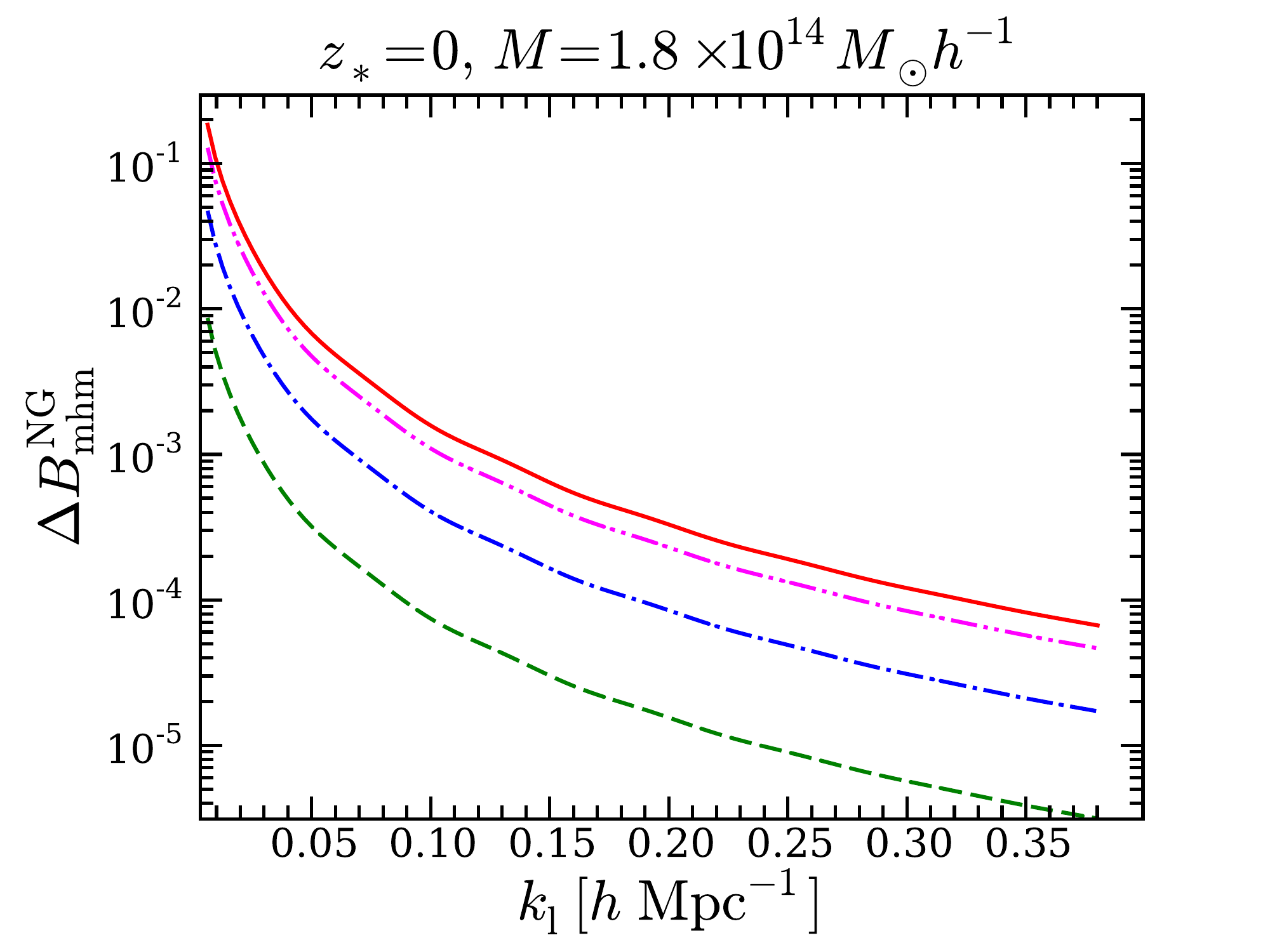}
\end{minipage}
\begin{minipage}[b]{0.49\textwidth}
\hspace{-.16in}\includegraphics[width=\textwidth]{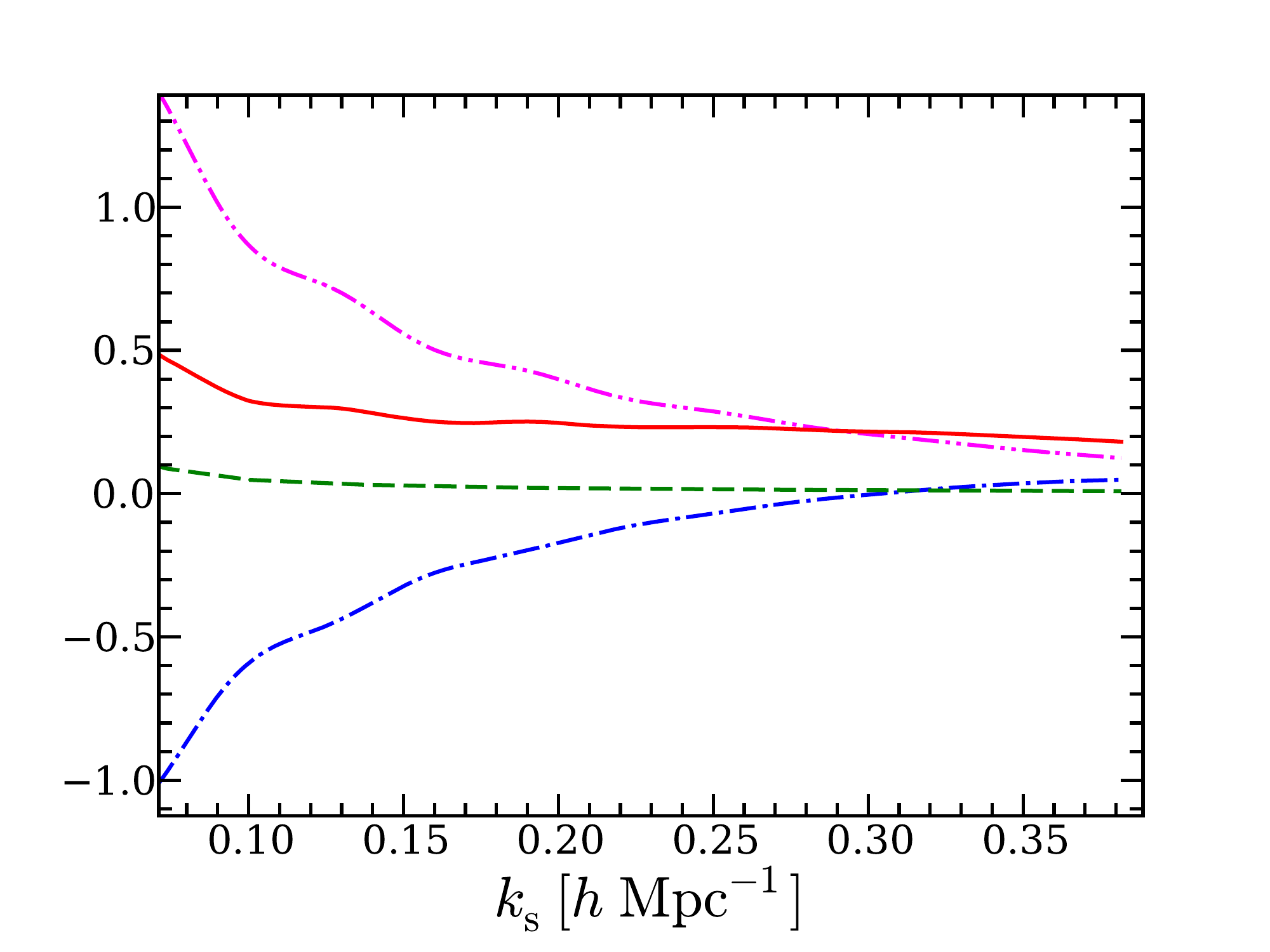}
\end{minipage}
\caption{The halo-matter-matter bispectrum, where the squeezed Fourier mode corresponds to the matter overdensity. The bispectra are calculated for the same non-Gaussian parameters and in the same configuration as in Fig \ref{fig:hmm1_th}. The bispectra are shown for halos at $z_\ast=0$. The top panel corresponds to the halos of mass $M=2.2\times 10^{13} \Msun$ while ones at the bottom are for halos of mass $M=1.8 \times 10^{14} \Msun$. Different lines correspond to the dominant contributions given in Eq. (\ref{eq:MSQ}).}
\label{fig:mhm2_th}
\end{figure}

Eqs.~(\ref{eq:Hsq}) and (\ref{eq:Msq}) are straightforwardly specialised to local PNG using the expressions for the 
bispectrum and trispectrum. For the halo squeezed case, we get 
\begin{align}
\label{eq:HSQ}
\lim_{k_{\rm l}\rightarrow 0}\Delta B_\text{hmm}^{NG}&(k_{\rm l},k_{\rm s},k_{\rm s};z_i)\approx   
\frac{4  f_{\rm NL}}{\mathcal{M}(\kL)}\frac{D^2(z_i)}{D^3(z_\ast)} P_0(\ks) P_0(\kL) c_1^L(\kL)\nonumber \\
&+ \frac{50\tau_{\rm NL}}{9\mathcal{M}^2(\kL)} \frac{D^2(z_i)}{D^4(z_\ast)} P_0(\kL)   P_0(\ks) 
\int\! \frac{d^3 q}{(2\pi)^3} \ c_2^L(\vq,-\vq) P_0(q)\,,
\end{align}
whereas, for the matter squeezed bispectrum, we obtain
\begin{align}
\label{eq:MSQ}
\lim_{k_{\rm l}\rightarrow 0}\Delta B_\text{mhm}^{NG}(k_{\rm l},k_{\rm s},k_{\rm s};z_i) &\approx  \frac{4  
f_{\rm NL}}{\mathcal{M}(\kL)} \frac{D^2(z_i)}{D^3(z_\ast)} P_0(\ks) P_0(\kL) c_1^L(\ks) \nonumber \\ 
& +  \frac{2  f_{\rm NL}}{\mathcal{M}(\kL)}  \frac{D^2(z_i)}{D^3(z_\ast)} \ P_0(\ks) P_0(\kL) \! 
\int\! \frac{d^3q}{(2\pi)^3}\, c_3^L(-\vks,\vq,-\!\vq)\, P_0(q) \nonumber \\
&+ \left(\frac{27 g_{\rm NL} +  25 \tau_{\rm NL}}{9\mathcal{M}(\kL)}\right) \frac{D^2(z_i)}{D^4(z_\ast)}  P_0(\kL)  
\int \frac{d^3 q}{(2\pi)^3}\  c_2^L (\vq,\vks-\vq)\nonumber \nonumber \\ 
&\times\left[\frac{\mathcal{M}(|\vks - \vq|)}{\mathcal{M}(\ks)\mathcal{M}(q)} P_0(\ks)P_0(q) 
+  \frac{\mathcal{M}(q)}{\mathcal{M}(\ks)\mathcal{M}( |\vks-\vq|)}P_0(|\vks-\vq|) P_0(\ks)\right. \nonumber \\
&+\left. \frac{\mathcal{M}(\ks)}{\mathcal{M}(q) \mathcal{M}(|\vks-\vq|)}P_0(|\vks-\vq|) P_0(q) \right]\,,
\end{align}
where we have only retained the dominant terms (neglecting those which involve $F^{(2)}$ since they are suppressed by a factor of  $D(z_i)/D(z_\ast)$) together with the trispectrum contribution.

In Figs. \ref{fig:hmm1_th} and \ref{fig:mhm2_th} we plot the different contributions to Eqs.~(\ref{eq:HSQ}) and 
(\ref{eq:HSQ}). Here and henceforth, we always divide the bispectrum by a factor of $(2\pi)^6$. 
For the halo squeezed bispectrum, we indeed see that the trispectrum loop dominates at larger scales due to its stronger divergence. 
At those large scales, we have also explicitly checked that the additional loop corrections are indeed suppressed. 
Higher order primordial correlation functions will have even stronger divergences, but these have to overcome a 
suppression of higher powers of $\Phi$. Therefore, we do not expect them to be significant on the scales probed
by surveys, even in the presence of large PNG. 
For the matter squeezed bispectrum, the loop involving the bispectrum is of the same order as the tree level. 
However, we expect higher order loops to be suppressed by the non-linearity of the DM at $z=z_i$.

Let us now turn to the halo bispectrum. An analogous calculation to that performed above gives 
\begin{align}\label{eq:huh}
&\lim_{\kL \rightarrow 0} \Delta B_\text{hhh}^{NG}(k_{\rm l},k_{\rm s},k_{\rm s};z_i) \approx 
\frac{1}{D(z_\ast)}\frac{4  f_{\rm NL}}{\mathcal{M}(\kL)} c_1^L(\kL)P_0(\kL)\, \left(c_1^L(\ks)\right)^2 P_0(\ks)\nonumber \\
& \qquad + \frac{1}{D(z_\ast)}\frac{4  f_{\rm NL}}{\mathcal{M}(\kL)} c_1^L(\kL) P_0(\kL) c_2^L(\vks,\vkL) P_0(\ks)
\int \frac{d^3 q}{(2 \pi)^3} \ c_2^L(\vq,-\vq) P_0(q) \nonumber \\
& \qquad + \frac{1}{D(z_\ast)}\frac{4  f_{\rm NL}}{\mathcal{M}(\kL)} \ c_1^L(\kL) P_0(\kL)\! \int\! \frac{d^3q}{(2\pi)^3}\, 
\left[c_2^L(\vq,\vks\!-\!\vq)\right]^2\, P_0(|\vks\!-\!\vq|)P_0(q) \nonumber \\
& \qquad +\frac{1}{D(z_\ast)}\frac{4  f_{\rm NL}}{\mathcal{M}(\kL)} \  c_1^L(\kL) P_0(\kL)\, c_1^L(\ks) P_0(\ks) \!
\int\! \frac{d^3q}{(2\pi)^3}\, c_3^L(\vks,\vq,-\!\vq)\, P_0(q) \nonumber \\
& \qquad + \frac{1}{D^2(z_\ast)} \frac{50\tau_{\rm NL}}{9\mathcal{M}^2(\kL)}P_0(\kL)P_0(\ks) \left[c_1^L(\ks) \right]^2 
\int \frac{d^3 q}{(2\pi)^3} \ c_2^L(\vq,-\vq) P_0(q) \nonumber \\
& \qquad + \frac{1}{D^2(z_\ast)}\left(\frac{6g_{\rm NL}}{\mathcal{M}(\kL)} + \frac{50\tau_{\rm NL}}{9\mathcal{M}(\kL)}\right)
P_0(\kL) c_1^L(\kL)c_1^L(\ks)\!  \int\!  \frac{d^3 q}{(2\pi)^3}\, c_2^L(\vq,-\vks-\vq) \nonumber \\
&\qquad\quad \times \left [\frac{\mathcal{M}(|\vks+\vq|)}{\mathcal{M}(\ks)\mathcal{M}(q)} P_0(\ks)P_0(q) 
+ \frac{\mathcal{M}(q)}{\mathcal{M}(\ks) \mathcal{M}(|\vks+\vq|)}  P_0(\ks)P_0(|\vks+\vq|) \right. \nonumber \\
&\qquad\qquad  + \left. \frac{\mathcal{M}(\ks)}{\mathcal{M}(q) \mathcal{M}(|\vks +\vq|)} P_0(q)P_0(|\vks+\vq|) \right]\,.
\end{align}
\begin{figure}
\centering
\begin{minipage}[b]{0.49\textwidth}
\hspace{.16in}\includegraphics[width=\textwidth]{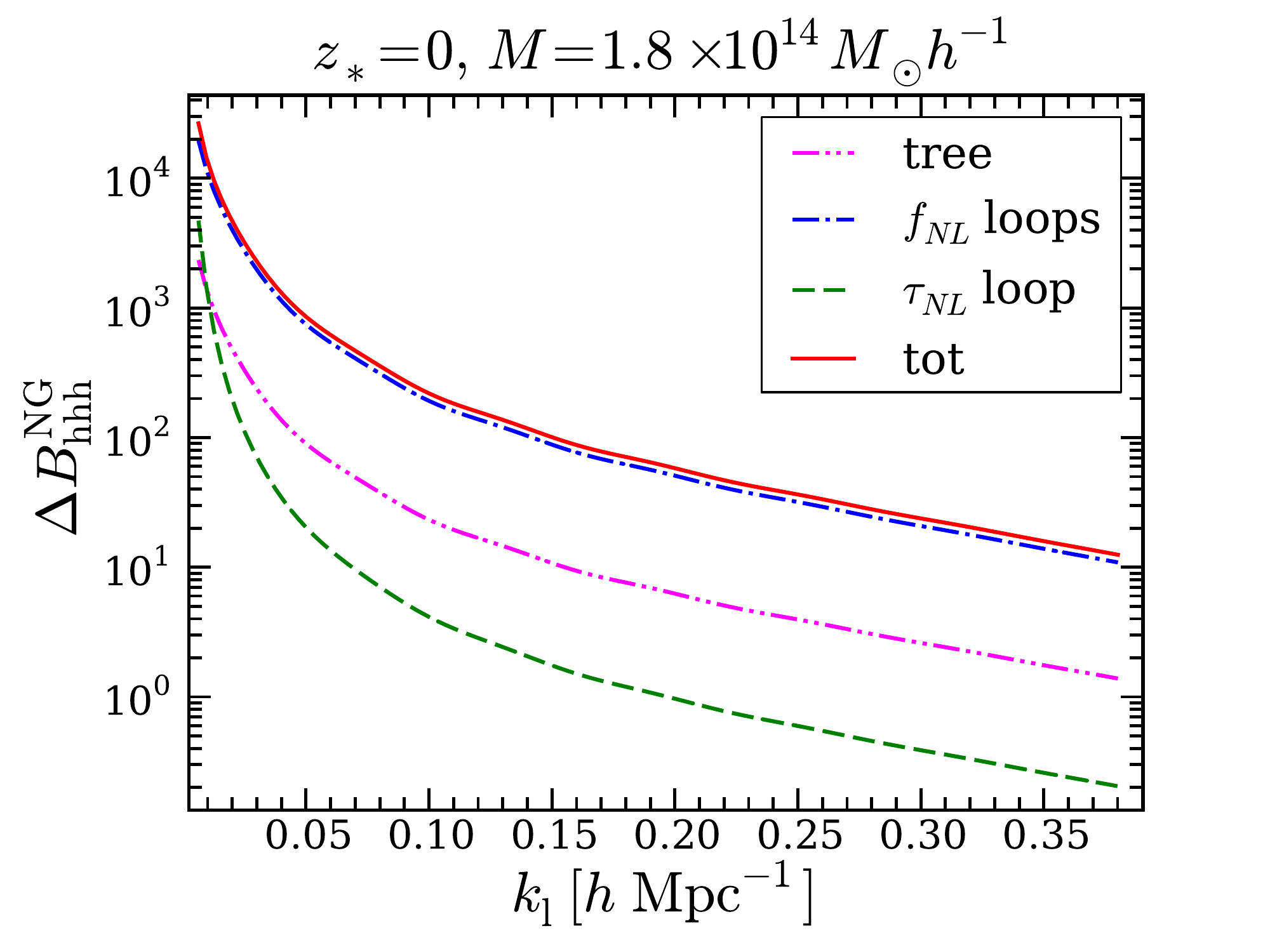}
\end{minipage}
\begin{minipage}[b]{0.49\textwidth}
\hspace{-.16in}\includegraphics[width=\textwidth]{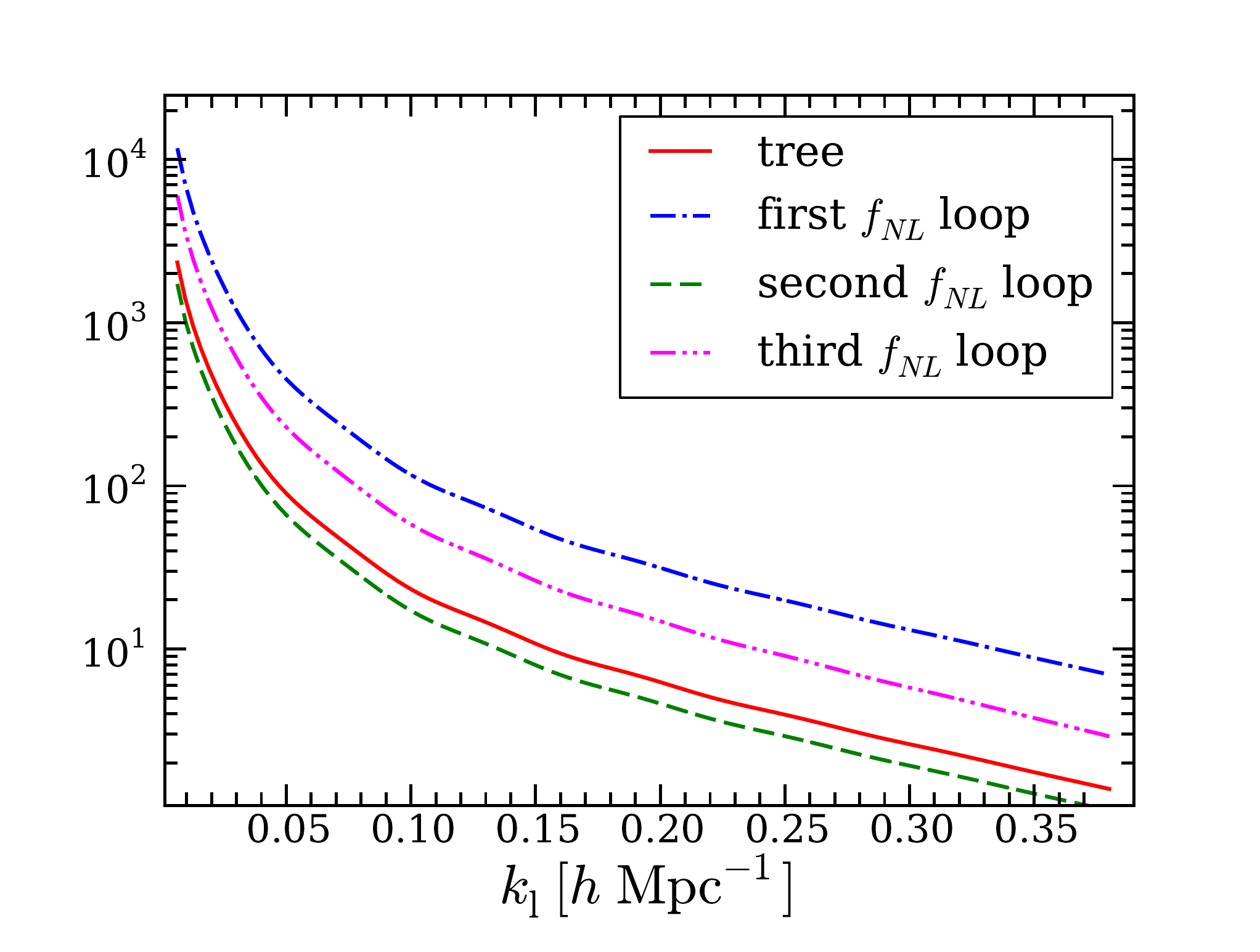}
\end{minipage}
 \caption{The squeezed limit of halo-halo-halo bispectrum. The bispectra are calculated for the same non-Gaussian parameters and in the same configuration as in Fig \ref{fig:hmm1_th}. The redshift and the halo mass is also the same as in that figure. The plot on the left shows dominant contributions given in Eq. (\ref{eq:huh}), while on the right, three loop contributions proportional to $f_{\rm NL}$.}
\label{fig:hhh_th}
\end{figure}

Let us make a few comments regarding this expression. 
Firstly, unlike $\Delta B_\text{hmm}^{NG}$, one-loop terms are 
not suppressed by any small parameter in Eq.~(\ref{eq:huh}). This is apparent in Fig. \ref{fig:hhh_th}, in which we 
separately plot the contributions of the tree level, the loops involving the primordial bispectrum and those involving 
the primordial trispectrum. One-loop terms involving a primordial bispectrum are clearly large compared to the tree 
level. This is compatible with the result of \cite{Yokoyama:2013mta}, who found that the situation is even worse at two 
loops. This suggests that there may be no perturbative expansion in this case. A possible cause of this problem is the 
fact that the bias parameters have not been appropriately defined in the presence of PNG,  i.e. they have not been 
appropriately ``renormalised''. In the Gaussian case, the renormalisation of the ESP bias coefficients $c_{ijkqlm}$ is 
naturally taken care of by the orthonormal polynomials. However, these polynomials are orthonormal only w.r.t. Gaussian
weights such as Normal or chi-square distributions. In the presence of PNG, the linear bias $c_1^L(k)$ will for instance
receive contributions from integrals involving the skewness. We leave a thorough treatment of this problem for future work.

Secondly, the term in the second line turns out to be proportional to the usual peak-background split non-Gaussian bias 
\cite{Slosar:2008hx}. In the peak formalism, the non-Gaussian bias is obtained from a one-loop integration of the 
primordial bispectrum and the second order bias, like in iPT. As shown in \cite{Desjacques:2013qx}, the relation
\be
\label{eq:c2NG}
\int\! \frac{d^3 q}{(2 \pi)^3} \ c_2^L(\vq,-\vq) P_0(q)= 
\left(\frac{\partial\ln\bar{n}}{\partial\ln\sigma_8}\right) \;.
\ee
holds exactly for a constant deterministic barrier, while it appears that it is not satisfied in the ESP approach with
moving barrier \cite{Biagetti:2015exa}. This issue will be addressed in future work.
Here, we note that, even in the case of a moving barrier, one still expects Eq.~(\ref{eq:c2NG}) to hold reasonably well 
for halos with a mass larger than a few $M_\star(z_\ast)$. All the measurements presented here satisfy this condition.

Thirdly, the tree-level non-Gaussian term (first line) and the second and third $f_{\rm NL}$ terms (third and forth lines) in Eq.~(\ref{eq:huh}) nearly add up to 
\be
\lim_{\kL\rightarrow 0} \Delta B_\text{hhh}^{NG} = 
\frac{1}{D(z_\ast)}\frac{4  f_{\rm NL}}{\mathcal{M}(\kL)} c_1^L(\kL) P_0(\kL) \, P_\text{h}(\ks) \;,
\label{eq:PP}
\ee
where $P_\text{h}(k)$ is the power spectrum of proto-halo centres at one loop. However, the term in the second line has 
the wrong coefficient for this simplification to take place (it should come with a factor of $1/2$).
There has been a claim in the literature \cite{Peloso:2013zw} that a result similar to \eqref{eq:PP} holds for the matter 
overdensity bispectrum. However, it is straightforward to realise that, at one loop in standard perturbation theory, the 
algebra for computing the matter bispectrum is the same as that is used here. Therefore, the result would be the same upon
replacing the bias coefficients by the DM non-linear evolution kernels. 
Hence, at the one loop level in standard perturbation theory, the non-Gaussian part of the result of \cite{Peloso:2013zw} 
does not hold. This is somewhat unsurprising since, unlike the Gaussian part of their result (which we agree with), there 
is no symmetry argument to relate the bispectrum to the power spectrum \cite{Kehagias:2013yd,Peloso:2013zw,Horn:2015dra}.

Finally, the discrete nature of the proto-halo centres induces shot-noise corrections. We have ignored them here since 
our main focus is on the cross-bispectrum $B_\text{hmm}$, which involves only one halo field. In $B_\text{hhh}$, these 
shot noise corrections could be modelled from first principle using peak theory, along the lines of e.g. 
\cite{Baldauf:2013hka}.

\subsection{Other theoretical approaches}

To emphasise the importance of Lagrangian $k$-dependent bias contributions, we will also compare our measurements to a 
local bias approach and to the S12i model. 

\subsubsection{The Local bias model} 
In the local bias approach, $\Delta B_\text{hmm}^{NG}$ and $\Delta B_\text{mhm}^{NG}$ are still given by Eqs.~(\ref{eq:HSQ})
and (\ref{eq:MSQ}), yet we turn off the scale-dependent terms in the Lagrangian bias functions $c_n^L(\vk_1,\dots,\vk_n)$.
Therefore, we have
\begin{equation}
c_n^L(\vk_1,\dots,\vk_n) \equiv b_n 
\end{equation} 
in the local bias model, where $b_n= b_{n00}$ is computed from the ESP approach. 

\subsubsection{The S12i model}

In addition to the ESP and local bias approach, we will also consider a model for the halo-matter bispectrum 
in which the NG bias parameters are explicitly obtained from a peak-background split (PBS).
In this model, the signatures of PNG are encoded both in the NG matter bispectrum and in the NG bias parameters 
predicted by the PBS.
Previous studies \cite{Giannantonio:2009ak, Baldauf:2010vn} have already derived the PBS bias parameters and 
used them to predict the halo bispectrum in the presence of PNG. 
Here, we will follow the derivation of \cite{Scoccimarro:2011pz}. 
Although our final prediction eventually agrees with the standard PNG bias formula, we will refer to this model 
as S12i, to signify that it was inspired  by the general derivation given in \cite{Scoccimarro:2011pz}.   
We will now summarise the main ingredients of this model. Details can be found in Appendix~\ref{sec:ESbias_PBS}.

For the halo-squeezed case,  we adopt the tree-level-only bispectrum 
\begin{align}
\lim_{k_{\rm l}\rightarrow 0}\Delta B^{NG}_{\text{hmm}} (k_{\rm l}, k_{\rm s},k_{\rm s};z_i)   
&=   \left(\frac{D(z_i)}{D(z_\ast)}\right)^2 \Big[ b_1^{(1)}(z_\ast) +  b_1^{(2)}( k_{\rm l}, z_\ast ) \Big]  
\tilde W_R(k_{\rm l})\,  B_0 ( k_{\rm s},k_{\rm s},k_{\rm l}) \nn\\     
& \quad + \left(\frac{D(z_i)}{D(z_\ast)}\right)^3 b_1^{(2)} (k_{\rm l},z_\ast ) \tilde W_R(k_{\rm l})\,  
B_{\rm m}^G( k_{\rm s},k_{\rm s},k_{\rm l}; z_\ast )\, , 
\end{align}
where  $b_1^{(1)} $  is the usual Gaussian PBS bias parameter, $b_1^{(2)} $ is the NG bias parameter and
$B_{\rm m}^G $ is the matter bispectrum induced by gravitational nonlinearities, and given by 
Eq.~(\ref{eq:BmNG_Gaussian}) at tree-level.
We will use the scale-independent bias parameter measured from the Lagrangian cross power spectrum between halo 
and matter as an estimate for  $b_1^{(1)}  $.   

In \cite{Scoccimarro:2011pz}, the NG bias parameter  $b_1^{(2)} $  is derived from the excursion set theory in 
a general setting (see the review in Appendix~\ref{sec:ESbias_PBS}).  As shown in \cite{Scoccimarro:2011pz}, 
under the assumption of Markovianity of the excursion set walk and the universality of the mass function, the 
general expression for  $b_1^{(2)}$ given in  Eq.~(\ref{eq:b1_2_massderivative_mfnweight}) reduces to the 
well-known formula Eq.~(\ref{eq:b1_NG_standard})  \cite{Dalal:2007cu,Slosar:2008hx,Matarrese:2008nc} in the 
low-$k$ approximation.    
Although  Eq.~(\ref{eq:b1_2_massderivative_mfnweight}) is quite general, it is technically more difficult to 
evaluate than Eq.~(\ref{eq:b1_NG_standard}) because an accurate measurement of the numerical mass function is 
 required.  In particular, as discussed in  Appendix~\ref{sec:ESbias_PBS},  for halos resolved with few 
particles (group 1 in our case),  it is numerically more  accurate to compute $b_1^{(2)} $ using  
Eq.~(\ref{eq:b1_NG_standard}) instead.  
Therefore we shall evaluate  $b_1^{(2)} $ using Eq.~(\ref{eq:b1_NG_standard}), while $b_1^{(1)} $ is obtained 
from the Gaussian simulations. 

In the matter-squeezed case, in addition to the tree level bispectrum, we also include the  1-loop correction 
proportional to $b_3$ 
\begin{align}
\label{eq:DeltaBNG_Msq_PBS}
\lim_{k_{\rm l}\rightarrow 0}\Delta B^{NG}_{\text{mhm}}(k_{\rm l},k_{\rm s},k_{\rm s};z_i)  
&=  \left(\frac{D(z_i)}{D(z_\ast)}\right)^2 \Big[ b_1^{(1)}(z_\ast) +  b_1^{(2)}( k_{\rm s}, z_\ast ) \Big]  
\tilde{W}_R(k_{\rm s})  B_0( k_{\rm s},k_{\rm l},k_{\rm s}) \\
&\quad +  \left(\frac{D(z_i)}{D(z_\ast)}\right)^3 b_1^{(2)} (k_{\rm s},z_\ast )  \tilde{W}_R(k_{\rm s})  
B_{\rm m}^G( k_{\rm s},k_{\rm l},k_{\rm s}; z_\ast )  \nonumber \\ 
& \quad +\left(\frac{D(z_i)}{D(z_\ast)}\right)^2\frac{ b_3^{\rm ESP}(z_\ast)}{2} \tilde{W}_R(k_{\rm s}) 
P_0(k_{\rm s}) \int\!  \frac{d^3 q}{ (2 \pi )^3}  \tilde{W}_R(q) \tilde{W}_R( | \mb{k}_{\rm l}  - \mb{q} |)  
\nonumber \\
&\qquad \times B_0(- \mb{k}_{\rm l}, \mb{q}, \mb{k}_{\rm l} - \mb{q} ) \nonumber \;.
\end{align}
The reason for including this $b_3$-loop correction is because in the matter-squeezed case, the halo field can be quite nonlinear and hence the high order bias parameters can be important; we find that the  analogous term in the peak model calculations is significant in the matter-squeezed case.  However,  the value of $b_3$  sensitively depends on  the prescriptions used to compute it. In  Eq.~(\ref{eq:DeltaBNG_Msq_PBS}), we will use the ESP result, i.e.  $b_3^{\rm ESP} = b_{300}  $. We have checked that  when $b_3 $ is computed using Mo $\&$ White (MW) \cite{Mo:1995cs} and Sheth $\&$ Tormen (ST) \cite{Scoccimarro:2000gm} bias parameters, the  $b_3$-loop  often worsens the agreement with the simulation results, while when the ESP result is used for  $b_3$  we  find the agreement often improved compared with tree-level-only results.   It is worth stressing that this is one of the few examples where we find the results are sensitive to $b_3$ and hence able to differentiate different schemes used to compute it (see also the measurements of  $b_3$ using cross-correlations or the separate universe simulations  \cite{Angulo:2007ex, Lazeyras:2015lgp}). 

\begin{table}
\begin{center}
\caption{Halo samples used in this paper.   The mean mass $\bar{M} $ and the cross bias parameter measured from the Lagrangian cross power spectrum  $b_{\rm c}^* $ are shown.}
\vspace{.2in}
\begin{tabular}{c  c  c  c}
\hline
 Mass range ($10^{12} \Msun $)   &    Bin 1: $3.8 - 9.2$    &   Bin 2: $9.2 -92 $   &   Bin 3: $92 - 920  $   \\
\hline  \hline 
$z_\ast=2$   &  $\bar{M}= 5.53$  &  $\bar M = 15.9$ & --- \\
 &   $  b_{\rm c}^* =  3.0 $  & $b_{\rm c}^* = 4.6   $  &   ---   \\
\hline
$z_\ast=1$  &  $\bar{M}= 5.68$  &  $\bar M = 20.0$ & --- \\
&   $  b_{\rm c}^* =  1.6 $   &  $b_{\rm c}^* = 2.5  $   & ---      \\
\hline
$z_\ast=0$ & ---  &  $\bar M = 23.6$ & $ \bar M = 193$ \\
 &  ---  &  $b_{\rm c}^* = 1.3  $  &  $b_{\rm c}^* = 2.5  $     \\
\hline
\label{tab:HaloSample}
\end{tabular} 
\end{center}
\vspace{-.3in}
\end{table}

\section{Comparison to numerical simulations}
\label{sec:sims}

In this section, we confront our model predictions based on the ESP approach to the numerical simulation 
results, and contrast the ESP predictions with those obtained with the local bias and S12i models.
We stress that none of the models considered here has remaining free parameters that can be fitted to the 
bispectrum data.

\subsection{$N$-body simulations and halo catalogues}

We use a series of cosmological simulations evolving $1536^3$ particles in cubic box of size $L_{\rm box} = 2000\MpcOh$. 
These simulations were run on the Baobab cluster at the University of Geneva.
\begin{figure}
\vspace{-.4in}
\centering
\includegraphics[width=0.75\textwidth,height=5.5in]{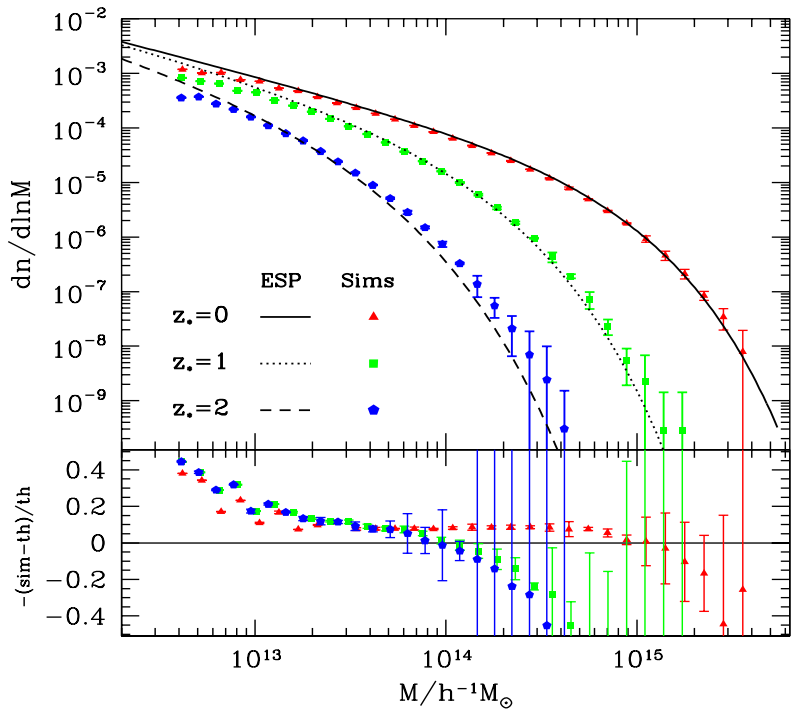}
\vspace{-1.2in} \caption{{\it Top panel}: Logarithmic mass function of SO halos for the 
Gaussian simulations. Different symbols refer to different redshifts as indicated in the figure. 
The solid, dotted and dashed curves represent the theoretical, ESP prediction at $z=0$, 1 and 2.
{\it Bottom panel}: Fractional deviation. In both panels, error bars denote the scatter among 
realisations.}
\label{fig:nmz0_z2}
\end{figure}
The cosmology is a flat $\Lambda$CDM model with $\Omega_{\rm m}=0.3$,  $\sigma_8 = 0.85 $, and $n_s =  0.967$.
The transfer function was obtained from the Boltzmann code {\small CLASS} \cite{Blas:2011rf}.
The initial particle displacements were implemented at $z_i=99$ using a modified version of the public code {\small 2LPTic} 
\cite{Crocce:2006ve,Scoccimarro:2011pz}. 
We shall use two different types of initial conditions for each of the four sets of realisations: 
Gaussian and PNG with $\fnl=250$. We note that this value is much higher than the latest Planck constraint 
on   $\fnl$, which reads $2.5 \pm 5.7$ for temperature data alone and $ 0.8 \pm 5.0 $ when combining the temperature with 
the polarisation data \cite{Ade:2015ava}. 
We use a large value of $\fnl$ to highlight the impact of the local PNG on the clustering statistics of DM halos more easily.  
The simulations were evolved using the public code {\small Gadget2} \cite{Springel:2005mi}, and the DM halos identified 
with the spherical overdensity (SO) code {\small AHF} \cite{Gill:2004km, Knollmann:2009pb}. 
We follow \cite{Tinker:2008ff} and adopt a threshold of $\Delta=200$ times the mean matter density of the Universe.
In what follows, all the results are the average over four realisations, while the error bars represent the 1$\sigma$ 
fluctuations among the realisations.

To construct the proto-halo catalogues, we trace the DM particles that belong to a virialized halo at redshift $z_\ast$
back to the initial redshift $z_i$. Their centre-of-mass position furnishes an estimate of the position of the proto-halo 
centre. We consider three different values of $z_\ast=0$, 1 and 2, and split the halo catalogues into three different 
mass bins.  The properties of the halo catalogues used in this paper is shown in Table  \ref{tab:HaloSample}. 

The proto-halo centres and the initial DM particles are interpolated  onto a regular cubical grid using the Cloud-in-Cell 
(CIC) algorithm to generate the fields $\delta_\text{h}(\vx,z_i)$ and $\delta_\text{m}(\vx,z_i)$. 
The cross-bispectra $B_\text{hmm}$ and $B_\text{mhm}$  are computed following \cite{Scoccimarro:2003wn}. 
We consider isosceles triangular configurations with one long mode and two short modes, $(\kL,\ks,\ks)$, up to a maximum 
wavenumber equal to $120 k_{\rm F}$, where  $k_{\rm F}= 2 \pi / L_{\rm box}$ is the fundamental mode of the simulations. 

\subsection{Halo mass function and local bias parameters}
The halo-matter cross-bispectra will be compared to those predicted by the ESP, local bias and the S12i model suitably averaged over halo mass. Namely, we will display
\begin{equation}
B_{\small \textsc{XYZ}}(\bar M,k_{\rm l},k_{\rm s},k_{\rm s};z_i) \equiv 
\frac{1}{\int_{M_\text{min}}^{M_\text{max}}\!dM\,\bar{n}_\text{h}(M)}
\int_{M_\text{min}}^{M_\text{max}}\!dM\, \bar{n}_\text{h}(M)\, B_{\small \textsc{XYZ}}(k_{\rm l},k_{\rm s},k_{\rm s};z_i) \;,
\end{equation}
where $M_\text{min}$ and $M_\text{max}$ are the minimum and maximum halo mass of a given bin.
For the excursion set peaks and local bias predictions, we will use the ESP mass function to perform the mass averaging.
The ESP mass function is given by 
\begin{equation}
\label{eq:nhalo}
\bar{n}_{\rm h}(M) = \frac{\bar{\rho}_m}{M^2}\, \nu_c f_\esp(B)\,\frac{d\log\nu_c}{d\log M} \;.
\end{equation}
where $\bar \rho_m$ is the mean comoving matter density, and $f_\esp(B)$ is the multiplicity function of excursion 
set peaks, which is generally a function of the collapse barrier $B$. 

Fig.~\ref{fig:nmz0_z2} shows the logarithmic mass function of our simulated SO halos at redshifts $z_\ast=0$, 1 and 
2. The error bars show the scatter among the realisations.
The curves represent the ESP theoretical predictions based on the square-root barrier $B=\delta_c+\beta\sigma_0$ with 
lognormal scatter $\beta$ described in Section \S\ref{sec:esp}.
While our ESP, parameter-free prediction is reasonably good at redshift $z_\ast=0$, it underestimates
the abundance of massive halos at higher redshift. 
Note, however, that we have used the same mean $\big\langle\beta\big\rangle$ and variance Var$(\beta)$ at all redshift,
even though these were inferred from halos which virialized at $z_\ast=0$ only (see \cite{Tinker:2008ff} for details).
It is plausible that the mean and variance in the linear collapse threshold depend on redshift. In particular, an 
increase in the variance with redshift would amplify the Eddington bias (i.e. more low mass halos would be scattered 
into the high mass tail) and, therefore, improve the agreement with simulations. 
Notwithstanding, we will not explore this issue any further here, and use the barrier shape at $z_\ast=0$ throughout.

For the S12i model, we use Eq.~(\ref{eq:b1_NG_standard}) with the Gaussian scale-independent bias directly measured 
in the simulations, so that the tree-level  results are already averaged over the mass range of a halo bin. 
Note, however, that the $b_3$-loop is evaluated at the mean mass of the halo bin.

We have not checked the extent to which the ESP predictions for the usual scale-independent, or local bias 
parameters $b_n\equiv b_{n00}$ agree with the simulations. 
Previous studies based on 1-point cross-correlation measurements \cite{Paranjape:2012jt,Paranjape:2013wna} and the 
separate Universe approach \cite{Lazeyras:2015lgp} have shown that the ESP predictions up to $b_{300}$ are accurate 
at the $\lesssim 10$\% percent level for massive halos. 
However, it is pretty clear that the density peak approximation eventually breaks down at low mass 
\cite{Ludlow:2010xd}. 
This is also reflected in the measurement of $\chi_1$ performed by \cite{Biagetti:2013hfa} which shows that, while
$\chi_1$ is still negative for $M\sim 10^{13}\Msun$, it does not assume the value $\chi_1=-3/(2\sigma_1^2)$ predicted
by the peak constraint.

\subsection{Halo-matter cross-bispectra}

\begin{figure}[H]
\centering
\begin{minipage}[]{0.49\textwidth}
\hspace{.1in}\includegraphics[width=3.1in,height=2.5in]{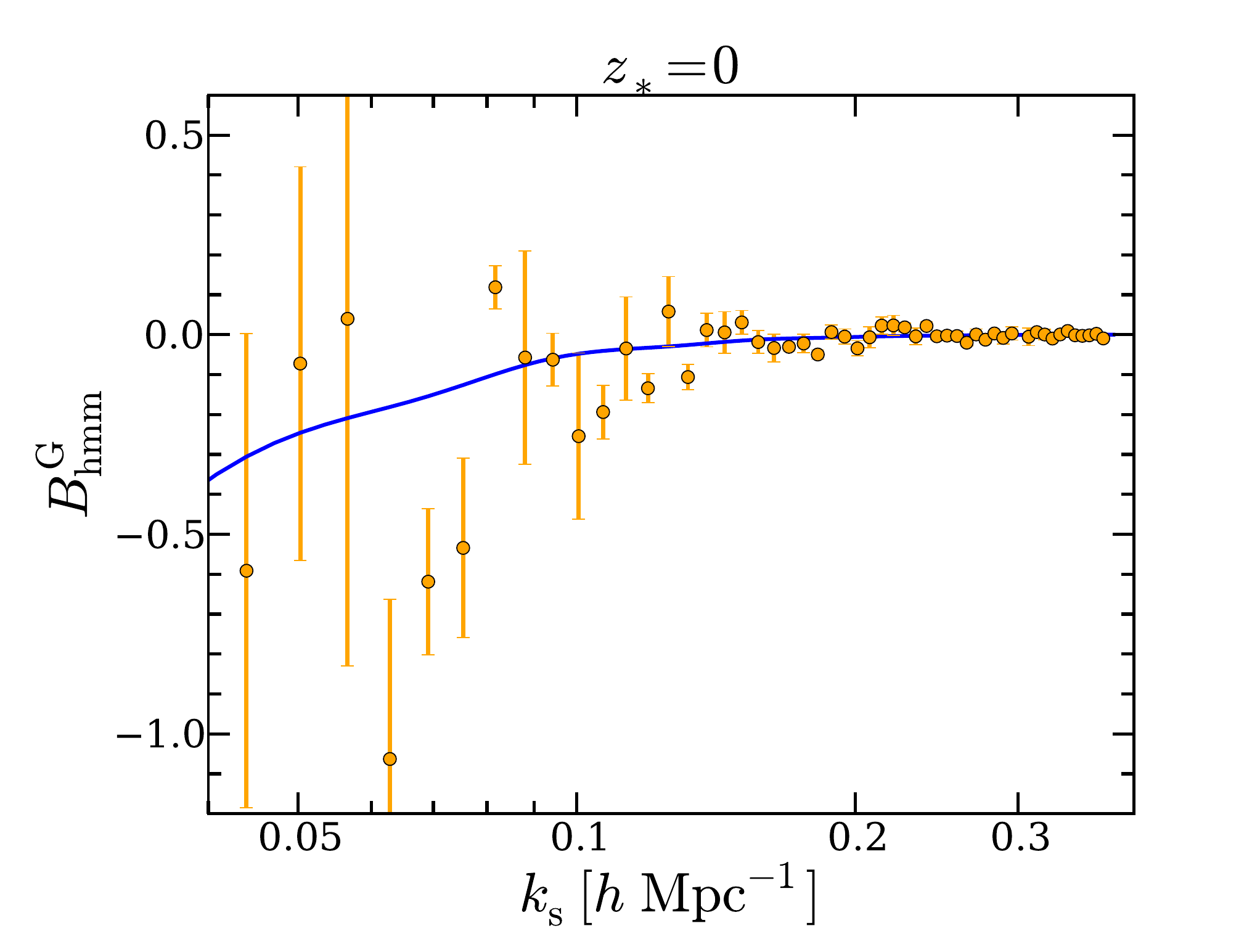}
\end{minipage}
\vspace{-.07in}
\begin{minipage}[]{0.49\textwidth}
\hspace{-.1in}\includegraphics[width=3.1in,height=2.5in]{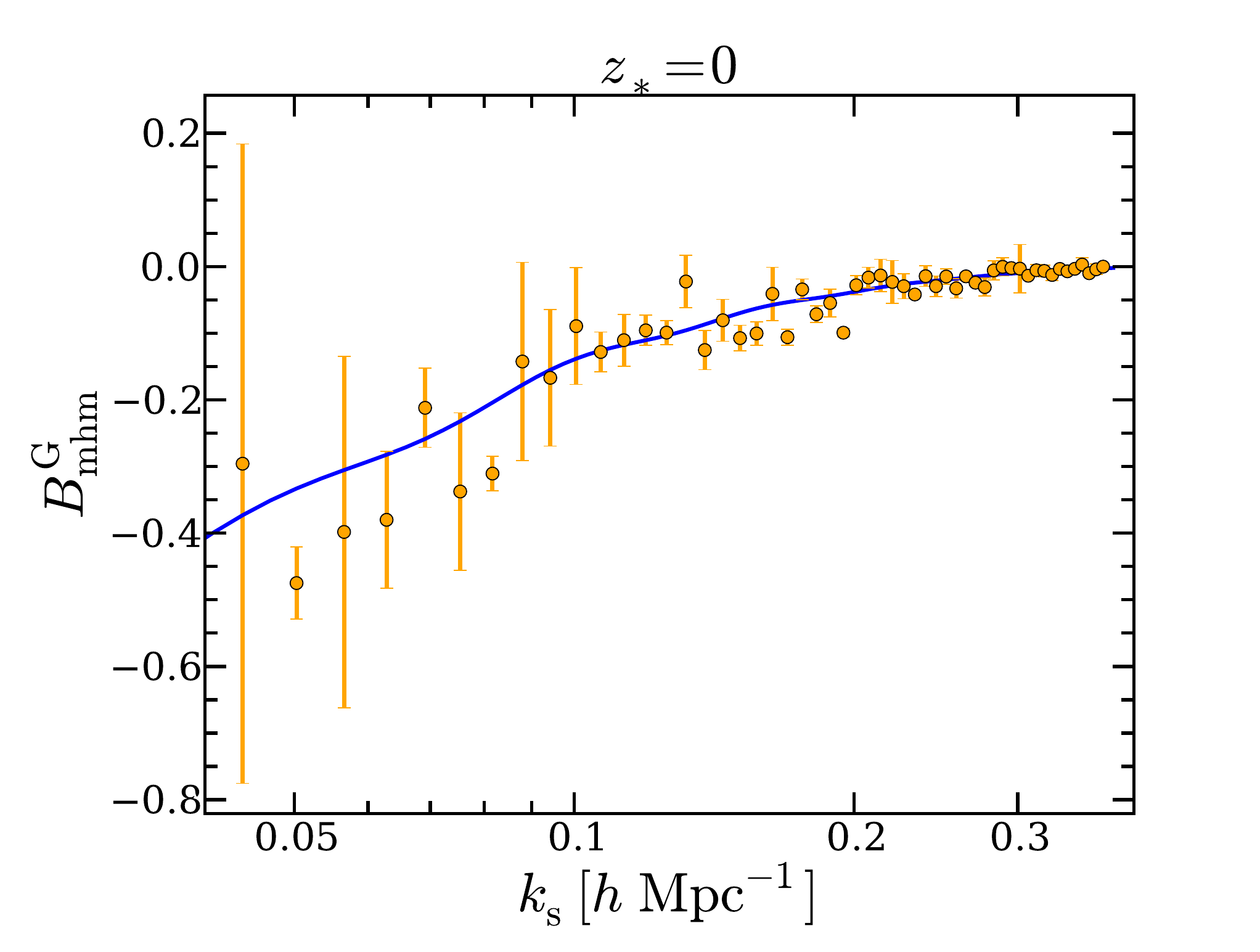}
\end{minipage}
\vspace{-.03in}
\begin{minipage}[]{0.49\textwidth}
\hspace{.1in}\includegraphics[width=3.1in,height=2.5in]{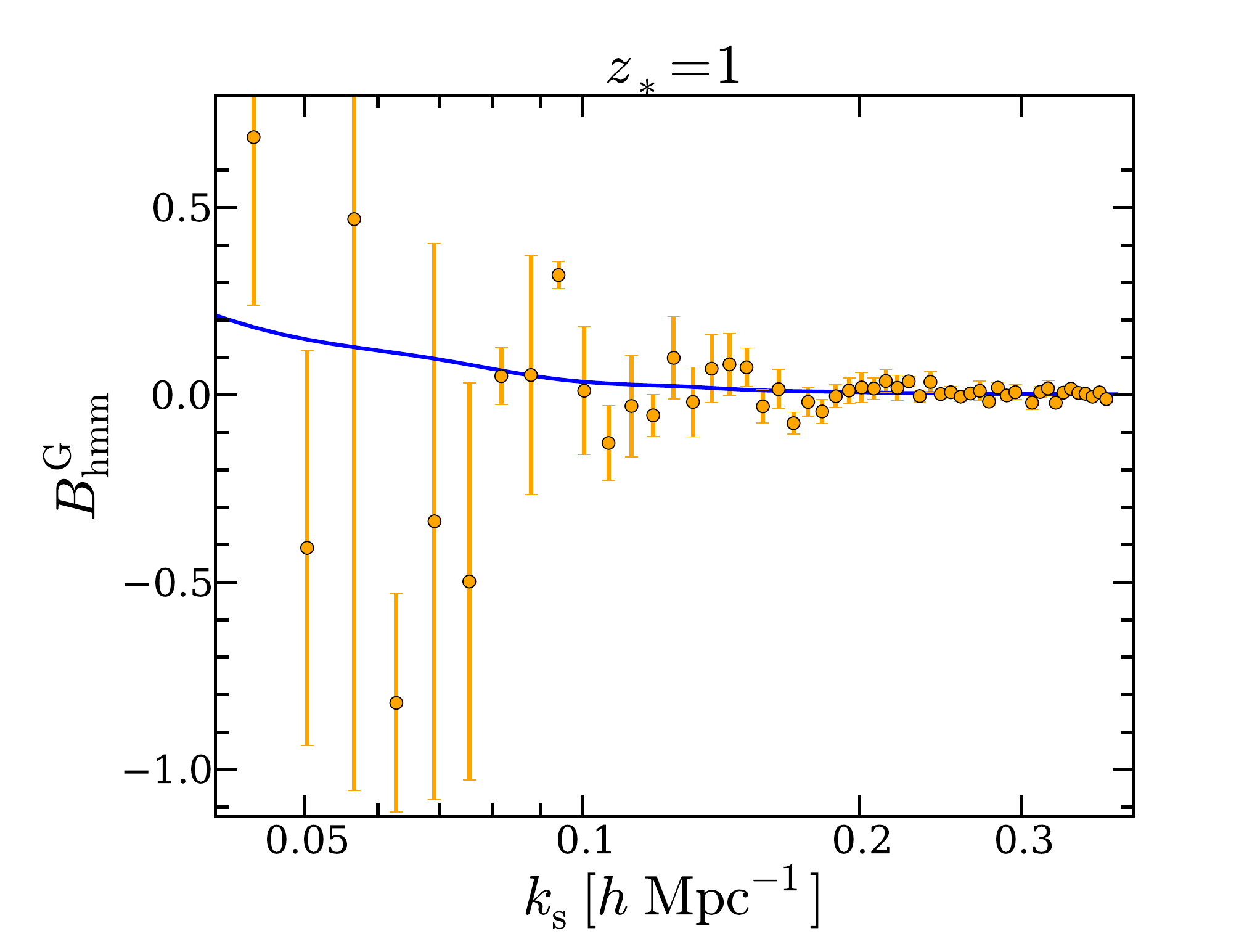}
\end{minipage}
\vspace{-.03in}
\begin{minipage}[]{0.49\textwidth}
\hspace{-.1in}\includegraphics[width=3.1in,height=2.5in]{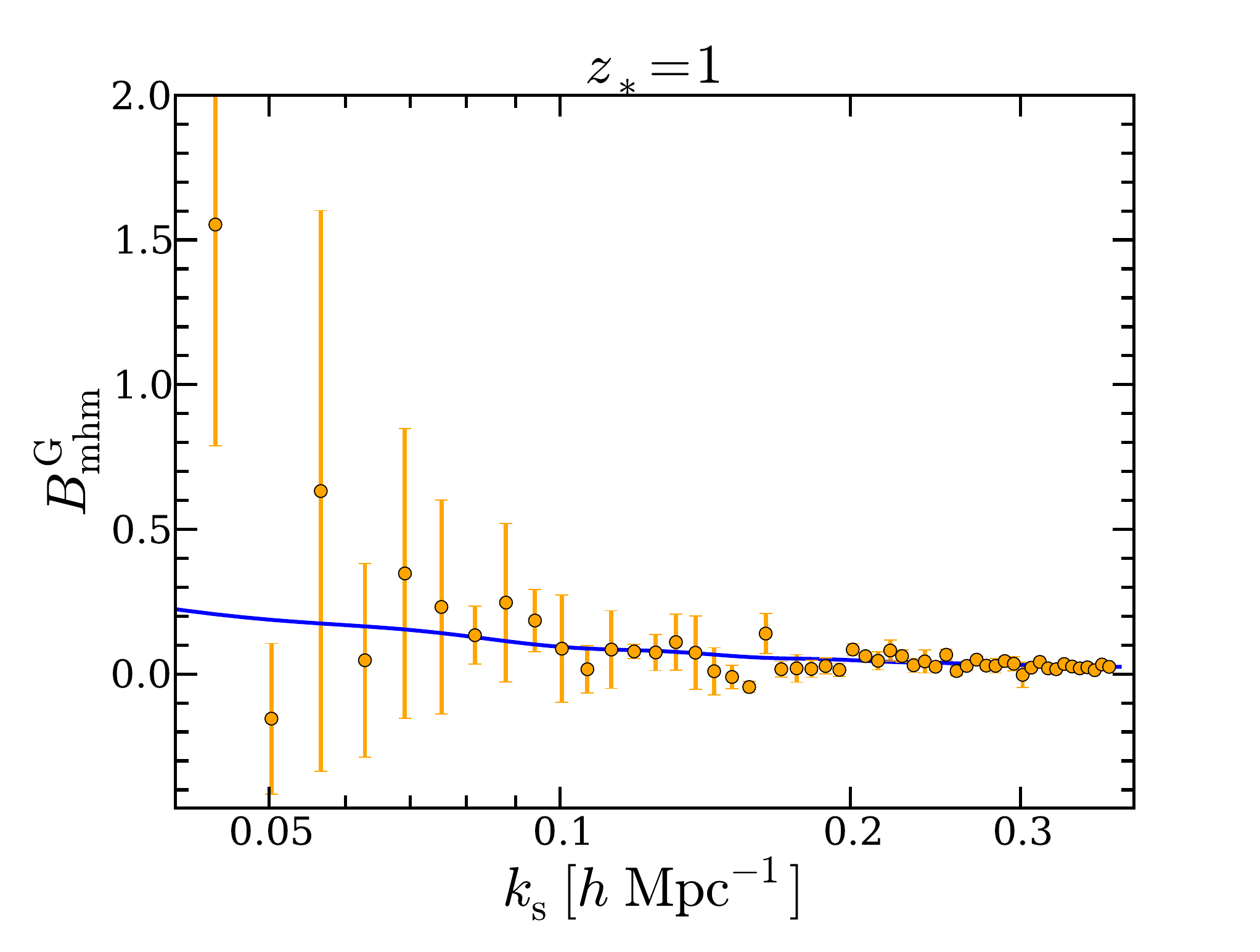}
\end{minipage}
\vspace{-.03in}
\begin{minipage}[b]{0.49\textwidth}
\hspace{.1in}\includegraphics[width=3.1in,height=2.5in]{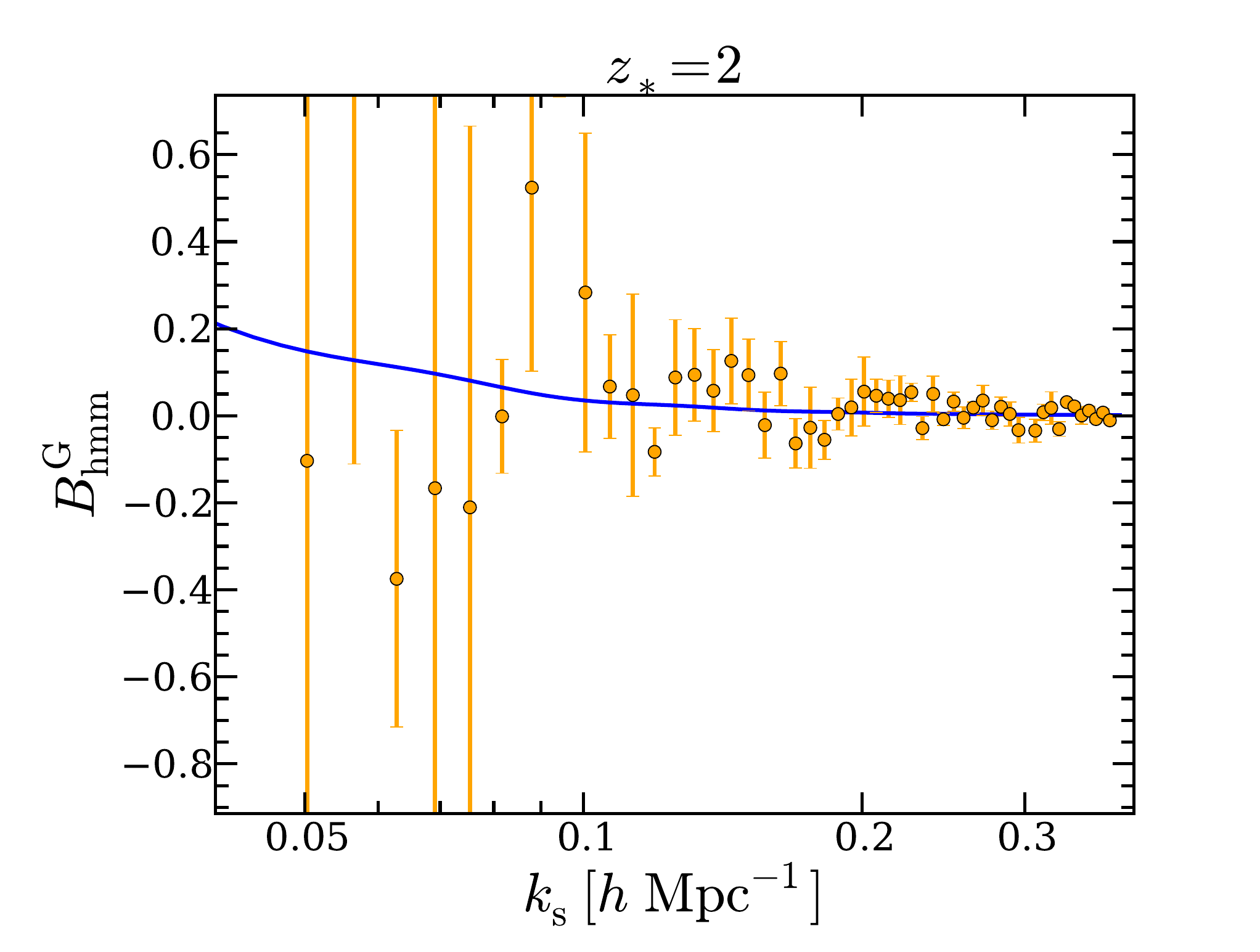}
\end{minipage}
\begin{minipage}[b]{0.49\textwidth}
\hspace{-.1in}\includegraphics[width=3.1in,height=2.5in]{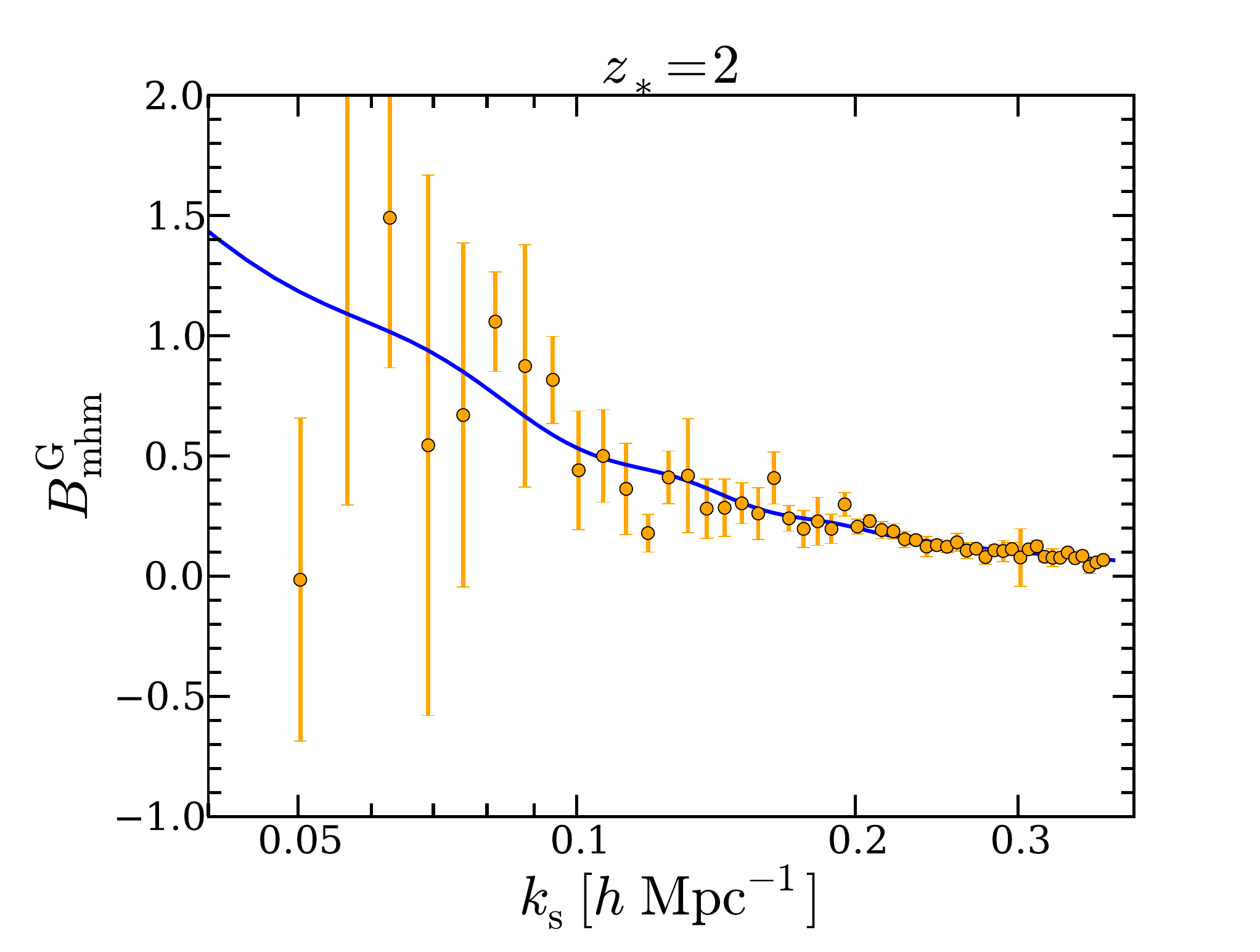}
\end{minipage}
\caption{The halo-matter-matter bispectrum for three redshifts $z_*= 0, 1, 2$ for Gaussian initial conditions. The halos are chosen from bin2 as defined in Table \ref{tab:HaloSample}. The plots on the left show the bispectra when the squeezed Fourier mode corresponds to the halo overdensity while on the right we show the matter-squeezed case. The solid line is the theoretical prediction of the ESP model including only the tree-level Gaussian contribution while the points are simulation measurements. The triangular configuration is chosen such that two sides are equal to $k_{\rm s}$ and the third equal to $k_{\rm l}$. The long-wavelength mode is fixed at $k_{\rm l}= 0.006 \ \mathrm{Mpc}^{-1} h$ while varying the short mode $k_{\rm s}$. }
\label{fig:G_sim}
\end{figure}

We begin with a consistency check and plot, in Fig.~\ref{fig:G_sim}, the measured halo-matter-matter bispectrum in the 
initial conditions ($z_i=99$) of the Gaussian simulations for the three collapse redshifts $z_\ast=0,1,2$. 
The halos are chosen from the second mass bin in our simulations with the mass range and mean mass given in Table \ref{tab:HaloSample}. The left and right panels show the case where the squeezed mode corresponds to the halo ($B_\text{hmm}^G$) and the matter overdensity ($B_\text{mhm}^G$), respectively.
We can understand the figure using the tree level halo-matter-matter cross-bispectrum, which consists of two terms: 
\begin{align}
\langle \delta_{\rm h} ( \mb{k}_1, z_i )   \delta_{\rm m} ( \mb{k}_2 , z_i) \delta_{\rm m} ( \mb{k}_3 ,z_i )  \rangle'_G  &\approx  \left(\frac{D(z_i)}{D(z_\ast)}\right)^2 c_2^L(\vk_2,\vk_3) P_0(k_2) P_0(k_3) \\
&  \hspace{-.6in}+  \left(\frac{D(z_i)}{D(z_\ast)}\right)^3 c_1^L(k_1)P_0(k_1) \Big[F_m^{(2)}(\vk_1,\vk_2)\, P_0(k_2)+ F_m^{(2)}(\vk_1,\vk_3)\,P_0(k_3)\Big] \nonumber \;,
\end{align}
where the prime denotes the fact that we have neglected the factors of $2\pi$ and the Dirac delta function. We have ignored the intrinsic non-linearity in the gravitational evolution of the halo number over-density (see Eq.~(\ref{eq:GamEspApprox})), which is always subdominant in the initial conditions. In Fig.~\ref{fig:G_sim}, we plot the ESP tree-level bispectrum. This bispectrum does not generally vanish away from the squeezed limit, i.e. for $\ks \lesssim 0.1 \hOMpc $. As $\ks$ increases however, the halo-squeezed or matter-squeezed bispectra rapidly decreases because of the power spectrum  suppression.  In the matter-squeezed case, the $c_2^{L}$-term is dominant over the $c_1^{L}$-term because of one less power of the growth factor.  In the halo-squeezed case, say $k_2, \,  k_3 \gg  k_1 $,  because we generally have  $P_0 (k_2), \, P_0 (k_3)  \ll P_0(k_1) $, the second term is boosted by a factor  $ P_0(k_1) / P_0 ( k_2)$. Thus as the short mode $k_s$ increases, the $c_2^{L}$-term dominates at first, but the  $c_1^{L}$-term takes over when the configuration is sufficiently squeezed. Overall, given large scatter of the data, the Lagrangian, halo-mass cross-bispectrum cannot easily distinguish between different models. 
 
\begin{figure}[H]
\centering
\begin{minipage}[]{0.49\textwidth}
\includegraphics[width=3.1in,height=2.5in]{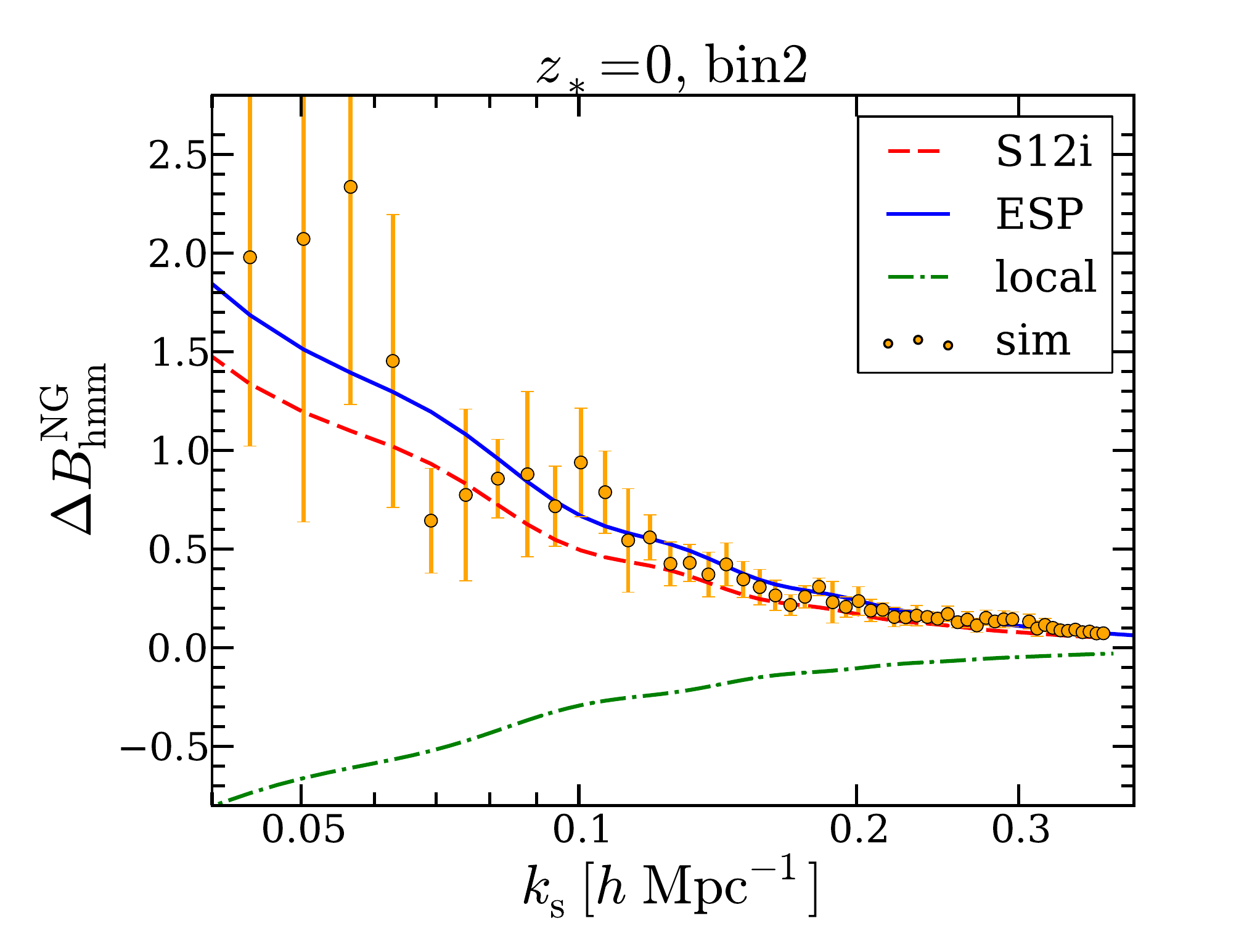}
\end{minipage}
\begin{minipage}[]{0.49\textwidth}
\hspace{-.2in}\includegraphics[width=3in,height=2.5in]{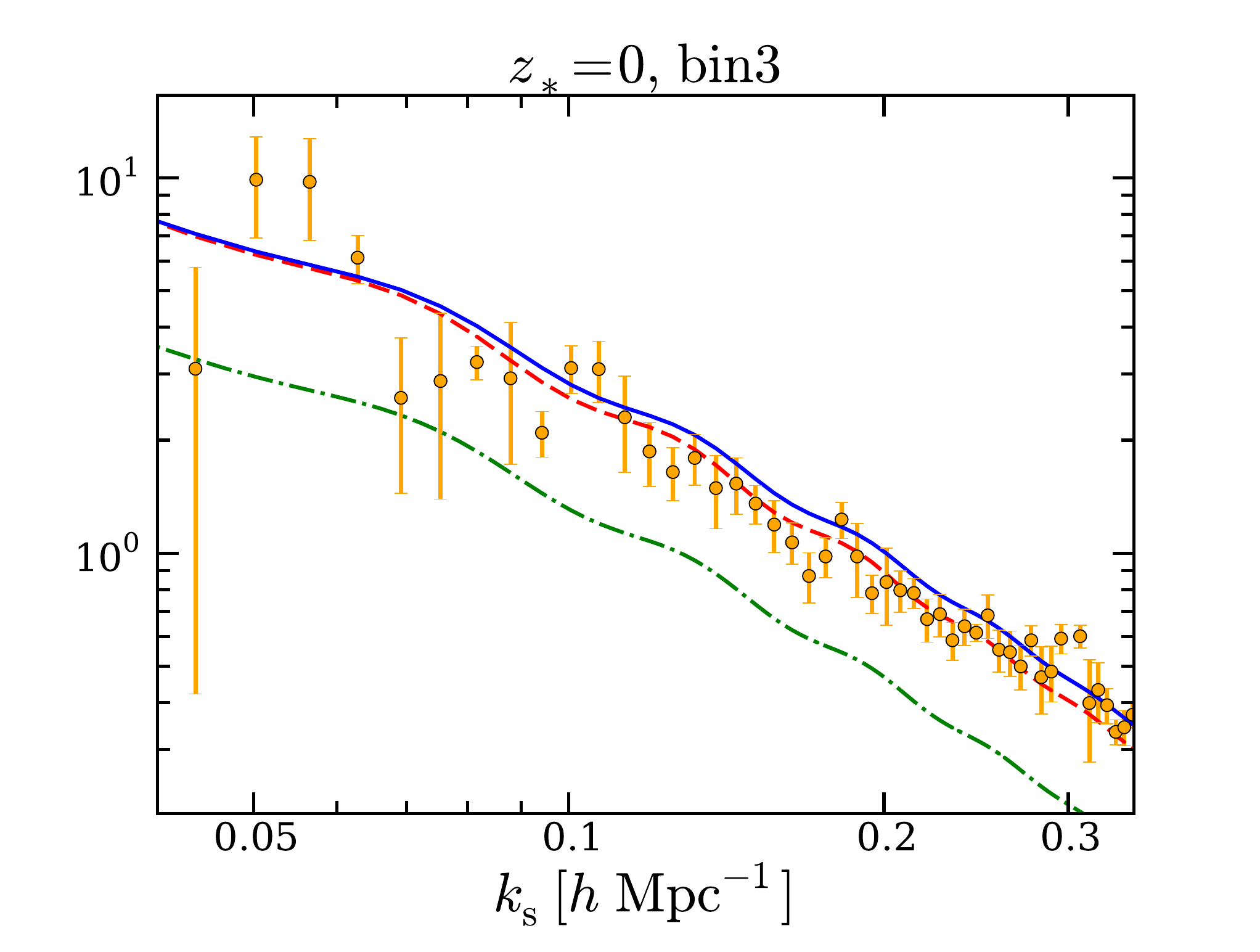}
\end{minipage}
\begin{minipage}[b]{0.49\textwidth}
\includegraphics[width=3.1in,height=2.5in]{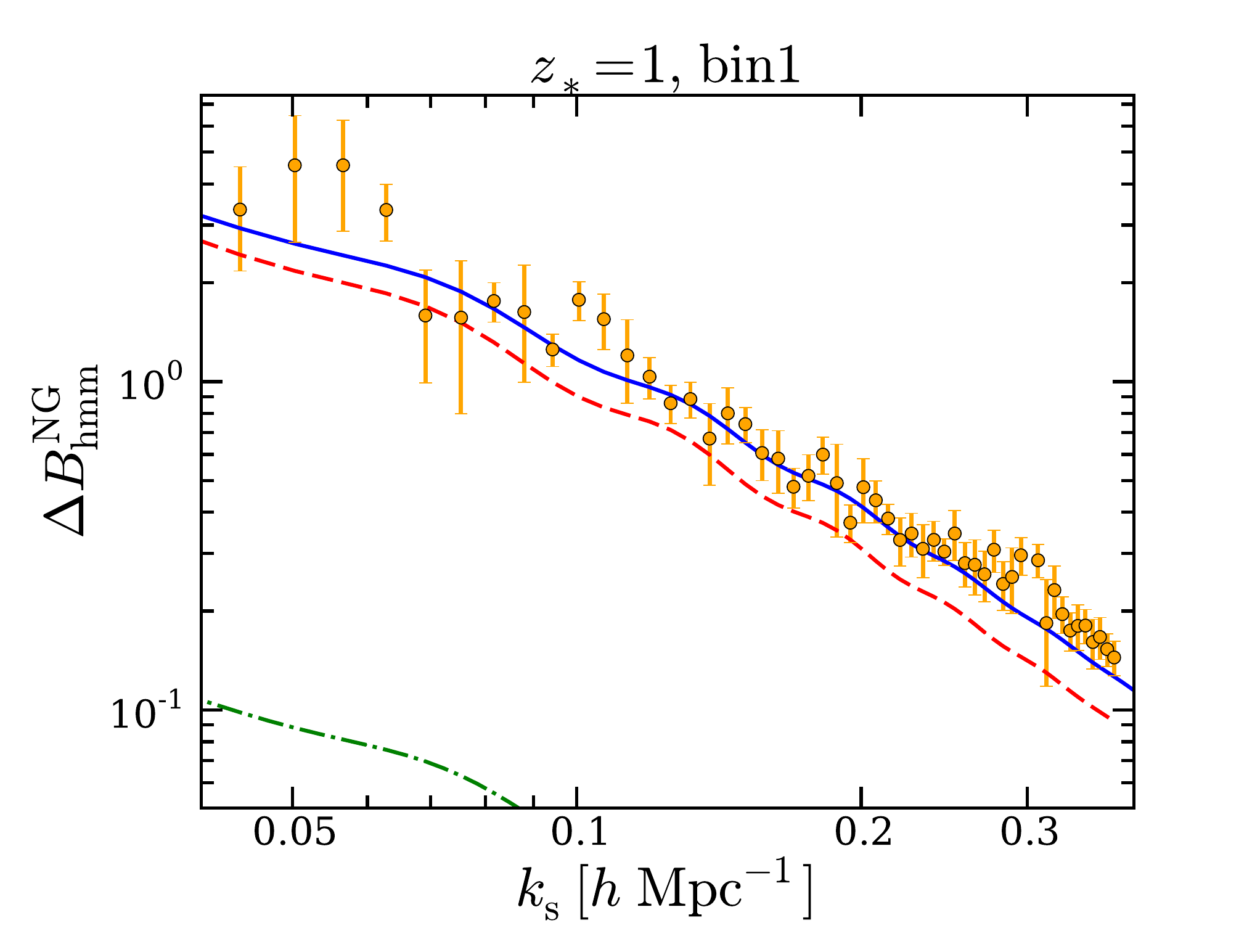}
\end{minipage}
\begin{minipage}[b]{0.49\textwidth}
\hspace{-.2in}\includegraphics[width=3in,height=2.5in]{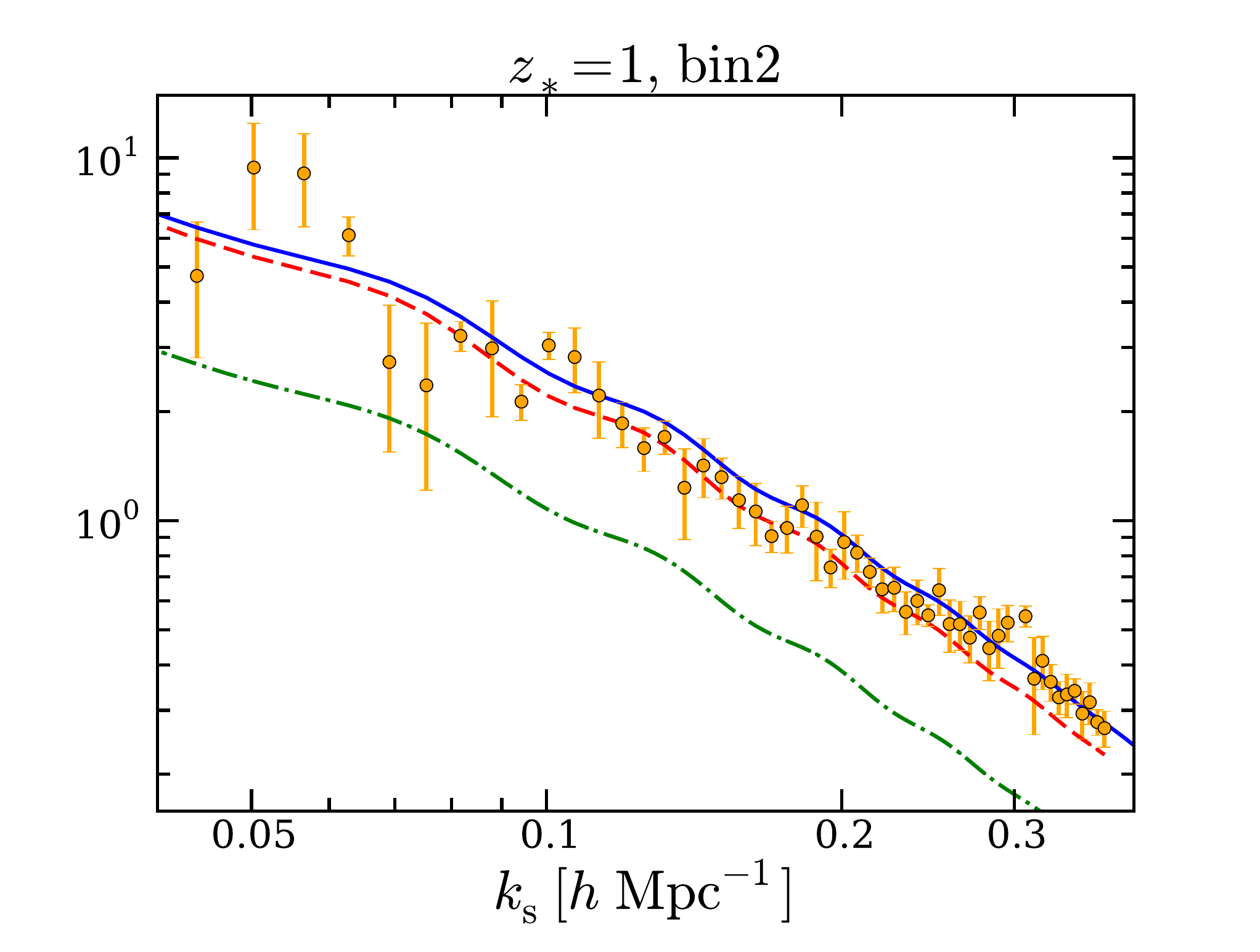}
\end{minipage}
\begin{minipage}[b]{0.49\textwidth}
\includegraphics[width=3.1in,height=2.5in]{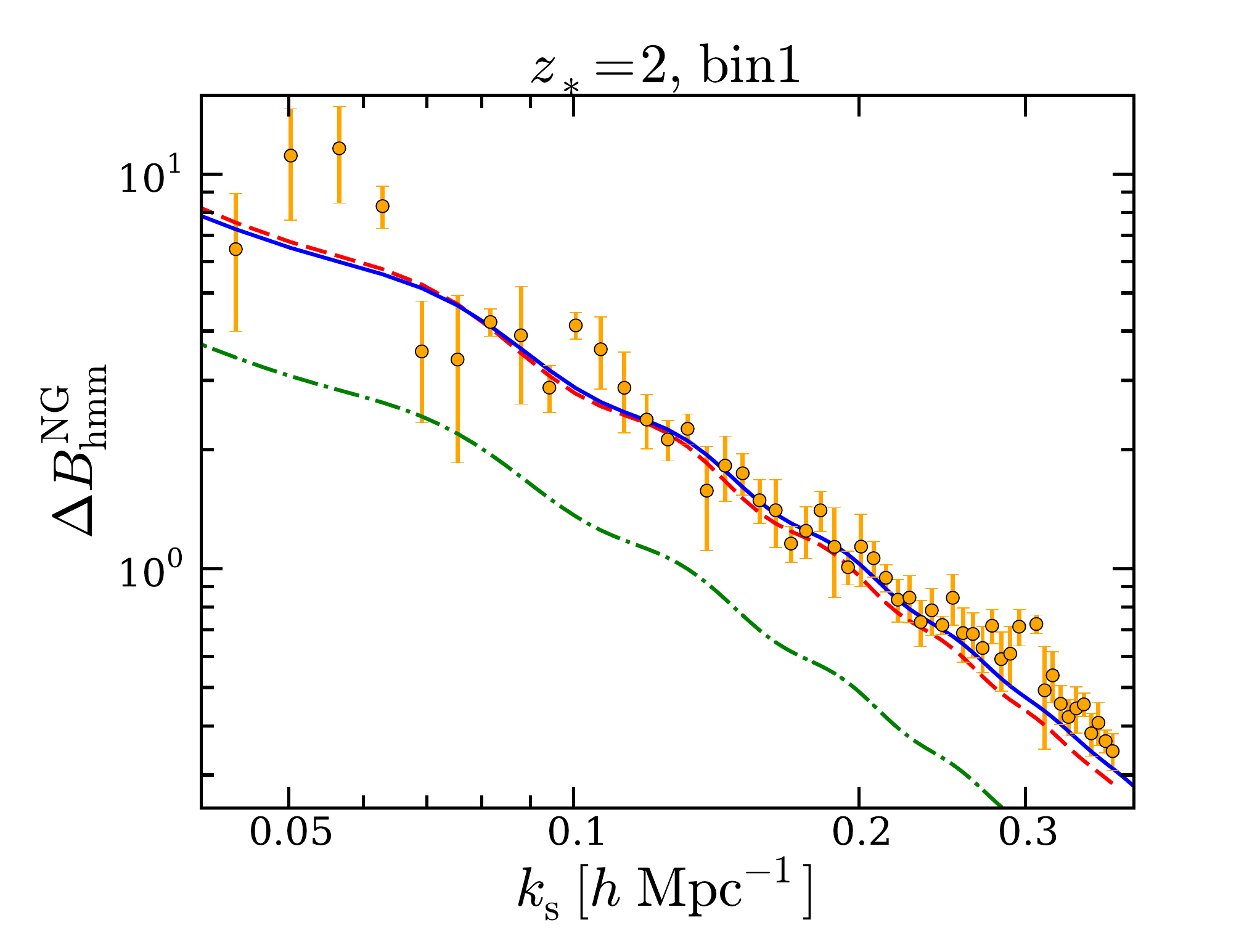}
\end{minipage}
\begin{minipage}[b]{0.49\textwidth}
\hspace{-.2in}\includegraphics[width=3in,height=2.5in]{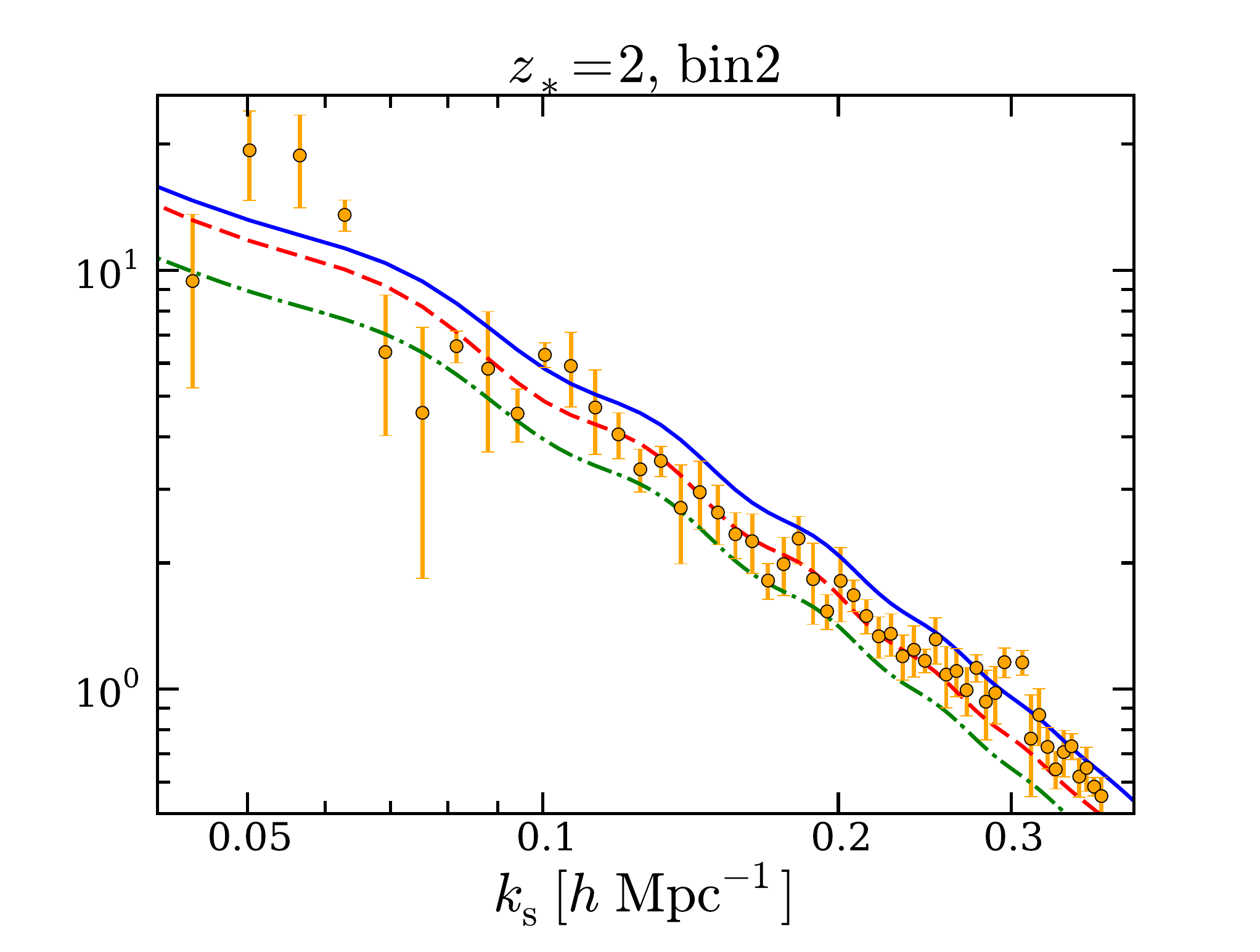}
\end{minipage}
\caption{Halo-matter-matter bispectrum, where the squeezed Fourier mode corresponds to the halo overdensity. The three rows correspond to redshifts $z_*= 0,1,2$. The mass range and mean masses of the three mass bins denoted as bin1, bin2 and bin3 are given in Table \ref{tab:HaloSample}. The solid line is the ESP model prediction while the dashed line is the S12i prediction and the dashed-dotted line is the prediction of the local model. The triangular configuration is the same as in Fig. \ref{fig:G_sim}}.
\label{fig:HSQ1_sim}
\end{figure}
\begin{figure}[H]
\centering
\begin{minipage}[]{0.49\textwidth}
\includegraphics[width=3.1in,height=2.5in]{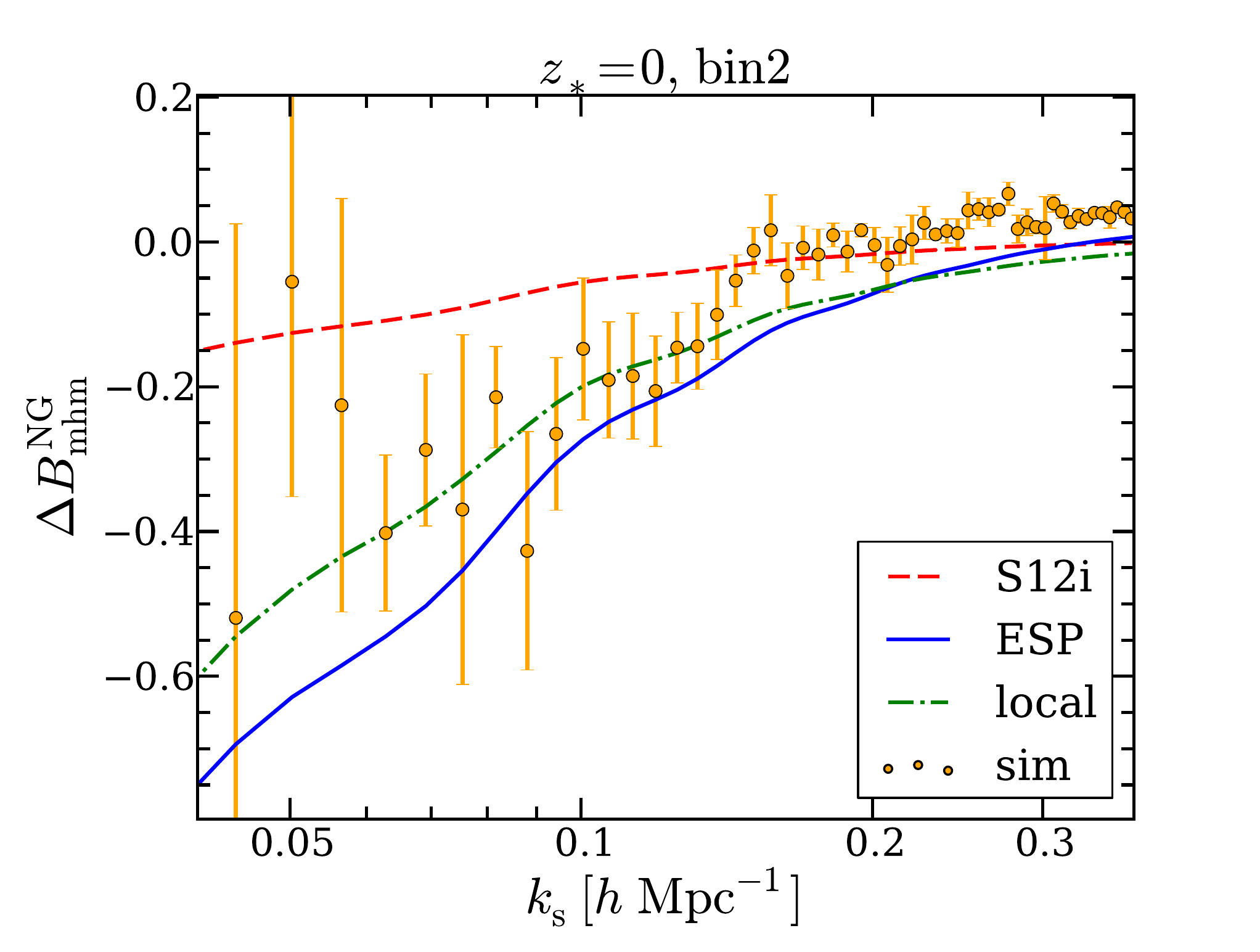}
\end{minipage}
\begin{minipage}[]{0.49\textwidth}
\hspace{-.2in}\includegraphics[width=3in,height=2.5in]{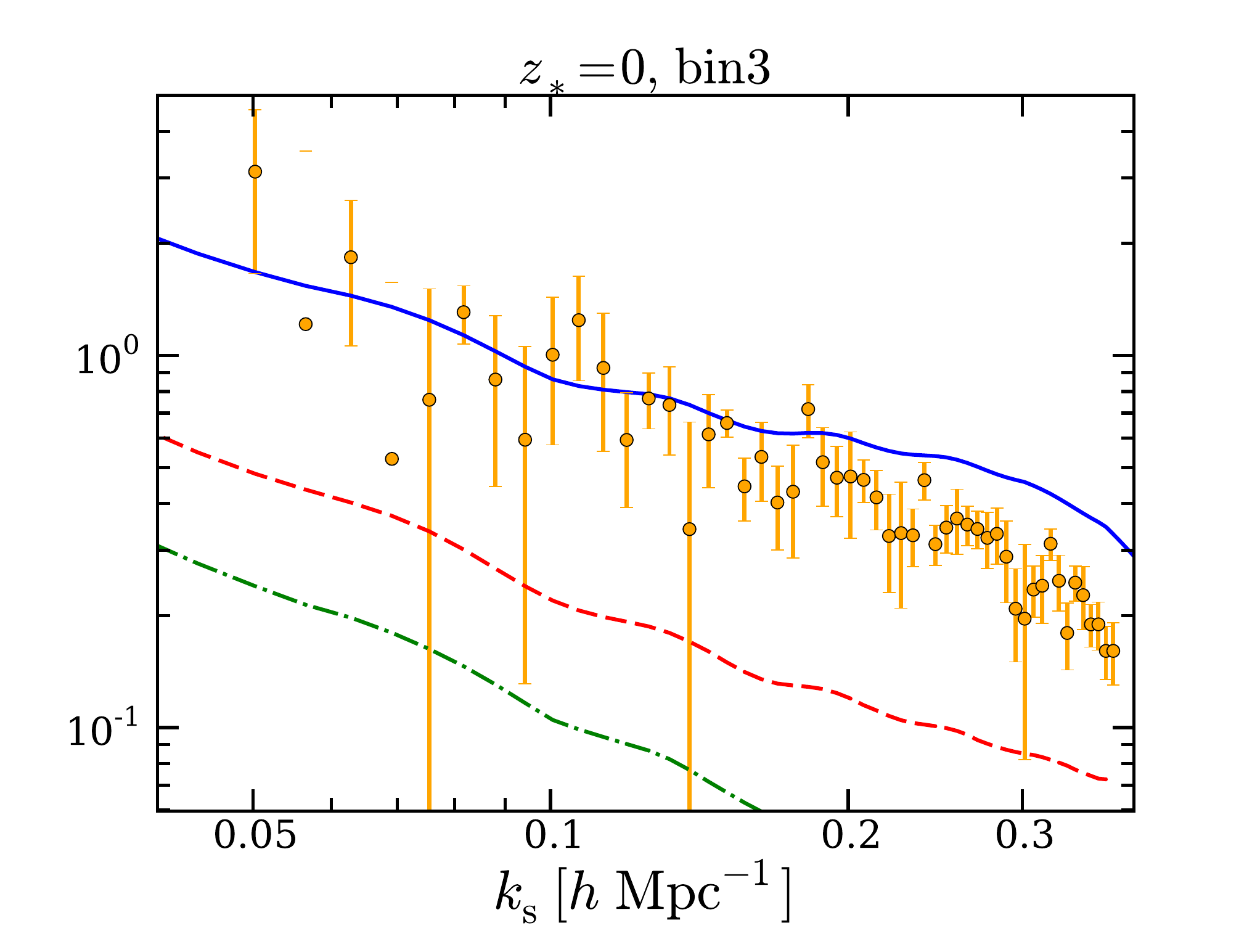}
\end{minipage}
\begin{minipage}[b]{0.49\textwidth}
\includegraphics[width=3.1in,height=2.5in]{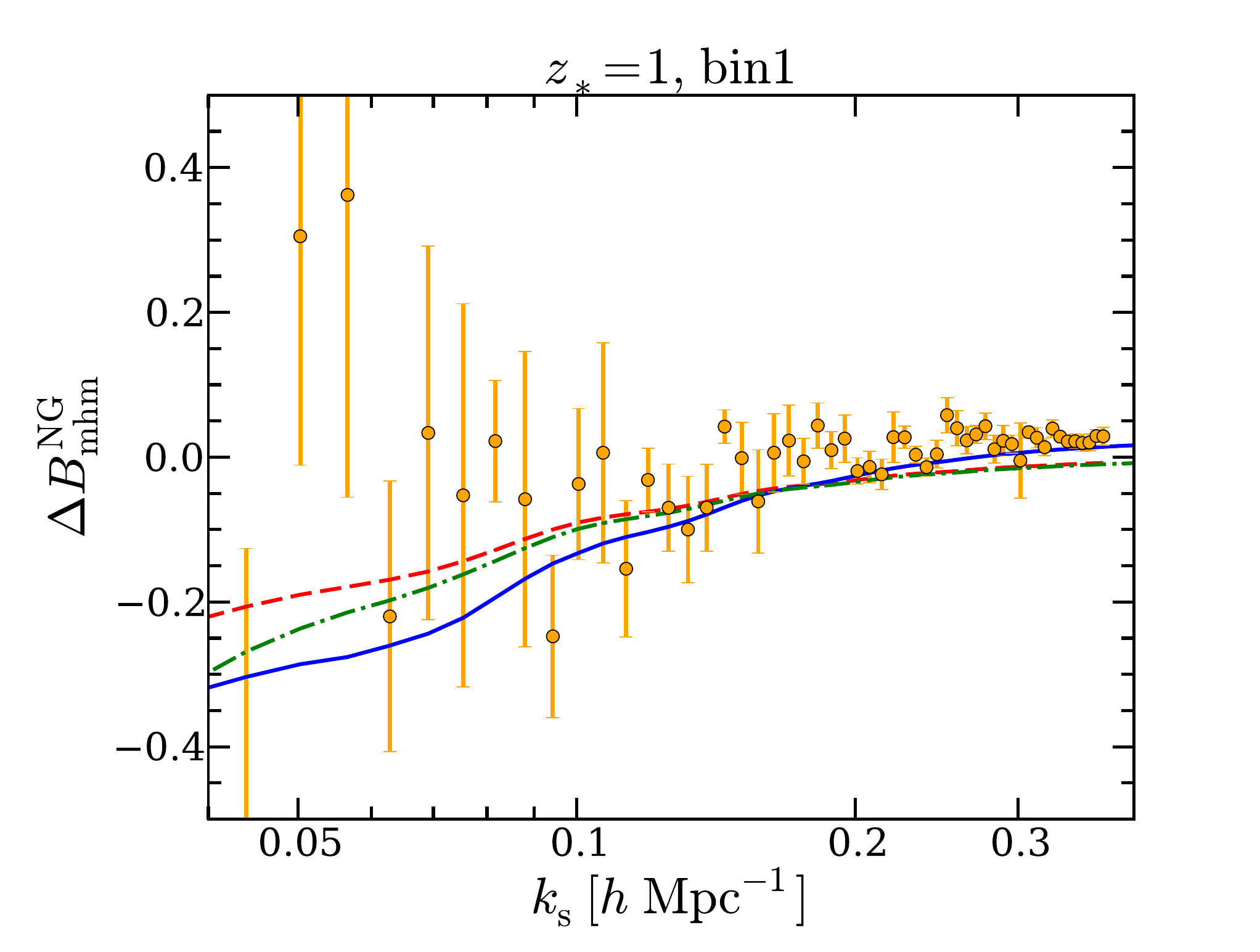}
\end{minipage}
\begin{minipage}[b]{0.49\textwidth}
\hspace{-.2in}\includegraphics[width=3in,height=2.5in]{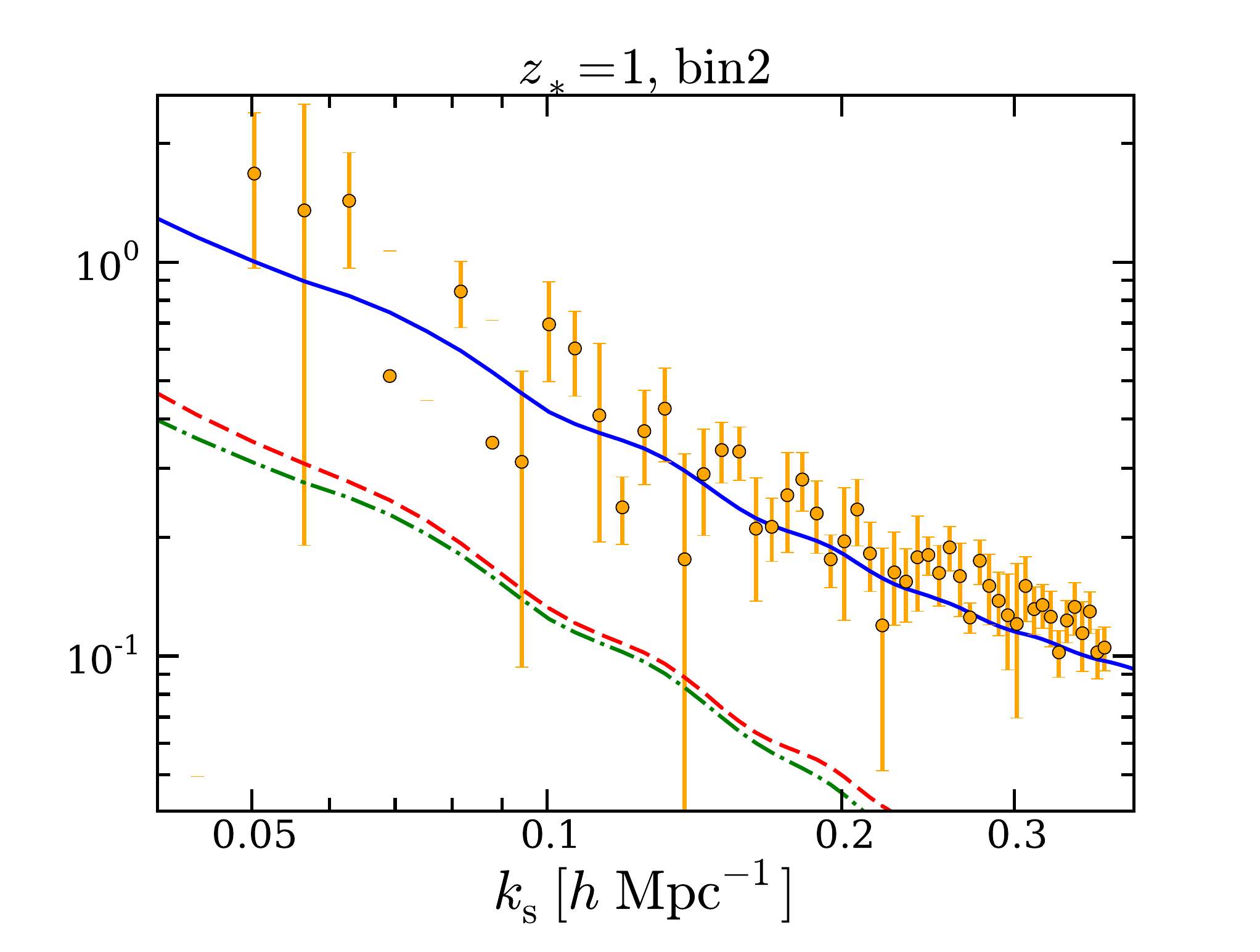}
\end{minipage}
\begin{minipage}[b]{0.49\textwidth}
\includegraphics[width=3.1in,height=2.5in]{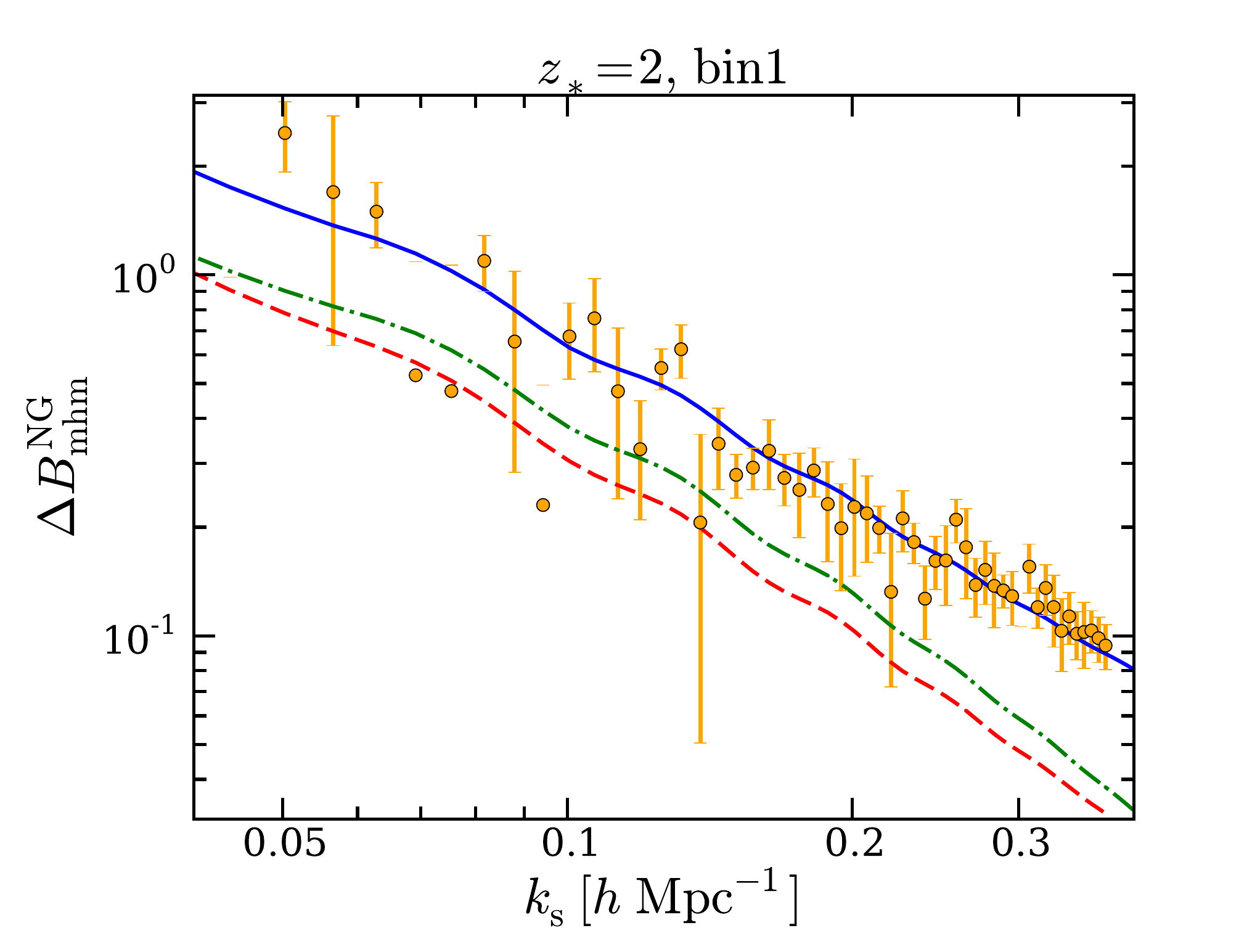}
\end{minipage}
\begin{minipage}[b]{0.49\textwidth}
\hspace{-.2in}\includegraphics[width=3in,height=2.5in]{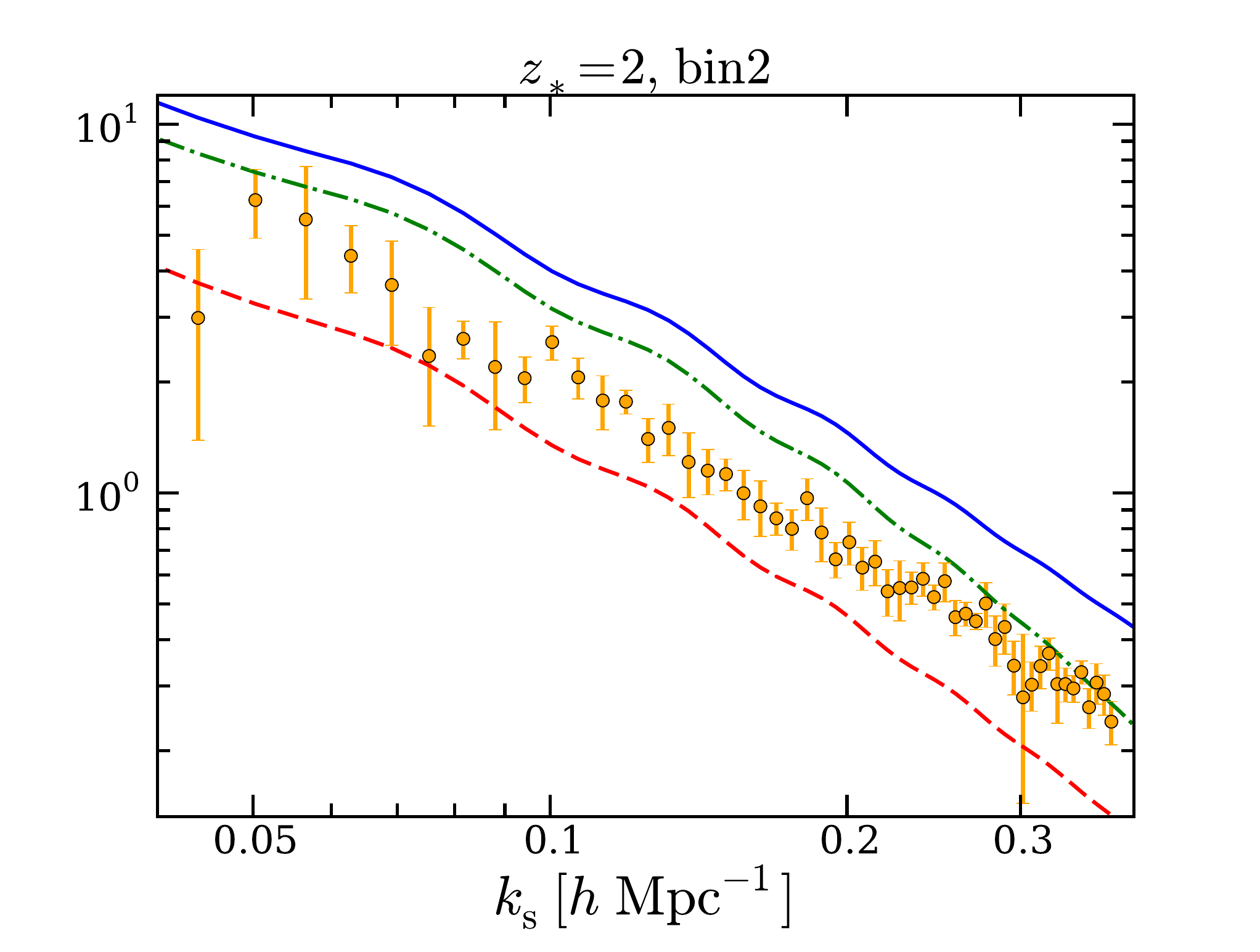}
\end{minipage}
\caption{Halo-matter-matter bispectrum, where the squeezed Fourier mode corresponds to the matter overdensity. The mass bins and the redshifts are the same as in Fig.~\ref{fig:HSQ1_sim}. The solid line is the ESP model prediction while the dashed line is the S12i model prediction and the dashed-dotted line is the prediction of the local model. The triangular configuration is the same as in Fig. \ref{fig:G_sim}}
\label{fig:MSQ2_sim}
\end{figure}
\hspace{-.35in}

We will now compare the non-Gaussian contributions to the cross-bispectra 
$\Delta B_\text{hmm}^{NG}$ and $\Delta B_\text{mhm}^{NG}$ measured in the simulations to those predicted by the ESP,
the local Lagrangian bias and the S12i model.
Figs. \ref{fig:HSQ1_sim} and \ref{fig:MSQ2_sim} display the results for the halo-squeezed and matter-squeezed case,
respectively.
To ensure that the halo mass $M$ is significantly larger than $M_\star(z_\ast)$ and, thus, comply with the spherical
collapse approximation, we consider the high mass bins (bin2 and bin3) at redshift $z_\ast=0$.
At $z_\ast=1$ and 2 however, $M_\ast\lesssim 10^{12}\hmsun$ such that we use the two lowest mass bins (bin1 and bin2),
whereas we discard bin3 which, owing to the low number density of these massive halos, does not provide good statistics.
We plot our results as a function of the wavenumber $\ks$ of the short mode while we keep the long mode fixed at 
$\kL = 2 k_F$.

For the halo-squeezed case shown in Fig. \ref{fig:HSQ1_sim}, our results indicate that, 
while there is a noticeable discrepancy between the prediction of the local bias model and simulations, the predictions of both the ESP approach and the S12i model are a better fit to simulations. In this case, the dominant loop contribution is proportional to the second order bias. 
For matter squeezed case shown in Fig. \ref{fig:MSQ2_sim}, the predictions from the S12i model and the local bias model are 
qualitatively similar and neither of the two is in a good agreement with our simulations.  In this case, the dominant loop contribution which is as large as the tree-level comes from the $b_3$-loop. In both models this contribution is included, assuming the third order bias $b_3$ to be constant. The fact that the S12i model works well in the halo-squeezed case, while it fails in the matter-squeezed limit suggests that the scale-dependence of $b_3$ can not be neglected. 

Notwithstanding, although the ESP model furnishes the best fit to all the numerical data, they appear to overestimate 
strongly the measurements for the larger mass bin at $z_\ast=2$. 
The discrepancy is more pronounced for $\Delta B_\text{mhm}^{NG}$ than for $\Delta B_\text{hmm}^{NG}$.
Even though we have not explored this issue in details, we suspect this might be due to the fact that, in our 
approximation, the bias coefficients are calculated from the Gaussian probability density whereas they should, in fact, be computed from the non-Gaussian PDFs. These ${\cal O}(\fnl^2)$ corrections are expected to increase with the halo mass $M$ and the value of $\fnl$. Furthermore, they have a sign opposite to the first-order non-Gaussian contribution and, e.g., would lower the theoretical predictions for $\fnl>0$.

\section{Conclusion}
\label{sec:conclusions}
Clustering statistics of the large scale structure provide a wealth of information both on the initial conditions of cosmic structure formation and on its subsequent gravitational evolution. Higher-order statistics such as galaxy bispectrum offer additional useful information to that which is accessible through power spectrum measurements. Furthermore, they provide a consistency check for the modelling of non-linearities in the power spectrum. Finally, they are natural observables for constraining primordial non-Gaussianity (PNG). However, extracting robust information from higher-order galaxy correlations presents several challenges, of which galaxy bias is one of the most significant. The relation between the observed galaxies and the underlying mass distribution generally is non-linear and scale-dependent. Therefore, it is essential to consider extensions to the simple local bias model, in which the bias parameters are assumed to be constant. 

In this work we have focused on halo bias and its impact on the halo-matter cross bispectrum, which involves one halo and two matter overdensity fields. Unlike the halo auto bispectrum, which is noisier and relatively difficult to model, the cross bispectrum offers a fairly clean probe of the halo bias. We have studied its squeezed limit in the presence of primordial non-Gaussianity (PNG) of the local type, which induces a characteristic $1/k^2$ scale-dependence. 
Our fiducial halo bias model is the excursion set peak (ESP) approach. In the squeezed limit, the ESP cross bispectrum can be easily computed using the integrated perturbation theory (iPT). We have also considered a simple local bias scheme, and a model (S12i) in which the NG bias factors are explicitly computed from a peak-background split. 
In all cases, the model parameters are constrained with statistics other than the cross bispectrum.
We have carried out the calculations and the measurements at the initial redshift of the simulations, in order to mitigate the contamination induced by the nonlinear gravitational evolution. This can be thought of as a good approximation for the Lagrangian space.

Interestingly, even the simple squeezed limit is not trivial. We have found that the results strongly depend on the scale-dependence of the quadratic and third-order bias functions. In the halo-squeezed limit (i.e. when the long mode corresponds to the halo fluctuation field), both the ESP and the S12i models are in reasonable agreement with the simulations (see Fig. \ref{fig:HSQ1_sim}). For the matter-squeezed case however, the ESP model fares significantly better than the S12i model (see Fig \ref{fig:MSQ2_sim}). This, however, does not call the peak-background split into question. After all, all the ESP bias factors can be derived from a peak-background split, as shown in 
\cite{Desjacques:2010gz, Desjacques:2012eb, Lazeyras:2015giz}. Rather, this suggests that the S12i bias prescription suffers from inconsistencies.
Finally, the local bias model, which is equivalent to keeping the scale-independent terms in the ESP bias functions, dramatically fails at reproducing the numerical results.
While the dominant loop contribution to the halo-squeezed case is proportional to the second-order bias, the matter-squeezed limit is controlled by the contribution from the third-order bias. Hence, each of them furnishes a test of the modelling of the second and third order halo bias, respectively. It should be noted that these loop corrections are as large as the tree level. Notwithstanding, this does not necessarily indicate the breaking of the perturbative expansion because the higher-order loops turn out to be suppressed (see Sec~\S\ref{sec:sqb}) . 

There are many directions in which this work can be extended. First of all, our measurements are not what will be 
actually measured in real surveys. Future work will take into account the nonlinear gravitational evolution, which 
was omitted here for sake of clarity. 
Still, we do not expect our conclusions to change noticeably because gravitational nonlinearities cannot generate 
a signal in the squeezed limit, as is apparent from Fig.\ref{fig:G_sim}.
In addition, we have found that, within the iPT, it is unclear whether a 
perturbative expansion in terms of small bias  parameters holds, suggested that this formalism could be improved. 
This will involve the renormalisation of the bias coefficients in the presence of PNG along, e.g., the lines of 
\cite{Assassi:2015jqa} (a recent treatment of this issue with a different formalism). 
To perform a similar analysis in Eulerian space, which would match more closely what is measured in galaxy surveys,
a self-consistent calculation of the halo bispectrum $B_\text{hhh}$ must be performed. 
Namely, this should include the effects of stochasticity induced, for example, by halo exclusion 
(see \cite{Baldauf:2015fbu}). Furthermore, it would certainly be interesting to explore other triangular shapes, and
consider PNG of the equilateral or orthogonal type. 
These shapes leave no imprint on the scale dependent bias, so that the galaxy bispectrum likely is our best hope for
for observing them in the future. 
Finally, the potential of the galaxy bispectrum for constraining PNG was investigated in previous works 
\cite{Sefusatti:2007ih, Jeong:2009vd, Sefusatti:2009qh}. 
In these studies however, the biasing relation is modelled using simple local model. It would be interesting to 
revisit their analysis using more sophisticated bias models such as the ESP approach.

\section*{Acknowledgement}

We are grateful to the anonymous referee for his/her helpful comments on an earlier version of this manuscript; and 
we thank Andreas Malaspinas and Yann Sagon for help while running the numerical simulations. 
M.B., K.C.C. and V.D. acknowledge support by the Swiss National Science Foundation. 
A.M. is supported by the Tomalla foundation for Gravity Research. 
K.C.C. also acknowledges support from the Spanish Ministerio de Economia y Competitividad grant  ESP2013-48274-C3-1-P. 
J.N. is supported by the Swiss National Science Foundation (SNSF), project ``The non-Gaussian Universe'' 
(project number: 200021140236).

\appendix

\section{Lagrangian ESP perturbative expansion and bias parameters}
\label{app:esp}

In this Appendix, we give explicit expressions for the Lagrangian perturbative bias expansion appropriate to excursion 
set peaks, $\delta_\esp^L(\vx)$, and for the third-order Lagrangian ESP bias function, $c_3^L(\vk_1,\vk_2,\vk_3)$, 
which is relevant to the evaluation of the matter-squeezed cross-bispectrum $\Delta B_\text{mhm}$ (see Section \ref{sec:locB}). 
The general methodology can be found in \cite{Desjacques:2012eb,Lazeyras:2015giz}.
We follow common practice and write the perturbative bias expansion in terms of the field $\delta$ and its derivative,
rather than the normalised variables introduced in Section \ref{sec:PC}. 
The effective or mean-field ESP overdensity, in Lagrangian space and up to third order in $\delta_R$ and its derivative, 
is
\begin{align}
\label{eq:loc_exp}
\delta_\esp^L(\vx) &= b_{100}\delta_R(\vx) - b_{010} \nabla^2\delta_R(\vx) - b_{001} \frac{d\delta_R}{dR}(\vx) \\
&+ \frac{1}{2} b_{200} \delta_R^2(\vx) + \frac{1}{2} b_{020} \big[\nabla^2\delta_R(\vx)\big]^2 
+ \frac{1}{2}b_{002} \left(\frac{d \delta_R}{dR}\right)^2\!\!(\vx) \nonumber \\
&-  b_{110}\delta_R(\vx)\nabla^2\delta_R(\vx) - b_{101} \delta_R(\vx) \frac{d\delta_R}{dR}(\vx)
+ b_{011} \nabla^2\delta(\vx)\frac{d\delta_R}{dR}(\vx) \nonumber \\
&+ \chi_1 \big(\nabla\delta_R)^2\!(\vx) + \frac{3}{2}\omega_{10}
\left[\partial_{ij}\delta_R-\frac{1}{3}\delta_{ij}\nabla^2\delta_R\right]^2\!\!\!\!(\vx)
\nonumber \\
&+ \frac{1}{3!} b_{300} \delta_R^3(\vx) - \frac{1}{3!} b_{030} {\left[\nabla^2 \delta_R(\vx)\right]}^3 
- \frac{1}{3!}b_{003} \left(\frac{d\delta_R}{dR}\right)^3\!\!(\vx) 
- \frac{1}{2} b_{210} \delta_R^2(\vx) \nabla^2 \delta_R(\vx)  \nonumber \\
&+  \frac{1}{2} b_{120} \delta_R(\vx)\big[\nabla^2 \delta_R(\vx)\big]^2 
- \frac{1}{2}b_{201} \delta_R^2(\vx)\frac{d \delta_R}{dR}(\vx) 
- \frac{1}{2}b_{021} \big[\nabla^2\delta_R(\vx)\big]^2\frac{d\delta_R}{dR}(\vx)\nonumber \\
&+ \frac{1}{2} b_{102} \delta_R(\vx) \left(\frac{d\delta_R}{dR}\right)^2\!\!(\vx)  
- b_{012}\nabla^2\delta_R(\vx) \left(\frac{d\delta_R}{dR}\right)^2\!\!(\vx) 
+ b_{111} \delta_R(\vx) \nabla^2\delta_R(\vx)\frac{d\delta_R}{dR}(\vx) \nonumber \\
&+ c_{100100}\delta_R(\vx) \big[\nabla \delta_R(\vx)\big]^2 - c_{010100}\big[\nabla \delta_R(\vx)\big]^2 
\nabla^2\delta_R(\vx) - c_{001100}\big[\nabla \delta_R(\vx)\big]^2 \frac{d\delta_R}{dR}(\vx) \nonumber \\ 
&+ \frac{3}{2}c_{100010}\left[\partial_{ij}\delta_R-\frac{1}{3}\delta_{ij}\nabla^2\delta_R\right]^2\!\!\!(\vx) 
\delta_R(\vx) -\frac{3}{2} c_{010010}\left[\partial_{ij}\delta_R-\frac{1}{3}\delta_{ij}\nabla^2\delta_R\right]^2\!\!\!(\vx) 
\nabla^2\delta_R(\vx) \nonumber \\
&- \frac{3}{2} c_{001010}\left[\partial_{ij}\delta_R-\frac{1}{3}\delta_{ij}\nabla^2\delta_R\right]^2\!\!\!(\vx) 
\frac{d \delta_R}{dR}(\vx) + \frac{45}{2 \sqrt{7}}\, \omega_{01}\, 
\left[\partial_{ij}\delta_R-\frac{1}{3}\delta_{ij}\nabla^2\delta_R\right]^3\!\!\!\!(\vx)
\nonumber \;.
\end{align}
A few comments are in order. 
Firstly, the Lagrangian perturbative expansion generally is a series in orthonormal polynomials 
\cite{Lazeyras:2015giz}.
Here, we have only written the term with highest power for simplicity, with the implicit rule that all zero-lag 
correlators should be discarded in the evaluation of 
$\big\langle\delta_\esp^L(\vx_1)\delta_\esp^L(\vx_2)\big\rangle$.
Secondly, we have taken advantage of the fact that the bias coefficients $c_{ijkqlm}$ sometimes simplify. 
In particular, $c_{ijk000}\equiv b_{ijk}$, $c_{000q00}\equiv \chi_q$ and $c_{0000lm}\equiv \omega_{lm}$, where
$b_{ijk}$, $\chi_q$ and $\omega_{lm}$ are given in Section \ref{sec:biases}.
Finally, note that the last term of this expression is proportional to the third invariant 
$J_3=(9/2){\rm tr}(\bar \zeta_{ij}^3)$, where $\bar \zeta_{ij}$ is the traceless part of the Hessian. 
The corresponding bias factor $\omega_{01}$ is given by Eq.~(\ref{eq:Flm}) with $\ell=0$ and $m=1$, i.e.
\be
\sigma_2^3\, \omega_{01} \equiv \frac{5}{3\sqrt{7}}\big\langle J_3\big\lvert {\rm pk} \big\rangle \;,
\ee
where the ensemble average is performed at the locations of density peaks. 

The first- and second-order bias functions $c_1^L(k)$ and $c_2^L(\vk_1,\vk_2)$ are given in Section 
\ref{sec:perturb_bias}. For sake of completeness, the third-order Lagrangian ESP bias function is
\begin{align}
c_3^L(\vk_1,\vk_2&,\vk_3) = \biggl\{b_{300} + b_{030} k_1^2 k_2^2 k_3^2 -b_{003}
\frac{d\ln\tilde{W}_R}{dR}(k_1)\frac{d\ln\tilde{W}_R}{dR}(k_2)\frac{d\ln\tilde{W}_R}{dR}(k_3) \\
&+ b_{210} \Big(k_1^2 + \mbox{2 cyc.}\Big) - b_{201}\bigg[\frac{d\ln\tilde{W}_R}{dR}(k_1) + \mbox{2 cyc.}\bigg]
-b_{021} \bigg[k_1^2 k_2^2 \frac{d\ln\tilde{W}_R}{dR}(k_3) + \mbox{2 cyc.}\bigg]
\nonumber \\
&+ b_{012}\bigg[k_1^2 \frac{d\ln\tilde{W}_R}{dR}(k_2) \frac{d\ln\tilde{W}_R}{dR}(k_3) + \mbox{2 cyc.}\bigg]
+ b_{102}\bigg[\frac{d\ln\tilde{W}_R}{dR}(k_2)\frac{d\ln\tilde{W}_R}{dR}(k_3) + \mbox{2 cyc.}\bigg]
\nonumber \\
&+ b_{120}\Big(k_1^2 k_2^2 + \mbox{2 cyc.}\Big)
- b_{111}\bigg[k_1^2 \frac{d\ln\tilde{W}_R}{dR}(k_2) + \mbox{5 perm.}\bigg]
-2 c_{100100}\Big(\vk_2\cdot\vk_3 + \mbox{2 cyc.} \Big)  \nonumber \\
& -2 c_{010100}\Big[k_1^2 \big(\vk_2\cdot\vk_3) + \mbox{2 cyc}\Big]
- 2 c_{001100}\bigg[\vk_1\cdot\vk_2\frac{d\ln\tilde{W}_R}{dR}(k_3) + \mbox{2 cyc.}\bigg] \nonumber \\
& + c_{100010} \Big[\big(3(\vk_1\cdot\vk_2)^2 - k_1^2 k_2^2\big) + \mbox{2 cyc.}\Big]
+ c_{010010} \Big[k_1^2 \big(3(\vk_2\cdot\vk_3)^2 - k_2^2 k_3^2\big) + \mbox{2 cyc.}\Big] \nonumber \\
& - c_{001010} \bigg[\big(3(\vk_1\cdot\vk_2)^2 - k_1^2 k_2^2\big) \frac{d\ln\tilde{W}_R}{dR}(k_3) + \mbox{2 cyc.}\bigg]
\nonumber \\
&- \frac{5\cdot 3^3}{\sqrt{7}} \omega_{01} 
\bigg[\big(\vk_1\cdot\vk_2\big)\big(\vk_2\cdot\vk_3\big)\big(\vk_3\cdot\vk_1\big)
-\frac{1}{3}\Big[\left(\vk_1\cdot\vk_2\right)k_3^2+\mbox{2 cyc.}\Big]+\frac{2}{9}k_1^2 k_2^2 k_3^2\bigg]\bigg\} 
\nonumber \\
& \times \tilde W_R(k_1)\, \tilde W_R(k_2)\, \tilde W_R(k_3) \nonumber \;.
\end{align} 
In the ESP implementation of \cite{Paranjape:2012jt}, tophat filters appear whenever the indices $i$ or $k$ are 
non-zero, whereas Gaussian filters arise whenever the remaining $j$, $q$, $l$ and $m$ are $\geq 1$.

\section{General expressions for the bispectra}
\label{app:bispectra}

In this appendix we show the general expression of the halo bispectrum $B_\text{hhh}$ and the halo matter matter 
bispectrum $B_\text{hmm}$ in the iPT formalism up to one loop. Our expressions hold in Lagrangian space, but the 
algebra in Eulerian space is analogous. The calculation of $B_\text{hhh}$ has been done in \cite{Yokoyama:2013mta}, 
and our results agree with theirs. We first write the bispectrum for Gaussian initial conditions. 
The halo bispectrum is
\begin{align}
&B_\text{hhh}^{G}(k_1,k_2,k_3;z_i) =c_1^L(k_1)c_1^L(k_2)c_2^L(\vk_1,\vk_2)P_0(k_1)P_0(k_2) + \mathrm{2\ perm.}\\
&+\frac{1}{2}c_1^L(k_1)P_0(k_1)\int\!\frac{\mathrm{d}\vq}{(2\pi)^3}c_3^L(\vk_1,\vk_2-\vq,\vq)c_2^L(\vk_2-\vq,\vq)P_0(q)P_0(|\vk_2-\vq|)+\mathrm{5\ perm.}\nonumber\\
&+\int\!\frac{\mathrm{d}\vq}{(2\pi)^3}c_2^L(-\vq,\vq+\vk_1)c_2^L(\vq+\vk_1,\vk_2-\vq)c^L_2(\vq,\vk_2-\vq)P_0(q)P_0(|\vk_1+\vq|)P_0(|\vk_2-\vq|) \nonumber \,.
\end{align}
An analogous expression holds for $B_\text{hmm}$ though some of the permutations are no longer symmetric
\begin{align}
&B_\text{hmm}^{G}(k_1,k_2,k_3;z_i) = \left(\frac{D(z_i)}{D(z_\ast)}\right)^2c_2^L(\vk_2,\vk_3)P_0(k_2)P_0(k_3)\nonumber\\
& + \left(\frac{D(z_i)}{D(z_\ast)}\right)^3 c_1^L(k_1)F_m^{(2)}(\vk_1,\vk_2)P_0(k_1)P_0(k_2) + (k_2 \leftrightarrow k_3)\nonumber\\
&+\frac{1}{2}\left(\frac{D(z_i)}{D(z_\ast)}\right)^3P_0(k_2)\int\!\frac{\mathrm{d}\vq}{(2\pi)^3}c_3^L(\vk_2,\vk_3-\vq,\vq)F_m^{(2)}(\vk_3-\vq,\vq)P_0(q)P_0(|\vk_3-\vq|)+(k_2\leftrightarrow k_3)\nonumber\\
&+\frac{1}{2}\left(\frac{D(z_i)}{D(z_\ast)}\right)^4P_0(k_2)\int\!\frac{\mathrm{d}\vq}{(2\pi)^3}F_m^{(3)}(\vk_2,\vk_1-\vq,\vq)c_2^L(\vk_1-\vq,\vq)P_0(q)P_0(|\vk_3-\vq|)+(k_2\leftrightarrow k_3)\nonumber\\
&+\frac{1}{2}\left(\frac{D(z_i)}{D(z_\ast)}\right)^5c_1^L(k_1)P_0(k_1)\int\!\frac{\mathrm{d}\vq}{(2\pi)^3}F_m^{(3)}(\vk_1,\vk_2-\vq,\vq)F_m^{(2)}(\vk_2-\vq,\vq)P_0(q)P_0(|\vk_3-\vq|)+(k_2\leftrightarrow k_3)\nonumber\\
&+\left(\frac{D(z_i)}{D(z_\ast)}\right)^4\int\!\frac{\mathrm{d}\vq}{(2\pi)^3}c_2^L(-\vq,\vq+\vk_1)F_m^{(2)}(\vq+\vk_1,\vk_2-\vq)F_m^{(2)}(\vq,\vk_2-\vq)\nonumber\\
&\phantom{D^2(z_h)D^4(z_\ast)\mathrm{d}\vq c_2^L(-\vq,\vq+\vk_1)F_m^{(2)}(\vq+\vk_1,\vk_2}\times P_0(q)P_0(|\vk_1+\vq|)P_0(|\vk_2-\vq|)\,.
\end{align}
The halo bispectrum in the presence of non-Gaussian initial conditions is modified by
\begin{align}
&\Delta B_\text{hhh}^{\mathrm{NG}}(k_1,k_2,k_3;z_i) = c_1^L(k_1) c_1^L(k_2) c_1^L(k_3)
\,B_0(k_1,k_2,k_3)  \nonumber \\ 
& + c_1^L(k_1)\! \int\!\frac{d^3q}{(2\pi)^3}\, 
c_2^L(\vq,\vk_2\!-\!\vq)\, c_2^L(\vk_1\!+\!\vq,\vk_2\!-\!\vq)\, P_0(|\vk_2\!-\!\vq|)\,
B_0(k_1,q,|\vk_1\!+\!\vq|) + 2 \, {\rm perm.}  \nonumber \\
& +\frac{1}{2} \,\ c_1^L(k_1) \,c_1^L(k_2) \,P_0(k_1)\! \int\! \frac{d^3q}{(2\pi)^3}\, 
c_3^L(\vk_1,\vq,\vk_2\!-\!\vq)\, B_0(k_2,q,|\vk_2\!-\!\vq|) + (k_1
\leftrightarrow k_2) + 2 \, {\rm perm.}  \nonumber \\
& +\frac{1}{2}\, P_0(k_1)\, c_1^L(k_1) \left[ c_2^L(\vk_1,\vk_2)\! \int \!\frac{d^3q}{(2\pi)^3}\, c_2^L(\vq,\vk_2\!-\!\vq)\,
B_0(k_2,q,|\vk_2\!-\!\vq|) + (k_2 \leftrightarrow k_3) \right] + 2 \, {\rm
  perm.} \nonumber \\
& +\frac{1}{2} \ c_1^L(k_1)\, c_1^L(k_2)\!  \int\!
\frac{d^3q}{(2\pi)^3}\,c_2^L(\vq,\vk_3\!-\!\vq)\, T_0(\vk_1,\vk_2,\vq,\vk_3\!-\!\vq) + 2 \, {\rm perm.}\,,
\label{eq:full_bispec}
\end{align}
An analogous expression holds for $\Delta B_\text{hmm}$
\begin{align}
&\Delta B_\text{hmm}^{NG}(k_1,k_2,k_3;z_i) = 
\left(\frac{D(z_i)}{D(z_\ast)}\right)^2 c_1^L(k_1)
\,B_0(k_1,k_2,k_3)  \nonumber \\ 
& + \left(\frac{D(z_i)}{D(z_\ast)}\right)^4\frac{1}{2}\, P_0(k_1)\, c_1^L(k_1) \left[ F^{(2)}_m(\vk_1,\vk_2)\! \int \!\frac{d^3q}{(2\pi)^3}\, F^{(2)}_m(\vq,\vk_2\!-\!\vq)\,
B_0(k_2,q,|\vk_2\!-\!\vq|) + (k_2 \leftrightarrow k_3) \right] \nonumber \\
& +\left(\frac{D(z_i)}{D(z_\ast)}\right)^3\frac{1}{2}\, \,\left[ P_0(k_3)c_2^L(\vk_2,\vk_3)\! \int \!\frac{d^3q}{(2\pi)^3}\, F^{(2)}_m(\vq,\vk_2\!-\!\vq)\,
B_0(k_2,q,|\vk_2\!-\!\vq|) + (k_2 \leftrightarrow k_3) \right] \nonumber \\
&  + \left(\frac{D(z_i)}{D(z_\ast)}\right)^3\frac{1}{2}\, \,\left[ P_0(k_3)F^{(2)}_m(\vk_1,\vk_2)\! \int \!\frac{d^3q}{(2\pi)^3}\, c_2^L(\vq,\vk_1\!-\!\vq)\,
B_0(k_1,q,|\vk_1\!-\!\vq|) + (k_2 \leftrightarrow k_3) \right] \nonumber \\
&  +\left(\frac{D(z_i)}{D(z_\ast)}\right)^4c_1^L(k_1)\! \int\!\frac{d^3q}{(2\pi)^3}\, 
F_m^{(2)}(\vq,\vk_2\!-\!\vq)\, F_m^{(2)}(\vk_1\!+\!\vq,\vk_2\!-\!\vq)\, P_0(|\vk_2\!-\!\vq|)\,
B_0(k_1,q,|\vk_1\!+\!\vq|)\nonumber \\
& + \left(\frac{D(z_i)}{D(z_\ast)}\right)^3\frac{1}{2}\left[\int\!\frac{d^3q}{(2\pi)^3}\, 
c_2^L(\vq,\vk_1\!-\!\vq)\, F_m^{(2)}(\vk_2\!+\!\vq,\vk_1\!-\!\vq)\, P_0(|\vk_1\!-\!\vq|)\,
B_0(k_2,q,|\vk_2\!+\!\vq|) + (k_2 \leftrightarrow k_3)\right]\nonumber \\
&  +\left(\frac{D(z_i)}{D(z_\ast)}\right)^3\frac{1}{2}\left[\int\!\frac{d^3q}{(2\pi)^3}\, 
F_m^{(2)}(\vq,\vk_2\!-\!\vq)\, c_2^L(\vk_3\!+\!\vq,\vk_2\!-\!\vq)\, P_0(|\vk_2\!-\!\vq|)\,
B_0(k_3,q,|\vk_3\!+\!\vq|) + (k_2 \leftrightarrow k_3)\right]\nonumber \\
& +\left(\frac{D(z_i)}{D(z_\ast)}\right)^4\frac{1}{2}c_1^L(k_1)P_0(k_1)\!\left[ \int\! \frac{d^3q}{(2\pi)^3}\, 
F_m^{(3)}(\vk_1,\vq,\vk_2\!-\!\vq)\, B_0(k_2,q,|\vk_2\!-\!\vq|) + (k_2
\leftrightarrow k_3)\right] \nonumber \\
& +\left(\frac{D(z_i)}{D(z_\ast)}\right)^4\frac{1}{2}c_1^L(k_1)P_0(k_2)\!\left[ \int\! \frac{d^3q}{(2\pi)^3}\, 
F_m^{(3)}(\vk_2,\vq,\vk_1\!-\!\vq)\, B_0(k_1,q,|\vk_1\!-\!\vq|) + (k_2
\leftrightarrow k_3)\right] \nonumber \\
&+\left(\frac{D(z_i)}{D(z_\ast)}\right)^2\frac{1}{2}P_0(k_3)\!\left[ \int\! \frac{d^3q}{(2\pi)^3}\, 
c_3^L(\vk_3,\vq,\vk_2\!-\!\vq)\, B_0(k_2,q,|\vk_2\!-\!\vq|) + (k_2
\leftrightarrow k_3)\right] \nonumber \\
& +\left(\frac{D(z_i)}{D(z_\ast)}\right)^3\frac{1}{2} \ c_1^L(k_1)\!  \left[\int\!
\frac{d^3q}{(2\pi)^3}\,F_m^{(2)}(\vq,\vk_3\!-\!\vq)\, T_0(\vk_1,\vk_2,\vq,\vk_3\!-\!\vq) + (k_2 \leftrightarrow k_3)\right] \nonumber \\
& +\left(\frac{D(z_i)}{D(z_\ast)}\right)^2\frac{1}{2}\int\!
\frac{d^3q}{(2\pi)^3}\,c_2^L(\vq,\vk_1\!-\!\vq)\, T_0(\vk_3,\vk_2,\vq,\vk_1\!-\!\vq) \,.
\label{eq:full_bispec_hmm}
\end{align}

\section{The S12i model: theoretical considerations}
\label{sec:ESbias_PBS}

In this section, we present a model for the halo bispectrum in which the NG bias parameters are 
explicitly derived from a peak-background split.  We will first review the derivation of the NG 
bias parameters following \cite{Scoccimarro:2011pz}, before discussing the bispectrum prescription
used in the analysis.  Because our model is inspired by \cite{Scoccimarro:2011pz}, we will refer 
to it as S12i.  

\subsection{Excursion set bias from a peak-background split}

Ref. \cite{Scoccimarro:2011pz} took advantage of the peak-background split to derive the PNG bias 
parameters within the excursion set theory.  For the Gaussian case, the biases can be obtained
upon considering the modulation of local density fluctuations by a long wavelength perturbation 
$\delta_{\rm l} $. The effect of $\delta_{\rm l} $ can be implemented through a position-dependent
offset in the collapse threshold \cite{Kaiser:1984sw,Mo:1995cs,Sheth:1999mn}.

In the local PNG model, the short mode is given by
\beq
\Phi_{\rm s} ( \mb{q} ) = \phi_{\rm s} ( \mb{q} ) +  K_{\mb{q }}^S [ \phi, \phi ] , 
\eeq
with the coupling term
\beq
\label{eq:Ks_coupling} 
K_{\mb{q }}^S [ \phi, \phi ] =  \fnl \int\! \frac{ d^3 p_1}{ ( 2 \pi )^3 } \frac{ d^3 p_2}{ ( 2 \pi)^3 }    \Ddel( \mb{q} - \mb{p}_{12} ) 
\big[  \phi_{\rm s} ( \mb{p}_1 )   \phi_{\rm s} ( \mb{p}_2 ) + 2  \phi_{\rm l} ( \mb{p}_1 )   \phi_{\rm s} ( \mb{p}_2 )    \big], 
\eeq 
where  $\phi_{\rm s} $ and $ \phi_{\rm l} $ denote the Gaussian short and long modes.  

We focus on the coupling $ \phi_{\rm l} \phi_{\rm s} $, which induces a modulation of the small-scale 
cumulants, such as the variance and skewness. 
This was used in Ref.~\cite{Scoccimarro:2011pz} to derive the bias in the presence of PNG.

We begin by summarising  the general rules for computing the $n^{\rm th}$ cumulant, 
$\langle \Phi^n \rangle_{\phi_{\rm l} } $, in the case of a PNG of the local quadratic type. 
Here, the subscript $ \phi_{\rm l} $ indicates that the long mode is held fixed in the 
expectation value. The rules are as follows:
\begin{itemize}
  \item Write down $n$ points representing the short modes. 
   
  The long modes can only arise from  the  $\phi_{\rm l} \phi_{\rm s}  $ coupling, 
  and, hence, can be thought of as arising from the short modes. 

  \item To each of the short mode, we can attach another short mode through the
    $\phi_{\rm s} \phi_{\rm s}$ coupling in Eq.~(\ref{eq:Ks_coupling}). If no short mode is attached, then 
    it is simply a random Gaussian wavemode. If another short mode is attached, then it contributes a 
    factor of $\fnl$.   

  \item Connect the short modes together with the power spectrum, so that the resulting diagram is connected. 
    Note that we do not need to care about the long modes because they are fixed in the ensemble average  
    $\langle \dots  \rangle_{\phi_{\rm l}} $. 

  \item Finally, we can also choose to attach a long mode to each Gaussian short mode. If so, then the 
    coupling contributes a factor of $\fnl$.  

\end{itemize}
From the above diagrammatic rule, we see that the topology of the diagrams is entirely determined by the 
short modes, while the number of long modes is determined by the number of free short modes at the end. 
We also note that powers of $\fnl$ increase for higher order cumulants because we need more short mode 
couplings to form a connected diagram.  

The leading PNG correction that is modulated by the long mode arises from the variance and reads 
\beqa
\label{eq:variance_phil}
\langle \delta_{\rm s}^2  (\mb{x} ) \rangle_{\phi_{\rm l}  }  &=& \int\! \frac{ d^3 q_1}{ (2 \pi)^3 }  \frac{d^3 q_2}{(2 \pi)^3 } e^{ i \mb{q}_{12} \cdot \mb{x} } \alpha( \mb{q}_1 )  \alpha( \mb{q}_2 )    \langle \Phi_{\rm s} ( \mb{q}_1 )  \Phi_{\rm s} ( \mb{q}_2 )  \rangle_{\phi_{\rm l}  }  \nn \\
&=&  4  \fnl \int\!  \frac{ d^3 q }{ (2 \pi )^3 }  \alpha (q ) P_\phi ( q )  \varphi_{\mb{q} } ( \mb{x} ) \;,
\eeqa
with  $ \alpha$  and  $ \varphi $ defined as
\beqa
\alpha( k, z ) & = &  \mathcal{M}(k)  D(z)  ,  \nn \\
 \varphi_{\mb{q} } ( \mb{x} ) & =&  \int\! \frac{ d^3 p}{ (2 \pi)^3 } e^{i \mb{p} \cdot \mb{x} } \alpha(|\mb{p} - \mb{ q} |) K^s_{ \mb{p} - \mb{ q} }  ( - \mb{q} , \mb{p} ) \phi_{\rm l}( \mb{p} ) \;. 
\eeqa
The density contrast of the halo fluctuation field can then be expanded in terms of the long mode by means of a functional derivative.  In particular, the linear halo bias term is given by
\beqa
&& \frac{ 1 }{ N } \partial_M \int_{-\infty}^{ \delta_{\rm c} }\! d \delta_{\rm s}  \int\! \frac{ d^3 k }{( 2 \pi )^3  }  \sum_{j=1}^{\infty}  \frac{ \mathcal{ D} \mathcal{C}_j(\mb{x}) }{ \mathcal{D} \phi_{\rm l}( \mb{k} ) } \frac{ \partial \Pi(\delta_{\rm s}, \sigma_M^2, \phi_{\rm l} ) }{ \partial  \mathcal{C}_j(\mb{x})  }  \Big|_{\phi_{\rm l} =  0} \phi_{\rm l}( \mb{k} ),  
 \eeqa  
where $\mathcal{C}_j $ denotes the $ j^{\rm th}$ cumulant,  $ \langle  \delta_{\rm s}^j  \rangle_{ \phi_{l }  }$, $\sigma_M$ is the rms variance of matter fluctuation on the halo mass scale $M$, and the normalisation factor $N$ is given by the first crossing distribution in the absence of the long mode $ \phi_{\rm l} $, 
\beq
N =  \partial_M \int_{- \infty}^{\delta_{\rm c} }\! d \delta_{\rm s} \Pi( \delta_{\rm s}, \sigma^2_M , 0 ) \;. 
\eeq
In this formalism, the linear halo bias already depends on all the cumulants. For  $j = 1$, we recover the standard, Gaussian peak-background split bias $b_1^{(1)} $. The leading PNG correction arises from the variance and the corresponding NG bias correction can be written
\beq
\label{eq:b1_2NG}
b_1^{(2)}(k)  =  \frac{ \partial_{ \sigma_M^2 } [ I (k) \mathcal{F} ]  }  { \alpha (k)  \mathcal{F}  } \;,
\eeq
where  $\mathcal{F}$ is the first-crossing distribution and the integral $I$ is defined as 
\beq
\label{eq:integral_I}
I(k)  =  4 \fnl \int\! \frac{ d^3 p}{ (2 \pi )^3 }  \alpha( p) \alpha ( | \mb{k} -  \mb{p} | )  K^s_{ \mb{p} }( \mb{k} - \mb{p} , - \mb{k}  )    P_{\phi}(| \mb{k} - \mb{p} |) \;.  
\eeq
In terms of the mass function, we can write  $b_1^{(2)} $ as
\beq
\label{eq:b1_2_massderivative_mfnweight}
b_1^{(2)} =   \frac{ \partial_{ M}\Big[  I \frac{d n }{ d \ln M } \big( \frac{d \sigma^2_M  }{ d M } \big)^{-1 }   \Big]   }{ \alpha(k)  \frac{ d n  }{ d \ln M } } . 
\eeq
As emphasised in \cite{Scoccimarro:2011pz}, under the assumption of Markovian random walks and a universal mass function, Eq.~(\ref{eq:b1_2NG}) can be written in the well-known form \cite{Dalal:2007cu,Slosar:2008hx,Matarrese:2008nc}
\beq
\label{eq:b1_NG_standard}
 b_1^{(2)}(z_\ast) =   \frac{3 \fnl \Omega_{\rm m} H_0^2 \delta_{\rm c} }{  k^2 T(k) }\frac{  b_1^{(1)}(z_\ast) }{ D(z_\ast) }
\eeq
upon taking the low-$k$ limit of $I$. 

The more general prediction Eq.~(\ref{eq:b1_2_massderivative_mfnweight}) can be computed  numerically using the mass function measured in an $N$-body simulation. In this work, we have used halos with at least 20 particles. However, the low mass end of this halo mass function significantly underestimates the true mass function even though the halo clustering properties, which are the focus of our analysis, are well reproduced. We have checked that, when the numerical mass function is accurately determined, Eq.~(\ref{eq:b1_2_massderivative_mfnweight}) often predicts the halo scale-dependent PNG bias more accurately than Eq.~(\ref{eq:b1_NG_standard}). Nevertheless, we use Eq.~(\ref{eq:b1_NG_standard}) for the computations in the main text, which is appropriate to the $M\gtrsim M_\star$ halos we consider.

\subsection{Analytic prediction for the halo-matter bispectrum}

We write the halo density as 
\beq
\delta_{\rm h}( \mb{k}) = \Big[b_1^{(1)} +  b_1^{(2)}( k ) \Big] \delta_{\rm m}^R( \mb{k} ) 
+  \frac{ \mathfrak{b}_2 }{  2 } \star \delta_{\rm m}^R \star \delta_{\rm m}^R ( \mb{k} ) \;, 
\eeq
where $ \mathfrak{b}_2 $ includes not only the local bias $b_2 $, but also nonlocal terms \cite{Sheth:2012fc}. For the model presented here, $ \delta_{\rm m}^R$ is the {\it nonlinear} matter density field smoothed with a top-hat filter.  In \cite{Chan:2015zjt}, the effective window function of halos was found to be  more extended than a top hat. 

We can write the tree-level cross bispectrum as
\beqa
\langle \delta_{\rm m} ( \mb{k}_1, z_i )   \delta_{\rm m} ( \mb{k}_2 , z_i)    \delta_{\rm h} ( \mb{k}_3 ,z_i )  \rangle'
& = &  \left(\frac{D(z_i)}{D(z_\ast)}\right)^2 \Big[ b_1^{(1)}(z_\ast) +  b_1^{(2)}( k_3, z_\ast ) \Big]  
\tilde W_R(k_3)  B_0(k_1,k_2,k_3) \nn \\
&& +  \left(\frac{D(z_i)}{D(z_\ast)}\right)^3 b_1^{(2)} (k_3,z_\ast ) \tilde  W_R(k_3)  B_{\rm m}^G( k_1,k_2,k_3; z_\ast )   \nn \\
& & +   B_{\rm mmh}^G (  k_1,k_2,k_3; z_i ) \;, 
\eeqa
where $  B_{\rm mmh}^G$ is the bispectrum in the Gaussian case, which we will assume is given by the Gaussian simulation,  
 and $ B_{\rm m}^G $ is the DM bispectrum in the case of Gaussian initial conditions, 
which is
\beqa
\label{eq:BmNG_Gaussian}
 B_{\rm m}^G(k_1,k_2,k_3; z_\ast) &=& 2 F_2( \mb{k}_1, \mb{k}_2 ) P_0(k_1) P_0(k_2) + \, 2 \, \mathrm{cyc.}
\end{eqnarray}
at tree-level.

The second order bias receives a contribution from PNG, $b_2^{(2)} $, which can be computed under the same assumptions that lead to Eq.~(\ref{eq:b1_NG_standard}). It is given by \cite{Scoccimarro:2011pz}
\beq
b_2^{(2) }(k_1, k_2 ) = \frac{ 4 \fnl^2  \delta_{\rm c}  b_1^{(1)} }{ \alpha(k_1)  \alpha(k_2) } . 
\eeq
However, we found that this contribution is at least two orders of magnitude smaller than the term proportional to $ b_1^{(1)} ( z_\ast )  B_{\rm m}^{NG} (k_1, k_2, k_3, z_\ast ) $, and we thus shall neglect it here. Note also that, although  Eq.~(\ref{eq:b1_NG_standard})  for $b_1^{(2)}  $ is obtained in the low-$k$ approximation (the low-$k$ limit of Eq.~(\ref{eq:integral_I})), we have checked that the effects are negligible in the final results even for the matter-squeezed cases in which the wavenumbers are not small.

Finally, as discussed in the main text, for the matter-squeeze case, the loop contribution due to the third order bias is the dominant one, and reads  
\begin{align}
\label{eq:b3S12i}
B_{b_3 \rm{ loop}  } &=  \langle \delta_{\rm m}( \mb{k}_1)  \delta_{\rm m}( \mb{k}_2)  \frac{b_3^{(1)} }{6 } \delta_{\rm m}^R \star\delta_{\rm m}^R \star\delta_{\rm m}^R ( \mb{k}_3)  \rangle'  \\
& =   \frac{ b_3^{(1)}}{2} \tilde W_R(k_1) P_0(k_1) \int\!  \frac{d^3 q}{ (2 \pi )^3} \tilde W_R(q) \tilde W_R( | \mb{k}_2  - \mb{q} |)   B_0(- \mb{k}_2, \mb{q}, \mb{k}_2 - \mb{q} ) + ( \mb{k}_1 \leftrightarrow   \mb{k}_2   ) \;.  \nonumber
\end{align} 
	 
\begin{figure}
\centering
\includegraphics[width=0.75\textwidth]{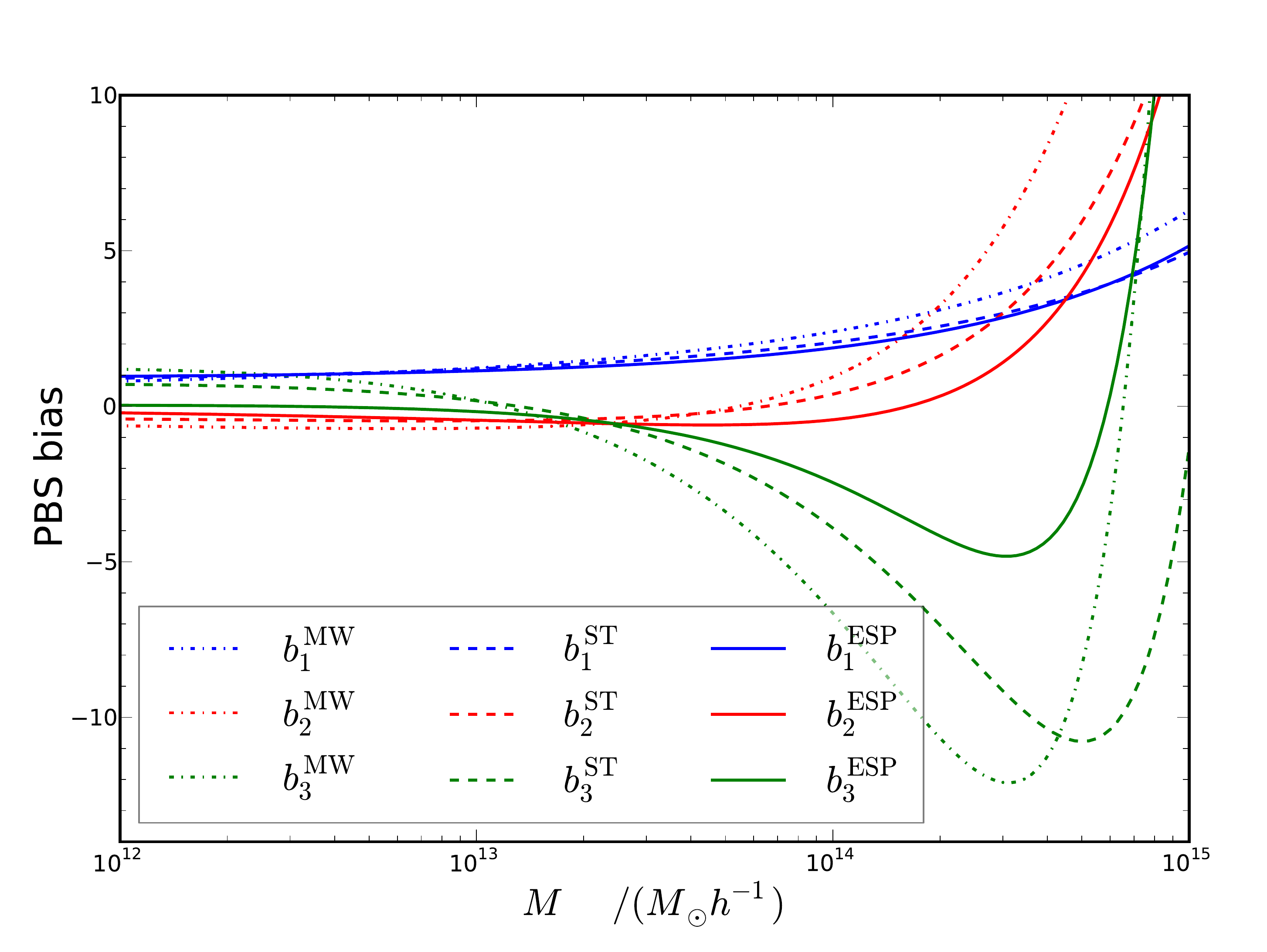}
\caption{   The MW (dotted-dashed), ST (dashed), and ESP (solid) bias parameters $b_1$ (blue),  $b_2$ (red), and $b_3$ (green)  are compared.   }
\label{fig:PBS_bias_MW_ST_ESP}
\end{figure}

Before computing the bispectrum terms, it is instructive to check the scale-independent bias parameters obtained from different bias schemes.  In Fig.~\ref{fig:PBS_bias_MW_ST_ESP}, we compare three prescriptions for the PBS Gaussian bias parameters: MW \cite{Mo:1995cs}, ST \cite{Scoccimarro:2000gm}, and the scale-independent ESP bias. The MW and ST bias are derived from the Press-Schechter mass function \cite{Press:1973iz} and the Sheth-Tormen  mass function \cite{Sheth:1999mn}, respectively.  For the ESP bias, we use $b_n^{(1)}\equiv b_{n00}$. We have assumed $z_\ast=0$. As can be seen, the ST results often fall between those of MW and ESP.  The difference between these prescriptions increases with the order of the bias parameter. Consequently, $B_{ b_3  \mathrm{ loop}} $ turns out to be fairly sensitive to the exact value of $ b_3^{(1)}$.  We also note that $b_3^{(1)}$ varies rapidly  in the high peak regime, which implies that a small error can lead to large differences in the prediction. This might explain why, for large halo mass, including the $b_3$-loop rarely improves the agreement with the simulations.
When the bias parameters are computed using MW and ST prescriptions, the inclusion of the $b_3^{(1)}$-loop often worsens the agreement with the simulations relative to the tree-level-only prediction. The ST results perform marginally better than the MW ones. On the other hand, the ESP results often lead to a better agreement with the numerical data. This is likely due to the fact that our ESP implementation is designed to reproduce the clustering of SO halos, which are the ones analysed here.

Finally we summarise the cross bispectrum model model used  in the main text. For the halo squeezed  case,  we adopt the tree-level only bispectrum 
\beqa
\lim_{k_{\rm l}\rightarrow 0}\Delta B^{NG}_{\text{hmm}} (k_{\rm l}, k_{\rm s},k_{\rm s};z_i)   
&= &  \left(\frac{D(z_i)}{D(z_\ast)}\right)^2\Big[ b_1^{(1)}(z_\ast) +  b_1^{(2)}( k_{\rm l}, z_\ast ) \Big]  \tilde{W}_R(k_{\rm l})  B_0 ( k_{\rm s},k_{\rm s},k_{\rm l}) \nn\\     
&&+ \left(\frac{D(z_i)}{D(z_\ast)}\right)^3 b_1^{(2)} (k_{\rm l},z_\ast )  \tilde{W}_R(k_{\rm l})  B_{\rm m}^G( k_{\rm s},k_{\rm s},k_{\rm l}; z_\ast ) \;,
\eeqa
while in the matter-squeezed case we include also the $b_3$-loop 
\begin{align}
\label{eq:DeltaBNG_Msq_PBS}
\lim_{k_{\rm l}\rightarrow 0}\Delta B^{NG}_{\text{mhm}}(k_{\rm l},k_{\rm s},k_{\rm s};z_i) 
&=\left(\frac{D(z_i)}{D(z_\ast)}\right)^2 \Big[ b_1^{(1)}(z_\ast) +  b_1^{(2)}( k_{\rm s}, z_\ast ) \Big]  
\tilde{W}_R(k_{\rm s})\,  B_0( k_{\rm s},k_{\rm l},k_{\rm s}) \\
&+  \left(\frac{D(z_i)}{D(z_\ast)}\right)^3 b_1^{(2)} (k_{\rm s},z_\ast )  \tilde{W}_R(k_{\rm s})\,  
B_{\rm m}^G( k_{\rm s},k_{\rm l},k_{\rm s}; z_\ast ) \nonumber \\
& +\left(\frac{D(z_i)}{D(z_\ast)}\right)^2\frac{ b_3^{\rm ESP}(z_\ast)  }{2} \tilde{W}_R(k_{\rm s}) 
P_0(k_{\rm s}) \int\!  \frac{d^3 q}{ (2 \pi )^3}  \tilde{W}_R(q) \tilde{W}_R( | \mb{k}_{\rm l}  - \mb{q} |)   
\nonumber \\
&\quad \times B_0(- \mb{k}_{\rm l}, \mb{q}, \mb{k}_{\rm l} - \mb{q} ) \;.
\end{align}
We note that we will use $b_3^{\rm ESP} $ in the loop calculations.   

\bibliography{SCINC}

\end{document}